\documentclass{article}

\usepackage{arxiv}

\usepackage[utf8]{inputenc} % allow utf-8 input
\usepackage[T1]{fontenc}    % use 8-bit T1 fonts
\usepackage{hyperref}       % hyperlinks
\usepackage{url}            % simple URL typesetting
\usepackage{booktabs}       % professional-quality tables
\usepackage{amsfonts}       % blackboard math symbols
\usepackage{nicefrac}       % compact symbols for 1/2, etc.
\usepackage{microtype}      % microtypography
\usepackage{lipsum,enumerate,bm,mathrsfs,yfonts}
\usepackage{graphicx,amsmath,multicol,multirow}
\usepackage{tabularx}

\usepackage{url}
\usepackage{amsthm}
\usepackage[style=apa, backend=biber, natbib=true]{biblatex}
\usepackage{adjustbox,xr,comment} 

\usepackage{moreverb}
\usepackage{bm}
\usepackage{hyperref,bm}
\usepackage{xr,import,rotating}
\usepackage{float}
\usepackage{spreadtab}
\usepackage{tikz}
\usetikzlibrary{shapes.geometric, arrows, shadows}
\usepackage{adjustbox} 
\usepackage{graphicx,makecell} 

\usepackage{siunitx}
\theoremstyle{definition}

\newtheorem{theorem}{Theorem}[section]   % numbered within sections
\theoremstyle{definition}

\newtheorem{proposition}[theorem]{Proposition}
\newtheorem{assumption}[theorem]{Assumption}

\theoremstyle{remark}

\makeatletter

\newcommand*{\addFileDependency}[1]{% argument=file name and extension
\typeout{(#1)}
\@addtofilelist{#1}
\IfFileExists{#1}{}{\typeout{No file #1.}}
}\makeatother
\title{Estimands and doubly robust estimation for cluster-randomized trials with survival outcomes}

\author{
 Xi Fang \\
    Department of Biostatistics \\
    Yale School of Public Health \\
    New Haven, CT, USA\\
   \And
Bingkai Wang\\
    Department of Biostatistics\\
    School of Public Health\\
    University of Michigan \\
    Ann Arbor, MI, USA
\And
Liangyuan Hu\\
Department of Biostatistics and Epidemiology\\
Rutgers School of Public Health \\
Piscataway, NJ, USA
\And
 Fan Li \\
Department of Biostatistics \\
Yale School of Public Health \\
New Haven, CT, USA\\
  \texttt{fan.f.li@yale.edu} \\
}

\begin{document}
\maketitle
\begin{abstract}
Cluster-randomized trials (CRTs) are experimental designs where groups or clusters of participants, rather than the individual participants themselves, are randomized to intervention groups. Analyzing CRT requires distinguishing between treatment effects at the cluster level and the individual level, which requires a clear definition of the estimands under the potential outcomes framework. For analyzing survival outcomes, it is common to assess the treatment effect by comparing survival functions or restricted mean survival times between treatment groups. In this article, we formally characterize cluster-level and individual-level treatment effect estimands with right-censored survival outcomes in CRTs and propose doubly robust estimators for targeting such estimands. Under covariate-dependent censoring, our estimators ensure consistency when either the censoring model or the outcome model is correctly specified, but not necessarily both. We explore different modeling options for the censoring and outcome models to estimate the censoring and survival distributions, and investigate a deletion-based jackknife method for variance and interval estimation. Extensive simulations demonstrate that the proposed methods perform adequately in finite samples. Finally, we illustrate our method by analyzing a completed CRT with survival endpoints.
\end{abstract}

% keywords can be removed
\keywords{Cluster randomized trial; estimands; double robustness; survival function; informative cluster size; restricted mean survival time}

\section{Introduction}\label{intro}

%% p1: background on two estimands in CRT
Cluster-randomized trials (CRTs) represent a distinct class of experimental designs wherein groups or clusters of individuals---such as in schools, hospitals, or communities, rather than the individuals themselves, are randomized to intervention groups \citep{murray1998design, turner2017review1}. This design is particularly useful in settings where individual randomization is impractical or when intervention naturally applies to the entire clusters. Typically, the data observations collected within the same cluster tend to be more correlated, resulting in a positive intracluster correlation coefficient (ICC). Although the classic literature on the design and analysis of CRTs has emphasized the importance of accounting for ICC, recent discussions have also highlighted the importance of estimands specification \citep{kahan2023estimands} and estimand-driven analysis \citep{kahan2023demystifying}. These discussions are consistent with the regulatory framework established in the ICH E9(R1) addendum \citep{european2020e9}, which addresses key attributes for estimands and summarizes different strategies for handling intercurrent events in clinical trials.

Due to the hierarchical data structure, different weighting schemes can yield different average treatment effect estimands in a CRT. For example, \cite{kahan2023estimands} explained that both the cluster-average treatment effect (c-ATE) and the individual-level average treatment effect (i-ATE) can be relevant, depending on the scientific question \citep{kahan2023informative}. Specifically, the i-ATE estimand gives equal weight to each individual across clusters, aiming to mimic the treatment effect had individual randomization been possible. In contrast, c-ATE gives equal weight to each cluster irrespective of its size to quantify average change for the population of clusters or providers. These two estimands carry different magnitudes in the presence of informative cluster size; that is, when the cluster size is marginally correlated with the cluster-specific treatment effect. In particular, it has been shown that, under informative cluster size, the treatment coefficients reported from the linear mixed model or generalized estimating equations with the exchangeable working correlation structure do not converge to either c-ATE or i-ATE, and one needs to align analytic methods with the estimand of interest \citep{kahan2023demystifying,wang2022two}. There is an emerging literature devoted to estimand-driven analysis of CRTs, with a central recommendation that estimand specification---reflecting the scientific question---should guide the choice of statistical analysis, rather than the other way around. When targeting c-ATE or i-ATE with a continuous or binary outcome, appropriate methods include independent generalized estimating equations (with proper cluster size weight adjustment as necessary) \citep{kahan2023estimands,kahan2023demystifying}, weighted g-computation \citep{wang2024model}, model-robust standardization \citep{li2025model}, de-biased machine learning \citep{wang2024model,wang2024handling} and targeted maximum likelihood estimation \citep{benitez2023defining,nugent2024blurring,balzer2023two}. 

%p2 Existing methods for survival outcomes in CRT and limitation
Survival outcomes are not uncommon in CRTs \citep{caille2021methodological}, but require approaches that appropriately address right censoring. Several methods have been developed to support the design and analysis of CRTs with a censored survival endpoint. For example, to analyze survival outcomes in CRTs, model-based methods such as frailty and marginal Cox models are often used \citep{caille2021methodological}. The frailty model incorporates a random effect, or ``frailty'', to account for unmeasured heterogeneity across clusters \citep{ripatti2000estimation, terry2000modeling}. Alternatively, the marginal Cox model typically assumes working independence and considers the robust sandwich variance estimator for inference \citep{wei1989regression}; \cite{wang2023improving} further proposed the bias-corrected sandwich variance estimators to offer small-sample adjustments. These model-based methods primarily estimate hazard ratio parameters. However, hazard ratio may have an ambiguous counterfactual interpretation due to the built-in selection bias \citep{martinussen2020subtleties,fay2024causal}. Alternative effect measures such as the survival probability and the restricted mean survival time (RMST), which are defined, respectively, as the marginal probability of surviving beyond time \(t\) and the expected survival time up to a specified time point \(t\), offer clearer causal interpretations \citep{fay2024causal}. 
With independent data, \cite{zhang2012contrasting} and \cite{bai2013doubly} proposed doubly robust methods to estimate treatment-specific survival probabilities in clinical trials. \cite{zhong2022restricted} and \cite{chen2023clustered} proposed model-based estimates for RMST under independent and clustered data, respectively. Relatedly, \cite{young2020causal} formalized estimands under a potential outcomes framework for competing risks with independent data. They also emphasized that estimands should be defined independently of estimation strategies, especially in time-to-event settings where traditional models often carry obscure causal interpretation.

% SPCE - cluster-average SPCE = cluster-level survival probability causal effect (c-SPCE vs i-SPCE)
% RMST = cluster-level restricted mean survival time (c-RMST)

%% p3: our contribution and the organization 
Despite the existing literature on survival analysis in CRTs, little attention has been given to estimands, nor robust methods for estimand-aligned methods with censored survival outcomes. To fill in this gap, we propose versions of cluster-average survival probability causal effect (c-SPCE) and individual-level survival probability causal effect (i-SPCE). Additionally, we introduce the cluster-level RMST (c-RMST) and individual-level RMST (i-RMST) estimands as summaries of their respective counterfactual survival functions. We then describe the doubly robust methods to consistently estimate both the cluster-average estimands and the individual-average estimands for each effect measure. Our estimators remain consistent to their respective target estimands even when only one of the models---the censoring model or the outcome model---is correctly specified, but not necessarily both. We explain how the doubly robust method standardizes the predictions from familiar survival models used for correlated data, including frailty and marginal Cox models, to target the estimand of interest. A cluster-jackknife is explored as a universal approach for statistical inference. To facilitate implementation, we provide an R package \texttt{DRsurvCRT} on GitHub \url{https://github.com/fancy575/DRsurvCRT}, with a short tutorial available in Web Appendix E.

%% p4 section organization
The remainder of this article is organized as follows. In {Section \ref{estimand}}, we introduce the cluster-average and individual-average estimands defined on the scale of the survival distribution and RMST. We also present the form of the doubly robust estimators for these quantities. In {Section \ref{reg_model}}, we offer strategies for modeling the survival outcomes in CRTs, and how regression predictions are integrated into the doubly robust estimator. In {Section \ref{jk_var}}, we describe the jackknife variance estimator. In Section \ref{simulation}, we carry out extensive simulation studies to examine the finite-sample performance of our proposed methods under various data generating processes, by considering different levels of ICC, degree of censoring, and model mis-specification scenarios. The simulation results provide empirical evidence to support the robustness and efficiency of our methods compared with traditional approaches, and shed light on improved practice for addressing censored survival outcomes in CRTs. Section \ref{data_ex} provides an illustrative analysis of the Strategies to Reduce Injuries and Develop Confidence in Elders (STRIDE) CRT using the proposed methods. {Section \ref{discussion}} concludes with a discussion on our findings and outlines potential directions for future research.

\section{Formalizing the treatment effect estimands in CRTs with survival outcomes} \label{estimand}

We consider a CRT with \( M \) clusters. Each cluster $i$ consists of $N_i$ participants, and the total number of participants across all clusters is  \( \sum_{i=1}^{M}N_i = N \). For each cluster $i$, a binary treatment indicator  \( A_i \in \{0,1\} \), \( i=1,\dots,M \) is randomized and specifies whether the cluster receives active intervention (\(A_i=1 \)) or usual care (\( A_i=0 \)). Additionally, each cluster is characterized by a vector of cluster-level covariates \( \bm{W}_i \), where \(\bm{W}_i\) is  \( q \)-dimensional vector. For each participant \( j \) in cluster \( i \), (\( i=1,\dots,M \) and \( j=1,\dots, N_i \)), let the vector \( \bm{Z}_{ij} \) be a \( p \)-dimensional set of individual-level covariates that may be associated with the event time of interest. Denote \(\bm{V}_{ij} = \{ \bm{W}_i,\bm{Z}_{ij} \}\) as the collection of all baseline covariates for participant \(j=1,\dots,N_i\) in cluster \(i=1,\dots, M\). In the analysis of survival data in CRT, each participant has a possibly latent event time \( T_{ij} \) and a censoring time \( C_{ij} \). 
Since not all participants experience the event due to right censoring, the observed time for each participant is \( U_{ij} = \min\{T_{ij}, C_{ij}\} \). The censoring indicator is defined as \( \Delta_{ij} = \mathbb{I}\{T_{ij} \leq C_{ij}\} \), which, following the convention, indicates whether the event occurred before censoring ($\mathbb{I}$ is the indicator function). 

To formalize treatment effect estimands in CRTs with survival outcomes, we pursue the potential outcomes framework. Let \( T^{(a)}_{ij} \) represent the potential event time for participant \( j \) in cluster \( i \), under the condition that the cluster receives treatment \( A_i = a \), where \( a \in \{0, 1\} \). Similarly, let \( C^{(a)}_{ij} \) denote the potential censoring time under the same treatment condition. To address the challenge of unequal and potentially informative cluster sizes in CRTs, we define the cluster-level counterfactual survival function as well as the individual-level counterfactual survival function as:
\[
S_C^{(a)}(t) = E \left\{ \frac{1}{N_i}\sum_{j=1}^{N_i} \mathbb{I} (T_{ij}^{(a)} \geq t) \right\}, \quad
S_I^{(a)}(t) = \frac{E \left\{ \sum_{j=1}^{N_i} \mathbb{I}(T_{ij}^{(a)} \geq t) \right\}}{E(N_i)},
\]
for treatment \( a \in \{0, 1\} \). Similar to the c-ATE and i-ATE estimand defined for non-survival outcomes \citep{kahan2023estimands,kahan2023demystifying}, the cluster-level counterfactual survival function describes the impact of intervention on the population of clusters on the mean event rate scale, whereas the individual-level counterfactual survival function describes the impact of intervention on the population of participants pooled across clusters and can be interpreted similarly as the analogous quantity in an individually randomized trial. Furthermore, the corresponding cluster-level and individual-level RMST estimands are defined as:
\[
\mu_C^{(a)}(t) = E \left\{ \frac{1}{N_i}\sum_{j=1}^{N_i} \min\left(T_{ij}^{(a)},t\right)\right\}, \quad
\mu_I^{(a)}(t) = \frac{E \left\{ \sum_{j=1}^{N_i}\min\left(T_{ij}^{(a)},t\right)\right\}}{E(N_i)}.
\]
Notably, under a super-population framework that we pursue here, the expectation is taken with respect to the super population of clusters \citep{wang2024model}. Because of this, the individual-level estimands $S_I^{(a)}(t)$ and $\mu_I^{(a)}(t)$ appear as ratios of two cluster-level expectations. These two estimands may be more intuitive to understand once we replace the cluster-level expectation operator with a cluster-level empirical average operator (i.e., to obtain their finite-population counterparts), such that $S_I^{(a)}(t)={\sum_{i=1}^M\sum_{j=1}^{N_i} \mathbb{I}(T_{ij}^{(a)} \geq t) }/{\sum_{i=1}^M N_i}$ and $\mu_I^{(a)}(t)={\sum_{i=1}^M\sum_{j=1}^{N_i} \min\left(T_{ij}^{(a)},t\right) }/{\sum_{i=1}^M N_i}$ are indeed individual-level averages; we refer to Table 2 in \cite{kahan2023demystifying} for additional comparisons between super-population and finite-population estimands in CRTs.

%Researchers can estimate either the 
Mathematically, the cluster-level or individual-level estimands differ in whether each cluster or each participant is given equal weight. That is, the cluster-level estimand assigns equal weight to each cluster regardless of its size, while the individual-level estimand assigns equal weight to each participant regardless of one's cluster membership. Importantly, when the cluster sizes \( N_i \) are only randomly varying and the potential survival outcomes are identically distributed given cluster size, these two estimands align, such that \( S_C^{(a)}(t) = S_I^{(a)}(t) \) and \( \mu_C^{(a)}(t) = \mu_I^{(a)}(t) \) \citep{wang2021mixed}. When cluster size is marginally associated with the cluster-specific treatment effect, the cluster-average and individual-average estimands carry different magnitude and diverge. In other words, cluster-average and individual-average estimands differ when cluster size is an effect modifier. The choice between these two types of estimands should be guided by the scientific question. That is, the individual-level estimands mimic the evidence that would have been generated from an individually-randomized trial, whereas the cluster-level estimands quantify the cluster-level changes attributed to the intervention (and hence are unique to CRTs). Finally, while our methodological development is motivated by the issue of informative cluster size \citep{kahan2023estimands}, we emphasize that cluster size is only one example of a cluster-level covariate that can drive the heterogeneity of the treatment effect between clusters in CRT. In principle, the proposed framework applies to cases with effect modification by any cluster-level covariates, but our focus on cluster size reflects its practical importance and the unique challenges it raises in the analysis of CRTs, as discussed extensively in the recent literature \citep{kahan2023estimands,kahan2023demystifying,kahan2023informative,wang2022two,wang2024model}.

Before discussing the estimators, we first point out an immediate result that connects the survival function estimand and RMST estimand at each level, with the proof outlined in the Web Appendix A.
\begin{proposition}
\label{prop1}
%Under regularity conditions, 
\emph{The cluster-level and individual-level RMST estimands can be expressed as the time integral of their corresponding survival function estimands. Specifically:
\[
\mu_C^{(a)}(t) = \int_{0}^{t} S_C^{(a)}(u) \, du, \quad
\mu_I^{(a)}(t) = \int_{0}^{t} S_I^{(a)}(u) \, du.
\]}
\end{proposition}

To quantify the treatment effects, we define the cluster-level survival probability causal effect (c-SPCE) and the individual-level survival probability causal effect (i-SPCE) as:
\[
\Delta_C^{\text{SPCE}}(t) = f\left\{S_C^{(1)}(t), S_C^{(0)}(t)\right\}, \quad 
\Delta_I^{\text{SPCE}}(t) = f\left\{S_I^{(1)}(t), S_I^{(0)}(t)\right\},
\]
where \( f(\cdot) \) is a pre-specified comparison function to define the scale of measure. For example, \( f(x, y) = x - y \) represents the estimand on the difference scale, and \( f(x, y) = x / y \) represents the estimand on the ratio scale. Similarly, the cluster-level and individual-level RMST treatment effects are defined as:
\[
\Delta_C^{\text{RMST}}(t) = f\left\{\mu_C^{(1)}(t), \mu_C^{(0)}(t)\right\}, \quad 
\Delta_I^{\text{RMST}}(t) = f\left\{\mu_I^{(1)}(t), \mu_I^{(0)}(t)\right\}.
\]

\section{A doubly robust estimator for analyzing CRTs with survival outcomes}
\label{dr_estimator}
\subsection{Constructing the estimator}
% Causal inference setup
In CRTs, one cannot fully observe the full pair of potential outcomes \( \{T_{ij}^{(1)}, T_{ij}^{(0)} \} \) or \( \{C_{ij}^{(1)}, C_{ij}^{(0)} \} \). Thus, we first state the following cluster-level Stable Unit Treatment Value Assumption (SUTVA).
\begin{assumption}{(Cluster-level SUTVA)}
\label{assum_1} (a) Each participant’s potential outcomes depend only on the treatment assigned to their own cluster. That is, for all \(i, j\), the potential event time \(T_{ij}^{(a)}\) and potential censoring time \(C_{ij}^{(a)}\) are functions only of the cluster-level treatment \(A_i\) and are unaffected by the treatment assignments to other clusters. (b) If cluster \(i\) receives treatment \(A_i = a\), then the observed outcomes equal the corresponding potential outcomes: \(T_{ij} = T_{ij}^{(a)}\) and \(C_{ij} = C_{ij}^{(a)}\) for all \(j = 1, \dots, N_i\).
\end{assumption}Part (a) of Assumption \ref{assum_1} assumes away between-cluster interference, and Part (b) of Assumption \ref{assum_1} is the usual causal consistency assumption. Under the cluster-level Stable Unit Treatment Value Assumption (SUTVA), the individual-level event time \( T_{ij} \) and censoring time \( C_{ij} \) can be expressed as:
\(
T_{ij} = A_i T_{ij}^{(1)} + (1-A_i) T_{ij}^{(0)},
\)
and
\(
C_{ij} = A_i C_{ij}^{(1)} + (1-A_i) C_{ij}^{(0)}. 
\)
In this section, we primarily focus on estimating the cluster-level and individual-level counterfactual survival distributions $S_C^{(a)}(t)$ and $S_I^{(a)}(t)$, respectively, as these quantities provide a basis for inferring the RMST estimands based on Proposition \ref{prop1}. The observed data for each participant is \( \bm{O}_{ij} =  (A_i, \bm{W}_i, N_i, \bm{Z}_{ij}, U_{ij}, \Delta_{ij}) \), where \( i = 1, \dots, M \) and \( j = 1, \dots, N_i \). Specifically, \( U_{ij}=T_{ij}^{(1)} \) is observed when \( A_i = 1 \) and \( \Delta_{ij} = 1 \), indicating an uncensored event when cluster $i$ is treated. Otherwise when \( A_i = 1 \) and \( \Delta_{ij} = 0 \), \( U_{ij}=C_{ij}^{(1)} \) is the observed censoring time when cluster $i$ is treated. A similar structure applies to observations in the usual care clusters. To construct our estimators based on the observed data, we make the following additional assumptions.

\begin{assumption}{(Independent clusters)}\label{assum_2}
The data from different clusters are independent, i.e., for any two distinct clusters \(i\) and \(i'\) (\(i \neq i'\)), the observed data \(\left(A_i, N_i, \bm{V}_i, \{T_{ij}, C_{ij}\}_{j=1}^{N_i} \right)\) and \(\left(A_{i'}, N_{i'}, \bm{V}_{i'}, \{T_{i'j}, C_{i'j}\}_{j=1}^{N_{i'}} \right)\) are independent realizations from a common joint distribution.
\end{assumption}
\begin{assumption}{(Cluster randomization)}\label{assum_3} \(A_i \perp \left\{ T_{ij}^{(1)}, T_{ij}^{(0)}, C_{ij}^{(1)}, C_{ij}^{(0)}; j = 1, \dots, N_i \right\}\), with \(P(A_i=1) \) being a constant that lies strictly within $(0,1)$.
\end{assumption}
\begin{assumption}{(Covariate-dependent censoring)}
\label{assum_4}
\(\left\{C_{i1}^{(a)},\ldots,C_{i,N_i}^{(a)}\right\} \perp \left\{T_{i1}^{(a)},\ldots,T_{i,N_i}^{(a)}\right\} \mid \bm{V}_i,N_i\) for \(a \in \{0, 1\}\). This and Assumption \ref{assum_3} implies that \(\left\{C_{i1},\ldots,C_{i,N_i}\right\} \perp \left\{T_{i1},\ldots,T_{i,N_i}\right\} \mid \left\{\bm{V}_i,N_i, A_i=a\right\}\) holds within each study arm $a$.
\end{assumption}

Assumption \ref{assum_2} establishes between-cluster independence, stating that observed data from distinct clusters are independent realizations from a common distribution; this is a common assumption for the analysis of CRTs. 
%Assumption \ref{assum_2} ensures no interference, requiring that each individual’s potential outcomes depend only on the treatment assigned to their own cluster and are unaffected by the treatment assignments to other clusters. Assumption \ref{assum_3} is the consistency assumption, equating observed outcomes with the potential outcomes under the assigned treatment. 
Assumption \ref{assum_3} assumes that treatment assignment for each cluster (\(A_i\)) is independent of all potential outcomes (\(T_{ij}^{(a)}\) and \(C_{ij}^{(a)}\)) in cluster \(i\). This is a special case of the usual conditional exchangeability assumption, due to the fact that the assignment process is determined by the cluster randomization design. The last part of Assumption \ref{assum_3} corresponds to the usual positivity assumption; that is the probability of treatment assignment lies strictly between 0 and 1. %Assumption \ref{assum_5}  This assumption often holds by the study design; 
Although Assumption \ref{assum_3} considers simple randomization, variations including stratified randomization and other restricted randomization can be stated and our estimators remain applicable \citep{wang2024asymptotic}. Assumption \ref{assum_4}, often referred to as the non-informative censoring assumption, states that within each cluster, the entire set of potential censoring times is independent of the entire set of potential failure times given all baseline covariates \(\bm{V}_i\) collected in that cluster. We adopt this assumption to accommodate scenarios where the individual censoring time (and the individual outcome) may depend not only on the individual’s own covariates but also on covariates of other members within the same cluster, for example through summary measures such as the cluster mean or median (see Section \ref{reg_model} for additional details). As we introduce Assumption \ref{assum_4}, it is important to distinguish informative cluster size from covariate-dependent censoring. Informative cluster size refers to heterogeneity of the treatment effects across clusters, emphasizing how cluster size impacts the survival outcome itself. By contrast, covariate-dependent censoring pertains to the censoring time, which acts as a nuisance process relative to the causal estimands of interest. These two concepts are related in that censoring can depend on cluster size just as outcomes can, but they operate on different parts of the data-generating process. In the ensuing simulation studies, we will examine cases where cluster size can simultaneously drive heterogeneity in outcomes and dependence in censoring. Collectively, Assumptions \ref{assum_1}-\ref{assum_4} are extensions of causal assumptions for independent data \citep{Hernan2020} to the CRT setting. To better describe the dependence among variables in our CRT setting with survival outcomes, in Web Appendix C, we also define the data-generating process using a nonparametric structural equation model \citep{pearl2009causality}, given by a series of equations that satisfy Assumptions \ref{assum_1}-\ref{assum_4}.

Since we observe \(T_{ij}^{(1)}\) only when \(A_i = 1\) (the cluster receives treatment) and \(\Delta_{ij} = 1\) (the event is uncensored), the observed data framework follows a monotone coarsening structure, as described by \cite{tsiatis2006semiparametric}. Under these assumptions, the probability of observing a complete case for any participant in a treated cluster (i.e., \(T_{ij}^{(1)}\)) can be expressed as:
\(
P(A_i=1, C_{ij}^{(1)} \geq T_{ij}^{(1)} | \bm{V}_i) = \pi^{(1)} P(C_{ij}^{(1)} \geq T_{ij}^{(1)} | \bm{V}_i) 
\)
where \(\pi^{(1)} = P(A_i = 1)\) is the randomization probability. By Assumptions \ref{assum_3} and \ref{assum_4}, we can express \(P(C_{ij}^{(1)} \geq T_{ij}^{(1)} \mid \bm{V}_i)\) as \(P(C_{ij} \geq T_{ij} \mid \bm{V}_i, A_i = 1)\), where the censoring mechanism is represented by the within-cluster baseline information. Substituting this expression into the probability of observing a complete case, we obtain 
\(
P(A_i = 1, C_{ij}^{(1)} \geq T_{ij}^{(1)} \mid \bm{V}_i) = \pi^{(1)} K_c^{(1)}(T_{ij}^{(1)} \mid \bm{V}_i),
\)
where 
\(
K_c^{(1)}(t \mid \bm{V}_i) = P(C_{ij} \geq t \mid \bm{V}_i, A = 1)
\)
represents the treatment-specific survival function of censoring at time \(t\), conditional on the covariate information for all participants in the cluster. The same reasoning applies symmetrically to \(a = 0\). In practice, with CRTs, the censoring process is often unknown and must be modeled based on observed data. %In clustered data, this modeling can be challenging due to the potential correlation among individuals within a cluster. 
We explain two common model choices for the censoring process in Section \ref{reg_model}.

The probability that a cluster is assigned to treatment \(a\), \(\pi^{(a)}\), is known as the propensity score \citep{rosenbaum1983central}. Propensity scores are widely used in observational studies to control for confounding by balancing covariates. In CRTs, the true propensity score \(\pi^{(a)}\) is determined by the study design. For instance, in simple randomization, where each cluster has an equal probability of being assigned to either treatment or usual care, the propensity scores are \(\pi^{(1)} = \pi^{(0)} = 1/2\). Under stratified or block cluster randomization, clusters are first grouped into strata based on covariates, and randomization is conducted within each stratum. Within each stratum, the propensity score \(\pi^{(a)}\) is typically constant and predetermined by the randomization procedure. Thus in general, there is no need to estimate the propensity score, as it is already known by the cluster randomization design. However, modeling the known propensity scores by incorporating prognostic covariates---that is, through modeling $\pi^{(a)} = P(A_i = a \mid \bm{V}_i)$---may lead to efficiency improvement as an implicit way to control for baseline chance imbalance, especially in small CRTs. As discussed by \cite{williamson2014variance}, and \cite{zeng2023propensity}, in individually randomized trials, estimating the propensity score using prognostic covariates can reduce the variance of the unadjusted treatment effect; and these results have been generalized to the analysis of CRTs \citep{wang2024handling,balzer2016adaptive,zhu2024leveraging}. Therefore, to implement the ensuing estimators, one could consider either using the known propensity score by study design, or estimate the known propensity score via commonly used logistic regression.

In the case of independent and identically distributed (i.i.d.) data, \cite{robins1992recovery} and \cite{tsiatis2006semiparametric} demonstrated that, under a correctly specified model for coarsening, which includes the propensity score and the treatment-specific censoring distribution conditional on covariates, all survival distribution estimators can be represented as augmented inverse probability weighted for coarsened data (AIPWCC) estimators. In CRTs, considering each cluster as an independent data unit, we denote $\bm{O}_i = \{O_{i1},\dots, O_{iN_i} \}$ as the set of observed data in cluster $i$, and adapt the AIPWCC influence function to define the cluster-level estimating function, as follows:

\begin{align}
     \phi_C(\bm{O}_i)^{(a)}(t) =  & \frac{1}{N_i}\sum_{j=1}^{N_i}\left[\frac{A_i^{a} (1-A_i)^{1-a} \Delta_{ij}}{\pi^{(a)}K_c^{(a)}(t\mid\bm{V}_i)} \left\{\mathbb{I}(U_{ij}\geq t) - S_{C}^{(a)}(t)
    \right\} - \left\{\frac{A_i^{a} (1-A_i)^{1-a}  - \pi^{(a)}}{\pi^{(a)}} \right\} E\left\{\mathbb{I}(T_{ij}^{(a)} \geq t) - S_C^{(a)}(t) \mid \bm{V}_i\right\}\right. \nonumber\\
    & \left.+ \int_{0}^{\infty} \frac{A_i^{a} (1-A_i)^{1-a}}{\pi^{(a)}} \frac{dM_c^{(a)}(u\mid\bm{O}_{ij})}{K_c^{(a)}(u\mid\bm{V}_i)} E\left\{\mathbb{I}(T_{ij}^{(a)} \geq t) - S_C^{(a)}(t) \mid T_{ij}^{(a)} \geq u,\bm{V}_i \right\}\right], \label{AIPWCC_cluster}
\end{align}
where \(
M_c^{(a)}(u \mid \bm{O}_{ij}) = N_{ij}^C(u) - \int_{0}^{u} \lambda_c^{(a)}(s \mid \bm{V}_i) Y_{ij}(s) \, ds\) 
is the marginal martingale increment for the censoring process, \( N_{ij}^C(u) = \mathbb{I}(U_{ij} \leq u, \Delta_{ij} = 0) \) denotes the counting process for censoring, \( Y_{ij}(u) = \mathbb{I}(U \geq u) \) is the at-risk indicator, and \(
\lambda_c^{(a)}(u \mid \bm{V}_i) = -{d \log K_c^{(a)}(u \mid \bm{V}_i)}/{du}\)
is the hazard function for censoring given covariates. In this representation, the first term represents the observed survival data, weighted by the inverse probability of censoring and treatment process. For the second and third terms, we adapt the formulation from \cite{bai2013doubly}, originally developed under the i.i.d. setting, to the clustered data structure. Following the framework of monotone coarsening \citep{tsiatis2006semiparametric}, the augmentation terms in the AIPWCC estimators are derived by projecting the influence function onto the tangent space of the nuisance parameters. This projection minimizes the asymptotic variance and yields the efficient influence function in the absence of within-cluster correlations. In the context of CRT, the within-cluster dependence introduces additional complexity into the tangent space structure for nuisance parameters. Consequently, identifying the appropriate Hilbert space projection becomes non-trivial, requires the inclusion of second-order terms, which often leads to practical challenges for implementation; also see \cite{wang2024handling} for a similar discussion in the context of missing data for CRTs. 

To balance the implementation feasibility and robustness, we interpret equation \eqref{AIPWCC_cluster} as the cluster-level estimating function under the working independence assumption. Under this working independence assumption, we adapt the form proposed by \cite{bai2013doubly} as shown in \eqref{AIPWCC_cluster} and propose estimators for \( S_C^{(a)}(t) \) and \( S_I^{(a)}(t) \). Specifically, the estimator for \( S_C^{(a)}(t) \) can be obtained by solving the estimating equation 
\(
\sum_{i=1}^{M} N_i^{-1} \sum_{j=1}^{N_i} \phi(\bm{O}_{ij})^{(a)}(t) = 0,
\)
and similarly, the estimator for \( S_I^{(a)}(t) \) is derived by solving:
\(
\sum_{i=1}^{M} \sum_{j=1}^{N_i} \Big\{ \phi(\bm{O}_{ij})^{(a)}(t) + S_C^{(a)}(t) - S_I^{(a)}(t) \Big\} = 0,
\)
where
\begin{align*}
    \phi(\bm{O}_{ij})^{(a)}(t) = &  \frac{A_i^{a} (1-A_i)^{1-a} \Delta_{ij}}{\pi^{(a)}K_c^{(a)}(t\mid\bm{V}_{i})} \mathbb{I}(U_{ij}\geq t) 
    - \left\{\frac{A_i^{a} (1-A_i)^{1-a}  - \pi^{(a)}}{\pi^{(a)}} \right\} P(T_{ij}^{(a)} \geq t|\bm{V}_i) \\
    & + \frac{A_i^{a} (1-A_i)^{1-a}}{\pi^{(a)}} \int_{0}^{\infty}  \frac{dM_c^{(a)}(u\mid\bm{O}_{ij})}{K_c^{(a)}(u\mid\bm{V}_{i})} \frac{P(T_{ij}^{(a)} \geq t|\bm{V}_i )}{P(T_{ij}^{(a)} \geq u|\bm{V}_i )} - S_{C}^{(a)}(t) .
\end{align*} 
In more explicit forms, the proposed estimators for the cluster-level and individual-level counterfactual survival functions can be expressed as follows:
\begin{align}
    \widehat{S}_{ij}^{(a)}(t) = & \frac{A_i^{a} (1-A_i)^{1-a} \mathbb{I}(U_{ij} \geq t) }{\pi^{(a)} K_c^{(a)}(t\mid\bm{V}_{i})} - \left\{ \frac{A_i^{a} (1-A_i)^{1-a} - \pi^{(a)}}{ \pi^{(a)}} \right\}P(T_{ij}^{(a)} \geq t\mid\bm{V}_{i})   \nonumber\\
    & + \frac{A_i^{a} (1-A_i)^{1-a} }{\pi^{(a)}} \int_{0}^{t} \frac{dM_{c}^{(a)}(u\mid\bm{O}_{ij}) }{K_c^{(a)}(u\mid \bm{V}_{i})} \frac{P(T_{ij}^{(a)} \geq t|\bm{V}_{i})}{P(T_{ij}^{(a)}\geq u\mid\bm{V}_{i} )}, \label{each_S}\\
    \widehat{S}_C^{(a)}(t) &= \frac{1}{M} \sum_{i=1}^{M} \frac{1}{N_i} \sum_{j=1}^{N_i} \widehat{S}_{ij}^{(a)}(t), \label{S_c} \\
    \widehat{S}_I^{(a)}(t) &= \frac{1}{\sum_{i=1}^{M}N_i} \sum_{i=1}^{M}  \widehat{S}_{ij}^{(a)}(t) \label{S_I}.
\end{align}
In CRTs where data collected from the same cluster can be correlated, the estimators presented above may not be fully locally efficient; however, Proposition \ref{dr_prop} formalizes the statement that they remain doubly robust estimators. The derivation is provided in Web Appendix B.
\begin{proposition}
\label{dr_prop}
\emph{In CRTs, the estimators \eqref{S_c} and \eqref{S_I} are consistent if at least one of the following conditions hold: (1) the model for the censoring mechanism, \(K_c^{(a)}(t \mid \bm{V}_i)\), is correctly specified; (2) the model for the survival outcome, \(P(T_{ij}^{(a)} \geq t \mid \bm{V}_i)\), is correctly specified. }
\end{proposition}
The motivation for using a doubly robust estimator is twofold. First, it offers a principled framework to clarify key components of the data-generating process and delineate the modeling assumptions required to identify the estimands of interest. Second, it provides a unified recipe that combines two specified models---one for the outcome and one for the censoring distribution. The censoring model is used to adjust for potential selection bias due to covariate-dependent censoring, while the outcome model accounts for baseline covariate imbalance and is a device for improving efficiency. The resulting estimator is doubly robust in the sense that it is consistent if either the outcome model or the censoring model is correctly specified, but not necessarily both. This contrasts with singly robust estimators, which rely solely on one of the two models and may be biased if that requisite model is mis-specified.
%The use of doubly robust estimation is motivated by the need to adjust for both covariate-dependent censoring and treatment assignment mechanisms, which may depend on cluster size and other baseline covariates. Our estimators combine a model for the outcome and a model for the censoring distribution to achieve consistency if either is correctly specified. While this approach addresses estimation bias from covariate-dependent censoring, it does not resolve the divergence between the two estimands caused by informative cluster size, which arises from differences in their target populations. However, the doubly robust estimator gives consistent estimates of the marginal survival function, which serves as the basis for constructing the estimators for both estimands.}
In practice, \(K_c^{(a)}(t \mid \bm{V}_i)\) and \(P(T_{ij}^{(a)} \geq t \mid \bm{V}_i)\) are unknown and must be estimated from the observed data. For \(K_c^{(a)}(t \mid \bm{V}_i)\), under Assumptions \ref{assum_3} and \ref{assum_4}, the estimation approach depends on the nature of the censoring mechanism. If the censoring process is completely at random, such that 
\(
K_c^{(a)}(t \mid \bm{V}_{i}) = P(C_{ij} \geq t \mid A_i = a),
\)
it can be estimated non-parametrically using the Kaplan-Meier estimator, as no additional covariate adjustments are required. If the censoring process for participant \(j\) in cluster \(i\) depends on the covariates \(\bm{V}_{i}\), the survival probability generalizes to 
\(
K_c^{(a)}(t \mid \bm{V}_{i}) = P(C_{ij} \geq t \mid \bm{V}_{i}, A_i = a).
\)
In this case, \(K_c^{(a)}(t \mid \bm{V}_{i})\) can be estimated through treatment-specific regression models. Similarly, \(P(T_{ij}^{(a)} \geq t \mid \bm{V}_{i})\) can also be estimated through regression models on the covariates \(\bm{V}_{i}\). In CRTs, it is often important to specify the regression models that account for the intra-cluster correlation; detailed examples will be discussed in Section \ref{reg_model}. 

Finally, given that we have specified and estimated models for the censoring and outcome processes, to estimate the SPCE estimand at the cluster level and individual level under an effect measure function such as \( f(x, y) = x - y \), the estimators can be expressed as:
\begin{align*}
    \widehat{\Delta}_{C}^{\text{SPCE}}(t) & = \frac{1}{M} \sum_{i=1}^{M}\frac{1}{N_i}\sum_{j=1}^{N_i} \left\{\widehat{S}_{ij}^{(1)}(t) - \widehat{S}_{ij}^{(0)}(t)\right\},\\
    \widehat{\Delta}_{I}^{\text{SPCE}}(t) & = \frac{1}{\sum_{i=1}^{M}N_i} \sum_{i=1}^{M} \sum_{j=1}^{N_i} \widehat{S}_{ij}^{(1)}(t) - \widehat{S}_{ij}^{(0)}(t).
\end{align*}
By Proposition \ref{prop1}, for RMST estimands $\mu_C^{(a)}(t)$ and the $\mu_I^{(a)}(t)$ can be derived with $\widehat{S}_{ij}^{(a)}(t)$ as
\begin{align*}
     \widehat{\Delta}_{C}^{\text{RMST}} & = \frac{1}{M} \sum_{i=1}^{M} \frac{1}{N_i} \sum_{j=1}^{N_i} \int_{0}^{t} \left\{\widehat{S}_{ij}^{(1)}(u) - \widehat{S}_{ij}^{(0)}(u) \right\} \, du \\
     \widehat{\Delta}_{I}^{\text{RMST}} & =  \frac{1}{\sum_{i=1}^{M}N_i} \sum_{i=1}^{M} \sum_{j=1}^{N_i} \int_{0}^{t} \left\{\widehat{S}_{ij}^{(1)}(u) - \widehat{S}_{ij}^{(0)}(u) \right\} \, du,
\end{align*}
where the integration is approximated using the trapezoidal rule. That is, for a discrete time grid \( \{ u_k \}_{k=1}^K \) with \( u_1 = 0 \) and \( u_K = t \), the integral can be practically approximated as:
\[
\int_{0}^{t} \left\{\widehat{S}_{ij}^{(1)}(u) - \widehat{S}_{ij}^{(0)}(u)\right\} \, du 
\approx \sum_{k=1}^{K-1} \frac{u_{k+1} - u_k}{2} \left[\left\{\widehat{S}_{ij}^{(1)}(u_k) - \widehat{S}_{ij}^{(0)}(u_k)\right\} + \left\{\widehat{S}_{ij}^{(1)}(u_{k+1}) - \widehat{S}_{ij}^{(0)}(u_{k+1})\right\}\right].
\]

Table \ref{tab:estimand_sum} summarizes the four estimands—survival probability and restricted mean survival time (RMST), defined at both the cluster and individual levels. Each estimand builds upon the individual-specific survival prediction \(\widehat{S}_{ij}^{(a)}(t)\) introduced in \eqref{each_S}. 

\begin{table}[htbp]
\centering
\caption{A concise summary of target estimands and proposed estimators in CRTs with a survival outcome.}
\label{tab:estimand_sum}
\begin{adjustbox}{max width=\textwidth}
\begin{tabular}{@{}p{2cm} p{4.5cm} p{5.5cm} p{5.8cm}@{}}
\toprule
\textbf{Level of Aggregation} & \textbf{Weighting Scheme} & \textbf{Estimand} & \textbf{Estimators} \\
\midrule

\multirow{4}{*}{Cluster} 
  & \multirow{4}{*}{Equal weight to each cluster} 
  & \(S_C^{(a)}(t) = E\Bigl[\tfrac{1}{N_i}\sum_{j=1}^{N_i} I(T_{ij}^{(a)} \ge t)\Bigr]\) 
    & \(\widehat{S}_C^{(a)}(t) = \tfrac{1}{M}\sum_{i=1}^{M}\tfrac{1}{N_i}\sum_{j=1}^{N_i} \widehat{S}_{ij}^{(a)}(t)\) \\

&& \(\Delta_C^{\mathrm{SPCE}}(t) = f\bigl(S_C^{(1)}(t),\,S_C^{(0)}(t)\bigr)\) 
    & \(\widehat{\Delta}_C^{\mathrm{SPCE}}(t) = f\bigl(\widehat{S}_C^{(1)}(t),\,\widehat{S}_C^{(0)}(t)\bigr)\) \\
\cmidrule(lr){3-4}
&& \(\mu_C^{(a)}(t) = \int_0^t S_C^{(a)}(u)\,\mathrm{d}u\) 
    & \(\widehat{\mu}_C^{(a)}(t) = \int_0^t \widehat{S}_C^{(a)}(u)\,\mathrm{d}u\) \\

&& \(\Delta_C^{\mathrm{RMST}}(t) = f\bigl(\mu_C^{(1)}(t),\,\mu_C^{(0)}(t)\bigr)\) 
    & \(\widehat{\Delta}_C^{\mathrm{RMST}}(t) = f\bigl(\widehat{\mu}_C^{(1)}(t),\,\widehat{\mu}_C^{(0)}(t)\bigr)\) \\
\midrule

\multirow{4}{*}{Individual} 
  & \multirow{4}{*}{Equal weight to each participant} 
  & \(S_I^{(a)}(t) = E(N)^{-1}\,E\Bigl[\sum_{j=1}^{N_i} I(T_{ij}^{(a)} \ge t)\Bigr]\) 
    & \(\widehat{S}_I^{(a)}(t) = \bigl(\sum_{i=1}^{M} N_i\bigr)^{-1}\sum_{i=1}^{M}\sum_{j=1}^{N_i} \widehat{S}_{ij}^{(a)}(t)\) \\

&& \(\Delta_I^{\mathrm{SPCE}}(t) = f\bigl(S_I^{(1)}(t),\,S_I^{(0)}(t)\bigr)\) 
    & \(\widehat{\Delta}_I^{\mathrm{SPCE}}(t) = f\bigl(\widehat{S}_I^{(1)}(t),\,\widehat{S}_I^{(0)}(t)\bigr)\) \\
\cmidrule(lr){3-4}
&& \(\mu_I^{(a)}(t) = \int_0^t S_I^{(a)}(u)\,\mathrm{d}u\) 
    & \(\widehat{\mu}_I^{(a)}(t) = \int_0^t \widehat{S}_I^{(a)}(u)\,\mathrm{d}u\) \\

&& \(\Delta_I^{\mathrm{RMST}}(t) = f\bigl(\mu_I^{(1)}(t),\,\mu_I^{(0)}(t)\bigr)\) 
    & \(\widehat{\Delta}_I^{\mathrm{RMST}}(t) = f\bigl(\widehat{\mu}_I^{(1)}(t),\,\widehat{\mu}_I^{(0)}(t)\bigr)\) \\

\bottomrule
\end{tabular}
\end{adjustbox}
\end{table}

\subsection{Specifying the censoring and outcome regression models}
\label{reg_model}
In constructing the doubly robust estimators \eqref{S_c} and \eqref{S_I}, both \(K_c^{(a)}(t\mid \bm{V}_i)\) and \(P(T_{ij}^{(a)} \geq t \mid \bm{V}_i)\) are essential components. 
Ideally, the correct specification of both models involves regressing on covariates across all participants in a cluster, which can be achieved by high-dimensional regression methods but may lack numerical stability and practicality. 
Therefore, we focus on the following models $ K_c^{(a)}(t \mid \bm{V}_{ij})$ and $P(T_{ij}^{(a)} \geq t \mid \bm{V}_{ij})$, where $\bm{V}_{ij} = \{N_{i}, \bm{W}_{i}, \bm{Z}_{ij}\}$. With this change from $\bm{V}_i$ to $\bm{V}_{ij}$, we largely simplify the model fitting but still allow certain cluster-level summaries of \(\bm{Z}_{ij}\)—--such as the cluster-specific mean \(\bm{\overline{Z}}_i\) for \(i = 1, \dots, M\)—--to be incorporated into \(\bm{W}_i\) to address contextual effects \citep{begg2003separation,seaman2014review}. This redefinition enables \(\bm{V}_{ij}\) to be further distinguished into cluster-level and individual-level components, capturing both aggregated effects and lower-level variations in cases when they are both of interest. If we assume $\bm{V}_{ij}$ captures all correlations between outcomes and $\bm{Z}_{ij'}$ for $j'\ne j$, then the models are correctly specified in the sense that \( K_c^{(a)}(t \mid \bm{V}_i) = K_c^{(a)}(t \mid \bm{V}_{ij}) \)
and \( P(T_{ij}^{(a)} \geq t \mid \bm{V}_i) = P(T_{ij}^{(a)} \geq t \mid \bm{V}_{ij}). \)
%Under covariate-dependent censoring, a frequent practice is to further assume the absence of \emph{covariate interference} such that the censoring and survival distribution of individual $j$ in cluster $i$ does not further depend on baseline information of other cluster members, beyond those included in $N_i$ and $\bm{W}_i$. 
With survival outcomes in CRTs, two common regression approaches are the marginal Cox model \citep{wang2023improving} and the frailty Cox model \citep{jahn2013sample}, which we elaborate below.
% We focus on these two commonly used regression techniques for addressing survival and censoring probabilities given baseline covariates. 

\subsubsection{Marginal Cox model}
The marginal Cox model is often used when the goal is to estimate population-level effects, without explicitly modeling the dependence between participants within clusters. This models assumes that the correlation between participants in the same cluster is a nuisance factor that can be accounted for through the robust sandwich variance technique \citep{lin1994cox}.
Under a marginal Cox framework, to estimate \(P(T_{ij}^{(a)} \geq t \mid \bm{V}_{ij})\), along with \(K_c^{(a)}(t \mid \bm{V}_{ij}) = P(C_{ij}^{(a)} \geq t \mid \bm{V}_{ij})\) and the martingale \(dM_c^{(a)}(u \mid \bm{O}_{ij})\) utilized in \eqref{S_c} and \eqref{S_I}, we first model the hazard functions: \( \lambda_{ij}^{(a),m}(t \mid \bm{V}_{ij}) \) for the outcome event time and \( h_{ij}^{(a),m}(t \mid \bm{V}_{ij}) \) for the censoring time, as follows:
\begin{align}
    \lambda^{(a),m}(t\mid\bm{V}_{ij}) & = \lambda_{0}^{(a),m} \exp\left(\left\{\bm{\beta}_{W}^{(a),m}\right\}^\top\bm{W}_i + \left\{\bm{\beta}_{Z}^{(a),m}\right\}^\top\bm{Z}_{ij}\right), \label{breslow_t}\\
    h^{(a),m}(t \mid \bm{V}_{ij}) & = h_0^{(a),m} \exp \left( \left\{ \bm{\alpha}_W^{(a),m} \right\}^\top \bm{W}_i + \left\{  \bm{\alpha}_Z^{(a),m}\right\}^\top\bm{Z}_{ij}  \right), \label{breslow_c}
\end{align}
where \(\lambda_{0}^{(a),m}(t)\) represents the unspecified treatment-specific baseline hazard function, and \(\bm{\beta}_{W}^{(a),m}\) and \(\bm{\beta}_{Z}^{(a),m}\) are the vectors of regression parameters for \(\bm{W}_i\) and \(\bm{Z}_{ij}\), respectively, in the outcome model. Similarly, \(h_0^{(a),m}(t)\) denotes the unspecified treatment-specific baseline hazard function for the censoring model, with \(\bm{\alpha}_W^{(a),m}\) and \(\bm{\alpha}_{Z}^{(a),m}\) being the corresponding regression parameter vectors for \(\bm{W}_i\) and \(\bm{Z}_{ij}\). The baseline hazard function can be estimated using the Breslow estimator, expressed as:
\begin{align*}
    \widehat{\lambda}_{0}^{(a),m}(t) & = \sum_{i=1}^{M} \sum_{j=1}^{N_i} \frac{dN_{ij}(t)}{\sum_{k=1}^{M} \sum_{l=1}^{N_k} Y_{kl}(t) \exp\left(\{\bm{\beta}_{W}^{(a),m}\}^\top \bm{W}_k + \{\bm{\beta}_{Z}^{(a),m}\}^\top \bm{Z}_{kl}\right)} ,\\
    \widehat{h}_{0}^{(a),m}(t) & = \sum_{i=1}^{M} \sum_{j=1}^{N_i} \frac{dN_{ij}^C(t)}{\sum_{k=1}^{M} \sum_{l=1}^{N_k} Y_{kl}(t) \exp\left(\{\bm{\alpha}_{W}^{(a),m}\}^\top \bm{W}_k + \{\bm{\alpha}_{Z}^{(a),m}\}^\top \bm{Z}_{kl}\right)},
\end{align*}
where \(N_{ij}(t) = \mathbb{I}(U_{ij} \leq t, \Delta_{ij} = 1)\) and \(N_{ij}^C(t) = \mathbb{I}(U_{ij} \leq t, \Delta_{ij} = 0)\) are the counting processes for the event time and censoring time, respectively, and \(Y_{kl}(t) = \mathbb{I}(U_{kl} \geq t)\) represents the risk set for \(k = 1, \dots, M\) and \(l = 1, \dots, N_i\). Let \(\widehat{\bm{\beta}}_{W}^{(a),m}\), \(\widehat{\bm{\beta}}_{Z}^{(a),m}\), \(\widehat{\bm{\alpha}}_{W}^{(a),m}\), and \(\widehat{\bm{\alpha}}_{Z}^{(a),m}\) denote the estimators of \(\bm{\beta}_{W}^{(a),m}\), \(\bm{\beta}_{Z}^{(a),m}\), \(\bm{\alpha}_{W}^{(a),m}\), and \(\bm{\alpha}_{Z}^{(a),m}\), respectively, which are obtained by solving the log pseudo-partial likelihood equations (specified under the working independence assumption; see equation (2) in \cite{wang2023improving} for an example) defined only on the subset of data where \(A_i = a\) \citep{lin1994cox} . The hazard functions \(\lambda(t \mid \bm{V}_{ij})\) and \(h(t \mid \bm{V}_{ij})\) can then be estimated as:
\begin{align*}
   \widehat{\lambda}^{(a),m}(t \mid\bm{V}_{ij}) & = \widehat{\lambda}_{0}^{(a),m}(t) \exp \left(  \left\{ \widehat{\bm{\beta}}_{W}^{(a),m}  \right\}^\top \bm{W}_i + \left\{\widehat{\bm{\beta}}_{Z}^{(a),m}  \right\}^\top \bm{Z}_{ij} \right) \\
    \widehat{h}^{(a),m}(t\mid \bm{V}_{ij}) & = \widehat{h}_{0}^{(a),m}(t) \exp \left(  \left\{ \widehat{\bm{\alpha}}_{W}^{(a),m}  \right\}^\top \bm{W}_i + \left\{\widehat{\bm{\alpha}}_{Z}^{(a),m}  \right\}^\top \bm{Z}_{ij} \right)
\end{align*}
The estimators for \(K_c^{(a)}(t \mid \bm{V}_{ij})\) and \(P(T^{(a)} \geq t \mid \bm{V}_{ij})\) can be formulated in terms of \(\widehat{\lambda}^{(a),m}(t \mid \bm{V}_{ij})\) and \(\widehat{h}^{(a),m}(t \mid \bm{V}_{ij})\) as follows:
\begin{align*}
    \widehat{P}(T_{ij}^{(a)} \geq t \mid \bm{V}_{ij}) & = \exp\left(-\int_{0}^{t} \widehat{\lambda}^{(a),m}(t \mid \bm{V}_{ij}) \, dt\right),\\
    \widehat{K}_c^{(a)}(t\mid \bm{V}_{ij}) & = \exp\left(-\int_{0}^{t} \widehat{h}^{(a),m}(t \mid \bm{V}_{ij}) \, dt\right).
\end{align*}
The martingale term \(M_c^{(a)}(t \mid \bm{O}_{ij})\) can be computed as \(\widehat{M}_c^{(a),m}(t \mid \bm{O}_{ij}) = N_{ij}^C(t) - \int_{0}^{t}Y_{ij}(u)\widehat{h}^{(a),m}(u \mid \bm{V}_{ij})\, du \).  The quantities \(\widehat{K}_c^{(a)}(t \mid \bm{V}_{ij})\), \(\widehat{P}(T_{ij}^{(a)} \geq t \mid \bm{V}_{ij})\), and \(d\widehat{M}_c^{(a)}(t \mid \bm{O}_{ij})\) can then be used as plug-ins for \eqref{each_S}, \eqref{S_c}, and \eqref{S_I}. The marginal Cox model can be conveniently implemented, for example, using the {\texttt{coxph}} function from the {\texttt{survival}} package in \texttt{R} \citep{terry2000modeling}.

\subsubsection{Frailty Cox model}
\label{frailty_cox}
In contrast, the frailty Cox model introduces random effects, known as frailties, to explicitly account for the correlation between participants within clusters, during the estimation process. The frailty terms modify the hazard function for each participant by a multiplicative factor, reflecting the cluster-specific effect on survival. This approach allows for more flexible modeling of dependencies within the cluster. We begin by modeling the event time and censoring time using treatment-specific conditional hazard functions given the latent cluster-level frailty. These models incorporate frailties \( B_i \) and \( R_i \) for \( i = 1, \ldots, M \), respectively, and are specified as follows and \( a = 0, 1 \):
\begin{align*}
    \lambda^{(a),f}(t \mid \bm{V}_{ij}, B_i) &= \lambda_0^{(a),f} B_i \exp\left( \left\{\bm{\beta}_W^{(a),f}\right\}^\top \bm{W}_i + \left\{\bm{\beta}_Z^{(a),f}\right\}^\top \bm{Z}_{ij} \right), \\
    h^{(a),f}(t \mid \bm{V}_{ij}, R_i) &= h_0^{(a),f} R_i \exp\left( \left\{\bm{\alpha}_W^{(a),f}\right\}^\top \bm{W}_i + \left\{\bm{\alpha}_Z^{(a),f}\right\}^\top \bm{Z}_{ij} \right),
\end{align*}
where \( B_i \) and \( R_i \) are random variables independent of \(\bm{V}_{ij}\) and follow distributions \( f_B(b) \) and \( f_R(r) \), respectively. The distributions \(f_B(b)\) and \(f_R(r)\) can be any distribution for the random effect, such as gamma, log-normal distribution, etc, providing flexibility to the frailty model \citep{balan2020tutorial}. The conditional outcome and censoring models, \( P(T_{ij}^{(a)} \geq t \mid \bm{V}_{ij}, B_i) \) and \( K_c^{(a)}(t \mid \bm{V}_{ij}, R_i) \), for an participant \( j \) in cluster \( i \) are then given by:
\begin{align*}
    P(T_{ij}^{(a)} \geq t \mid \bm{V}_{ij}, B_i) &= \exp\left\{ -\int_{0}^{t} \lambda^{(a),f}(u \mid \bm{V}_{ij}, B_i) \, du \right\}, \\
    K_c^{(a)}(t \mid \bm{V}_{ij}, R_i) &= \exp\left\{ -\int_{0}^{t} h^{(a),f}(u \mid \bm{V}_{ij}, R_i) \, du \right\}.
\end{align*}
However, to implement our doubly robust estimators, we instead need to induce  \(P(T_{ij}^{(a)} \geq t \mid \bm{V}_{ij})\) and \(K_c^{(a)}(t\mid \bm{V}_{ij})\) by integrating over the fraities in the conditional models. This involves determining the marginal distribution by taking the expectation of \(P(T_{ij}^{(a)} \geq t \mid \bm{V}_{ij},B_i )\) with respect to \(B_i\) and taking the expectation of \(K_c^{(a)}(t\mid \bm{V}_{ij},R_i)\) with respect to \(R_i\). That is, we need to obtain
\begin{align*}
    P(T_{ij}^{(a)} \geq t \mid \bm{V}_{ij} ) & = E_B\left\{ P(T_{ij}^{(a)} \geq t \mid \bm{V}_{ij},B_i )  \right\}  \\
    K_c^{(a)}(t\mid \bm{V}_{ij}) & = E_R \left \{ K_c^{(a)}(t\mid \bm{V}_{ij},R_i)  \right\},
\end{align*}
which can be derived through the Laplace transformation \citep{spiegel1965laplace,balan2020tutorial}.  Denote \( \lambda^{(a),f}(t \mid \bm{V}_{ij}) \) and \( h^{(a),f}(t \mid \bm{V}_{ij}) \) as the marginal hazard functions, which are calculated from the derivatives of the negative logarithms of the corresponding marginal distributions. Specifically, we have:
\begin{align}
    \lambda^{(a),f}(t \mid \bm{V}_{ij}) 
    & = \frac{d\{-\log P(T_{ij}^{(a)} \geq t \mid \bm{V}_{ij})\}}{dt} = E[B_i \mid T_{ij}^{(a)} \geq t] \, \lambda_0^{(a),f} \exp\left( \{\bm{\beta}_W^{(a),f}\}^\top \bm{W}_i + \{\bm{\beta}_Z^{(a),f}\}^\top \bm{Z}_{ij} \right), \label{frailty_marginal_t} \\
    h^{(a),f}(t \mid \bm{V}_{ij}) 
    & = \frac{d\{-\log K_c^{(a)}(t \mid \bm{V}_{ij})\}}{dt}  = E[R_i \mid C_{ij}^{(a)} \geq t] \, h_0^{(a),f} \exp\left( \{\bm{\alpha}_W^{(a),f}\}^\top \bm{W}_i + \{\bm{\alpha}_Z^{(a),f}\}^\top \bm{Z}_{ij} \right). \label{frailty_marginal_c}
\end{align}
This marginalization is necessary to align with marginal hazards and their associated quantities in the proposed doubly robust estimators under the working independence assumption. 
%with the working independence assumption in clustered survival data. 
Under this assumption, the marginal filtration constructed from the marginal distributions also ensures the martingale property. %The marginal and conditional hazards are equal only if \(E\{B_i \mid T_{ij}^{(a)} \geq t\} = 1\)  in outcome model and  \(E\{R_i \mid C_{ij}^{(a)} \geq t\} = 1\) in censoring model for all \(t\). 
When the frailty terms \(B_i\) and  \(R_i\) follow gamma distributions with shape and rate parameters \(\theta\) and  \(\gamma \), i.e., \(B_i \sim \text{Gamma}(\theta, \theta)\) and  \(R_i \sim \text{Gamma} (\gamma,\gamma) \), we can derive the following closed-form relationships:
\begin{align*}
     E[B_i|T_{ij}^{(a)} \geq t] & = \frac{\theta}{\theta + \int_{0}^{t} \lambda^{(a),f}(u \mid \bm{V}_{ij},B_i=1) \, du  },\\
     E[R_i|C_{ij}^{(a)} \geq t] &= \frac{\gamma}{\gamma + \int_{0}^{t} h^{(a),f}(u \mid \bm{V}_{ij},R_i=1) \, du  }.
\end{align*}
Similarly, the marginal distribution for outcome and censoring model can be derived as 
\begin{align}
     P(T_{ij}^{(a)} \geq t \mid \bm{V}_{ij} ) & = \left\{\frac{\theta}{ \theta + \int_{0}^{t} \lambda^{(a),f}(u \mid \bm{V}_{ij},B_i=1) \, du } \right\}^{\theta}, \label {frailty_t}\\
     K_c^{(a)}(t\mid \bm{V}_{ij}) & = \left\{\frac{\gamma}{ \gamma + \int_{0}^{t} h^{(a),f}(u \mid \bm{V}_{ij},R_i=1) \, du } \right\}^{\gamma} \label{frailty_c}.
\end{align}

The marginal distributions discussed above are specific to gamma frailty. In practice, however, a variety of distributions can be used for \(f_B(b)\) and \(f_R(r)\), collectively known as infinitely divisible distributions with tractable Laplace transformations. Examples include the Hougaard distribution and the inverse Gaussian distribution. The gamma distribution, as a member of this family, is particularly advantageous due to its mathematical simplicity in calculations \citep{hougaard2000analysis}. Another key benefit of gamma frailty lies in its direct relationship with a measure of the intracluster correlation, also known as Kendall's \(\tau\) in survival data. In our frailty model for outcome events and censoring, where \(B_i \sim \text{Gamma}(\theta, \theta)\) and \(R_i \sim \text{Gamma}(\gamma, \gamma)\), Kendall's \(\tau\) is given by \(1/(2\theta + 1)\) and \(1/(2\gamma + 1)\), respectively \citep{li2022comparison, meng2023simulating}. 
Denote \(\widehat{\bm{\beta}}_W^{(a),f}\), \(\widehat{\bm{\beta}}_Z^{(a),f}\), \(\widehat{\bm{\alpha}}_W^{(a),f}\), and \(\widehat{\bm{\alpha}}_Z^{(a),f}\) as the estimators of \(\bm{\beta}_W^{(a),f}\), \(\bm{\beta}_Z^{(a),f}\), \(\bm{\alpha}_W^{(a),f}\), and \(\bm{\alpha}_Z^{(a),f}\), respectively. Similarly, let \(\widehat{\theta}\) and \(\widehat{\gamma}\) be the estimators of \(\theta\) and \(\gamma\). Denote \(\widehat{\lambda}_0^{(a),f}(t)\) and \(\widehat{h}_0^{(a),f}(t)\) as the estimators of the baseline hazards \(\lambda_0^{(a),f}(t)\) and \(h_0^{(a),f}(t)\), respectively, which can be obtained using the Breslow estimators similar to equations \eqref{breslow_t} and \eqref{breslow_c}. The estimators for the outcome distribution \(P(T_{ij}^{(a)} \geq t \mid \bm{V}_{ij})\) and the censoring distribution \(K_c^{(a)}(t \mid \bm{V}_{ij})\) are derived by replacing \(\lambda^{(a),f}(u \mid \bm{V}_{ij}, B_i = 1)\) and \(h^{(a),f}(u \mid \bm{V}_{ij}, R_i = 1)\) with \(\widehat{\lambda}^{(a),f}(u \mid \bm{V}_{ij}, B_i = 1)\) and \(\widehat{h}^{(a),f}(u \mid \bm{V}_{ij}, R_i = 1)\), and by substituting \(\theta\) and \(\gamma\) with their respective estimators in equations \eqref{frailty_t} and \eqref{frailty_c}. Here, \(\widehat{\lambda}^{(a),f}(u \mid \bm{V}_{ij}, B_i = 1)\) and \(\widehat{h}^{(a),f}(u \mid \bm{V}_{ij}, R_i = 1)\) are obtained by replacing \(\bm{\beta}_W^{(a),f}\), \(\bm{\beta}_Z^{(a),f}\), \(\bm{\alpha}_W^{(a),f}\), and \(\bm{\alpha}_Z^{(a),f}\) with their corresponding estimators from the frailty models, and by replacing \(\lambda_0^{(a),f}(t)\) and \(h_0^{(a),f}(t)\) with \(\widehat{\lambda}_0^{(a),f}(t)\) and \(\widehat{h}_0^{(a),f}(t)\), respectively. The estimator of the martingale term \(dM_c^{(a)}(t \mid \bm{O}_{ij})\) can be derived from the estimators in equation \eqref{frailty_marginal_c} (denoted as \(\widehat{h}^{(a),f}(t \mid \bm{V}_{ij})\)), and the corresponding estimator for the martingale term is given by \(d\widehat{M}_c^{(a)}(t \mid \bm{O}_{ij}) = dN_{ij}^C(t) - Y_{ij} \widehat{h}^{(a),f}(t \mid \bm{V}_{ij})\). The frailty Cox model can be implemented, for example, using the {\texttt{frailtyEM}} package in \texttt{R} \citep{frailtyEM}, which provides flexible options for specifying the frailty distribution.

\subsection{Statistical inference via cluster jackknifing}
\label{jk_var}

There are multiple approaches to estimating the variance of the proposed doubly robust estimators in CRTs. One approach relies on the influence function, which formally accounts for the uncertainty in estimating all model parameters. For example, \cite{zhang2012double} derived such influence functions for independent survival data. However, as they noted, the resulting expressions involve complex martingale integrals, and because it is unknown which specified model is correctly specified in practice, all terms must be estimated. Direct plug-in variance estimators based on these influence functions can therefore accumulate substantial estimation error. To address this, \cite{zhang2012double} recommended bootstrap resampling. In the CRT setting, the situation is further complicated, as the influence function must be re-derived under each modeling choice (e.g., marginal Cox versus frailty Cox), which limits its practical convenience. Motivated by these considerations, we employ the cluster jackknife method, a resampling approach that iteratively omits one cluster (with all its individuals) at a time and recomputes the estimator. The variance is then estimated as the sample variance of these leave-one-cluster-out estimates across all iterations \citep{efron1981jackknife}.

The cluster jackknife has been previously used for statistical inference in linear models and generalized estimating equations for clustered data \citep{mancl2001covariance,mackinnon2023fast,hansen2022jackknife}, and has well-established theoretical connections to finite-sample bias corrections in variance estimation \citep{bell2002bias}. A theoretical investigation of the cluster jackknife approach in linear models can be found in \cite{hansen2022jackknife}, and the approach has also been applied in \cite{li2025model} for computing standardized treatment effect estimates for i-ATE and c-ATE in CRTs with non-survival outcomes. Compared with cluster bootstrap or permutation methods, the cluster jackknife is computationally more efficient when the number of clusters (i.e. independent units) is not excessively large and is particularly attractive when analytic variance estimators are difficult to derive or implement. Thus, the procedure is general enough to obviate any major modifications when different censoring or outcome models (e.g., marginal Cox or frailty models) are used for computing the doubly robust estimators, beyond refitting these models to the leave-one-cluster-out data sets. By removing one entire cluster at a time, the method also naturally accounts for the within-cluster correlation inherent in CRTs, ensuring that the resulting variance estimator reflects the unknown within-cluster dependence structure. Finally, we emphasize that the cluster jackknife approach is used only for computing the variance and the associated confidence intervals for uncertainty quantification; %and should not change the original point estimates. 
this procedure hence does not affect point estimation nor compromise causal validity under cluster randomization.%Because each deletion removes the entire randomization unit, and by Assumption 1, dropping a cluster does not alter the potential outcomes or the remaining clusters, the procedure preserves the causal analysis. Inference is therefore based on the effect sample size \(M\).}

To proceed, for $g \in \{1,\dots, M\}$, the leave-one-cluster-out estimator for $S_C^{(a)}$ and $S_I^{(a)}$ are defined as
\begin{align*}
    \widehat{S}_{ij}^{(a),-g}(t) =& \frac{A_i^{a} (1-A_i)^{1-a} I(U_{ij} \geq t) }{\pi^{(a)} \widehat{K}_c^{(a),-g}(t\mid\bm{V}_{i})} - \left\{ \frac{A_i^{a} (1-A_i)^{1-a} - \pi^{(a)}}{ \pi^{(a)}} \right\}\widehat{P}(T_{ij}^{(a)} \geq t\mid\bm{V}_{i})  \\
    & + \int_{0}^{t} \frac{A_i^{a} (1-A_i)^{1-a} }{\pi^{(a)}} \frac{d\widehat{M}_{c,ij}^{(a),-g}(u,\bm{V}_{i}) }{\widehat{K}_c^{(a),-g}(u\mid\bm{V}_{i})} \frac{\widehat{P}(T_{ij}^{(a),-g} \geq t\mid\bm{V}_{i})}{\widehat{P}(T_{ij}^{(a),-g}\geq u\mid\bm{V}_{i} )}, \\
    \widehat{S}_C^{(a),-g}(t) &= \frac{1}{M-1} \sum_{i \neq g} \frac{1}{N_i} \sum_{j=1}^{N_i} \widehat{S}_{ij}^{(a),-g}(t),\\
    \widehat{S}_I^{(a),-g}(t) &= \frac{1}{\sum_{i\neq g}N_i} \sum_{i\neq g} \sum_{j=1}^{N_i}\widehat{S}_{ij}^{(a),-g}(t).   
\end{align*}
In this leave-one-cluster-out estimator, participants in cluster \( g \) are iteratively removed, and outcomes are recalculated using the remaining participants. The outcome regression for \( K_c^{(a)}(t\mid \bm{V}_{i}) \) and \( P(T^{(a)} \geq t \mid \bm{V}_{i}) \) is then re-fitted with the remaining participants as well, denoted as \( \widehat{K}_c^{(a),-g}(t\mid \bm{V}_{i}) \) and \( \widehat{P}(T_{ij}^{(a),-g} \geq t \mid \bm{V}_{i}) \), respectively. 
The point-wise jackknife variance estimators \( \widehat{\Sigma}_{C,JK} \) and \( \widehat{\Sigma}_{I,JK} \) are then obtained from these leave-one-cluster-out estimators and are given by:
\begin{align*}
    \widehat{\Sigma}_{C,JK} & = \frac{M-1}{M} \sum_{g=1}^{M} 
    \begin{pmatrix}
    \{\widehat{S}_C^{(1),-g}(t) - \bar{S}_C^{(1)}(t) \}^2 & \{\widehat{S}_C^{(1),-g}(t) - \bar{S}_C^{(1)}(t) \}\{\widehat{S}_C^{(0),-g}(t) - \bar{S}_C^{(0)}(t) \}\\
    \{\widehat{S}_C^{(1),-g}(t) - \bar{S}_C^{(1)}(t) \}\{\widehat{S}_C^{(0),-g}(t) - \bar{S}_C^{(0)}(t) \} & \{\widehat{S}_C^{(0),-g}(t) - \bar{S}_C^{(0)}(t) \}^2 
    \end{pmatrix},\\
    \widehat{\Sigma}_{I,JK} & = \frac{M-1}{M} \sum_{g=1}^{M} 
    \begin{pmatrix}
    \{\widehat{S}_I^{(1),-g}(t) - \bar{S}_I^{(1)}(t) \}^2 & \{\widehat{S}_I^{(1),-g}(t) - \bar{S}_I^{(1)}(t) \}\{\widehat{S}_I^{(0),-g}(t) - \bar{S}_I^{(0)}(t) \}\\
    \{\widehat{S}_I^{(1),-g}(t) - \bar{S}_I^{(1)}(t) \}\{\widehat{S}_I^{(0),-g}(t) - \bar{S}_I^{(0)}(t) \} & \{\widehat{S}_I^{(0),-g}(t) - \bar{S}_I^{(0)}(t) \}^2 
    \end{pmatrix},
\end{align*}
where \(\bar{S}_C^{(a)}(t) = \frac{1}{M} \sum_{g=1}^{M} \widehat{S}_{C}^{(a),-g}(t)\) and \(\bar{S}_I^{(a)}(t) = \frac{1}{M} \sum_{g=1}^{M} \widehat{S}_{I}^{(a),-g}(t)\) for \(a=0,1\). To further address potential small-sample bias, we construct the confidence intervals using the \(t(M-2)\) distribution; this was adapted from the between-within degrees of freedom considered in prior simulations based on a model with an intercept and a cluster-level treatment indicator \citep{li2017evaluation}.

\section{Empirical validation of the proposed estimators}\label{simulation}
\subsection{Simulation design}\label{simulation1}

We simulate parallel-arm CRTs with a binary treatment and survival outcome to evaluate the finite-sample performance of the proposed methods. Each cluster \(i\) is independently randomized to either the treatment (\(A_i = 1\)) or usual care (\(A_i = 0\)) arm, with equal probability \(\pi^{(1)} =\pi^{(0)}= 0.5\); that is, a set of Bernoulli trials are simulated. We assumed \(M=50\) or \(M=26\) clusters, and the cluster size \(N_i\) followed a discrete uniform distribution with mean equal to \(110\), and coefficient of variation for cluster sizes equal to \(0.475\). For each participant \(j\) in cluster \(i\), the covariate vector \(\bm{V}_{ij}\) comprises two cluster-level covariates \(W_{i1},W_{i2}\) and two individual-level covariates \(Z_{ij1},Z_{ij2}\). Specifically, $W_{i1} \sim \text{Bernoulli}(0.5)$, $W_{i2} \sim \text{Normal}(\text{mean}=N_i/50,\text{sd}=1.5)$, $Z_{ij1} \sim \text{Normal}(\text{mean}=\log(N_i)/5, \text{sd}=1)$ and $Z_{ij2} \sim \text{Bernoulli}(0.5)$. Both $W_{i2}$ and \(Z_{ij1}\) depend on cluster size \(N_i\). 

Event and censoring times are generated on the scale of years since enrollment from arm-specific frailty Cox models of the general form described in Section \ref{frailty_cox}. That is,
\[\lambda^{(a)}(t \mid \mu_{ij}^{(a)}) = \lambda_0^{(a)} B_i^{(a)} \exp(\mu_{ij}^{(a)}) ,\quad  h^{(a)}(t \mid \mu_{ij}^C) = h_0^{(a)} R_i \exp(\upsilon_{ij}^{(a)} ), \]
for \(A_i=a\), where \(\mu_{ij}^{(a)}\) and \(\upsilon_{ij}^{(a)}\) denote the linear predictors for the event and censoring times, respectively. The event linear predictor was specified as
\[
\mu_{ij}^{(a)} = \beta_a a + \bm{\beta}^\top \bm{Q}_{ij} + \beta_{aN}\, aN_i/50,
\]
and the censoring linear predictor as
\[
\upsilon_{ij}^{(a)} = \bm{\alpha}^\top \bm{Q}_{ij},
\]
where $\bm{Q}_{ij}=(W_{i1},\,W_{i2},\,Z_{ij1},\,Z_{ij2},\,Z_{ij1}Z_{ij2},\,N_i/50)^\top$. Here $\bm{\beta}$ and $\bm{\alpha}$ are vectors of regression coefficients, and $\beta_a,\beta_{aN}$ control the direct effect of treatment and its interaction with cluster size.  The baseline hazards were parameterized as
\[
\lambda_0^{(a)} = \{\rho_0 - \rho_1(1-a)\}\, g_\lambda(N_i),
\qquad
h_0^{(a)} = \delta_0\, g_h(N_i),
\]
where $(\rho_0,\rho_1,\delta_0)$ denote scenario-specific parameters, and $g_\lambda(\cdot), g_h(\cdot)$ are either constant $1$ or proportional to $N_i/100$, depending on the design. The cluster-level frailty for event time, \(B_i^{(1)} \sim \text{Gamma}(\text{shape}=2, \text{rate}=2)\) for \(a=1\) (corresponding to Kendall's \(\tau=0.2\)), and \(B_i^{(0)} \sim \text{Gamma}(\text{shape}=4.5, \text{rate}=4.5)\) for \(a=0\) (corresponding to Kendall's \(\tau=0.1\)). This arm-specific frailty distribution induces differential association structures among the event times in treated and control clusters. Such a structure reflects realistic scenarios in pragmatic trials where treatment may have an impact on the degree of correlation beyond its impact on the event time marginal distribution. The frailty for censoring \(R_i \sim  \text{Gamma}(\text{shape}=9.5, \text{rate}=9.5)\), corresponds to Kendall's \(\tau=0.05\) and is assumed identical across arms for simplicity. Under this setup, the true censoring process is assumed to vary with baseline characteristics but is unaffected by treatment assignment. 
%That is, we allowed the true censoring process to vary with covariates but not with treatment, so as to focus on the main complication keep the focus of the simulation on outcome processes and informative cluster size. 
We imposed administrative censoring at 5 years as the maximum follow-up. The censoring mechanism is empirically calibrated such that, together with different covariate-dependent censoring generation, the overall censoring is approximately $25\%$, $50\%$, and $75\%$, with baseline hazard parameters tuned through Monte Carlo approximation to match the desired marginal censoring rate. To assess the robustness of our main simulation findings, we considered three main scenarios together with three additional variations, differing in the degree of informative cluster size, censoring, and number of clusters. The specifications are summarized in Table \ref{tab:scenarios}.  
% Requires: \usepackage{booktabs,makecell}
% Requires: \usepackage{booktabs,tabularx,makecell}
% Requires: \usepackage{booktabs,tabularx,makecell}
\begin{table}[htbp]
\centering
\renewcommand{\arraystretch}{1.2}
\caption{A brief summary of simulation scenarios. Vectors $(\beta_a,\bm{\beta},\beta_{aN})$ specify the event linear predictor, $\bm{\alpha}$ specifies the censoring linear predictor, and $(\rho_0,\rho_1,\delta_0)$ are parameters in baseline hazards $\lambda_0^{(a)}$ and $h_0^{(a)}$. Functions $g_\lambda(\cdot)$ and $g_h(\cdot)$ determine cluster-size dependence in baseline hazards. Administrative censoring was set at 5 years.}
\label{tab:scenarios}
{\footnotesize
\begin{tabularx}{\textwidth}{@{}lccccc c@{}}
\toprule
Simulation & \multicolumn{2}{c}{Log hazard ratios} &  \multicolumn{2}{c}{Baseline hazards parameters} & Sample size & Marginal\\
{scenario}
& \centering $(\beta_a,\ \bm{\beta},\ \beta_{aN})$
& \centering $\bm{\alpha}$
& \centering $(\rho_0,\ \rho_1,\ \delta_0)$
& \centering $g_\lambda(N_i),\ g_h(N_i)$
& {$M$}
& {Censoring rate} \\
\midrule
1
& \makecell[l]{$(-1.5,\ 0.5,\ 0.8,\ 0.4,$\\$0.3,\ 1,\ 0,\ 0)$}
& \makecell[l]{$(0.5,\ 0.3,\ 0.3,\ 0.5,$\\$0.5,\ 0)$}
& $(0.5,\ 0.2,\ 0.2)$
& \makecell[l]{$g_\lambda(N_i)=1$\\$g_h(N_i)=1$}
& 50
& 50\% \\
2
& \makecell[l]{$(0.5,\ 0.5,\!-0.2,\ 0.4,$\\$0.3,\ 1,\ 0.4,\!-1.5)$}
& \makecell[l]{$(0.3,\ 1,\ 1,\ 0.5,$\\$1,\ 0)$}
& $(0.6,\ 0.2,\ 0.001)$
& \makecell[l]{$g_\lambda(N_i)=N_i/100$\\$g_h(N_i)=1$}
& 50
& 50\% \\
3
& \makecell[l]{$(0.5,\ 0.5,\!-0.2,\ 0.4,$\\$0.3,\ 1,\ 0.4,\!-1.5)$}
& \makecell[l]{$(0.3,\ 0.8,\ 0.6,\ 0.5,$\\$1,\ 0.4)$}
& $(0.6,\ 0.2,\ 0.001)$
& \makecell[l]{$g_\lambda(N_i)=N_i/100$\\$g_h(N_i)=N_i/100$}
& 50
& 50\% \\
3(a)
& \makecell[l]{$(0.5,\ 0.5,\!-0.2,\ 0.4,$\\$0.3,\ 1,\ 0.4,\!-1.5)$}
& \makecell[l]{$(0.3,\ 0.8,\ 0.6,\ 0.5,$\\$1,\ 0.4)$}
& $(0.6,\ 0.2,\ 0.001)$
& \makecell[l]{$g_\lambda(N_i)=N_i/100$\\$g_h(N_i)=N_i/100$}
& 26
& 50\% \\
3(b)
& \makecell[l]{$(0.5,\ 0.5,\!-0.2,\ 0.4,$\\$0.3,\ 1,\ 0.4,\!-1.5)$}
& \makecell[l]{$(0.3,\ 0.8,\ 0.6,\ 0.5,$\\$1,\ 0.4)$}
& \makecell[l]{$(0.8,\ 0.2, 0.0005) $}
& \makecell[l]{$g_\lambda(N_i)=N_i/100$\\$g_h(N_i)=N_i/100$}
& 50
& 25\% \\
3(c)
& \makecell[l]{$(0.5,\ 0.5,\!-0.2,\ 0.4,$\\$0.3,\ 1,\ 0.4,\!-1.5)$}
& \makecell[l]{$(0.3,\ 0.8,\ 0.6,\ 0.5,$\\$1,\ 0.4)$}
& \makecell[l]{$(0.6,\ 0.2, 0.04)$}
& \makecell[l]{$g_\lambda(N_i)=N_i/100$\\$g_h(N_i)=N_i/100$}
& 50
& 75\% \\
\bottomrule
\end{tabularx}
}
\end{table}

The three simulation scenarios differ in the extent of informative cluster size for event process and treatment effect, and censoring mechanism. In Scenario 1 (where we change $W_{i2} \sim \text{Normal}(\text{mean}=1,\text{sd}=1.5)$ and $Z_{ij1} \sim \text{Normal}(1, \text{sd}=1)$), neither the event time nor the censoring time depends on the cluster size (i.e., no informative cluster size). Scenario 2 introduces informative cluster size for the event time and the treatment effect only while the censoring time remains independent of cluster size. Scenario 3 extends this structure by allowing informative cluster size to affect the event time, treatment effect, as well as the censoring time. Under this setup, Scenario 1 does not induce divergence between the cluster-average and individual-average estimands, while Scenario 2 and 3 are designed to illustrate this divergence under increasing degrees of informative cluster size. It is worth mentioning that, to the extent feasible, our choice of simulation parameters was intended to approximate typical practice in CRTs. For example, the effect sizes were chosen such that the corresponding hazard ratios for each covariate fall within a range commonly observed in real-world applications (between approximately 0.2 and 2.7). Commonly reported ICC values in CRTs typically do not exceed 0.2 \citep{eldridge2004lessons}, and \cite{meng2023simulating} indicated that the ICC is generally greater in magnitude than the corresponding Kendall’s $\tau$, especially at the lower tail. We therefore calibrated frailty variances to reflect moderate within-cluster correlation (e.g., Kendall’s \(\tau\) ranging from 0.05 to 0.2).

\begin{comment}

%, which reflects the consideration that loss to follow-up often depends on patient attributes rather than treatment itself. 
To assess the robustness of our main simulation findings, we consider three simulation scenarios with different \(\mu_{ij}^{(a)}\), \(\upsilon_{ij}^{(a)}\) and \(\lambda_0^{(a)}\), \(h_0^{(a)}\):
\begin{itemize}
    \item \textbf{Scenario 1}: 
    \begin{align*}
        &\mu_{ij}^{(a)} = -1.5a + 0.5W_{i1} + 0.8 W_{i2} + 0.4 Z_{ij1} + 0.3Z_{ij2} + Z_{ij1}Z_{ij2},\\
        & \upsilon_{ij}^{(a)} = 0.5W_{i1}+0.3W_{i2} + 0.3Z_{ij1} + 0.5Z_{ij2} + 0.5Z_{ij1}Z_{ij2},\\
        & \lambda_{0}^{(a)} = 0.5 - 0.2(1-a), \quad h_0^{(a)} = 0.2.
    \end{align*}
    \item \textbf{Scenario 2}:
    \begin{align*}
        &\mu_{ij}^{(a)} = 0.5a + 0.5W_{i1} - 0.2 W_{i2} + 0.4 Z_{ij1} + 0.3Z_{ij2} + Z_{ij1}Z_{ij2} + 0.4 N_i/50 - 1.5a N_i/50,\\
        & \upsilon_{ij}^{(a)} = 0.3W_{i1} + W_{i2} + Z_{ij1} + 0.5Z_{ij2} + Z_{ij1}Z_{ij2},\\
        & \lambda_{0}^{(a)} = \{0.6 - 0.2(1-a)\} N_i/100, \quad h_0^{(a)} = 0.001.
    \end{align*}
    \item \textbf{Scenario 3}:
    \begin{align*}
        &\mu_{ij}^{(a)} = 0.5a + 0.5W_{i1} - 0.2 W_{i2} + 0.4 Z_{ij1} + 0.3Z_{ij2} + Z_{ij1}Z_{ij2} + 0.4 N_i/50 - 1.5a N_i/50,\\
        & \upsilon_{ij}^{(a)} = 0.3W_{i1} + W_{i2} + Z_{ij1} + 0.5Z_{ij2} + Z_{ij1}Z_{ij2} + 0.4N_i/50,\\
        & \lambda_{0}^{(a)} = \{0.6 - 0.2(1-a)\} N_i/100, \quad h_0^{(a)} = 0.001N_i/100.
    \end{align*}
\end{itemize}
\end{comment}

Our target estimands include $S_{C}^{(a)}(t)$, $S_{I}^{(a)}(t)$, \(\mu_C^{(a)}(t)\), \(\mu_I^{(a)}(t)\) as well as the associated c-SPCE, i-SPCE, c-RMST and i-RMST on the difference scale. The true values of the estimands were obtained through a large-scale Monte Carlo procedure based on the known data-generating mechanism. For an individual $j$ in cluster $i$ assigned to arm $a$, the conditional survival probability is
\[S_{ij}^{(a)}(t \mid \bm{V}_{ij}, B_i) 
= \exp\!\left\{ - \lambda_0^{(a)} \, t \, B_i \, \exp\!\big( \mu_{ij}^{(a)} \big) \right\}.\]
We generated $10^6$ synthetic clusters, simulated covariates and frailty terms, and evaluated $S_{ij}^{(a)}(t \mid \bm{V}_{ij}, B_i)$ accordingly. These conditional survival functions were averaged within clusters to form cluster-level survival probabilities, which in turn were used to compute the true marginal estimands $\mu_C^{(a)}(t)$, $\mu_I^{(a)}(t)$, c-SPCE, i-SPCE, c-RMST, and i-RMST based on definitions in Section \ref{estimand}.

We compare 13 estimation strategies in the simulation study, and there specifications are given below.
\begin{enumerate}
\item (marginal-o1c1) the proposed method with correct specification of \(K_c^{(a)}(t\mid \bm{V}_{ij})\) and \(P(T_{ij}^{(a)} \geq t \mid \bm{V}_{ij} )\) using marginal Cox models that are compatible with the true data generating process.
\item (marginal-o1c0) the proposed method with mis-specified \(K_c^{(a)}(t\mid \bm{V}_i)\) and correctly specified \(P(T_{ij}^{(a)} \geq t \mid \bm{V}_{ij} )\) using marginal Cox models; here and throughout, censoring model mis-specification refers to omitting the interaction term \(Z_{ij1}Z_{ij2}\) and the cluster size \(N_i\) (when applicable) in estimating \(K_c^{(a)}(t\mid \bm{V}_i)\).
\item (marginal-o0c1) the proposed method with correctly specified  \(K_c^{(a)}(t\mid \bm{V}_{ij})\) and mis-specified \(P(T_{ij}^{(a)} \geq t \mid \bm{V}_{ij} )\) using marginal Cox models; here and throughout, outcome model mis-specification refers to omitting the interaction term \(Z_{ij1}Z_{ij2}\) and the cluster size \(N_i\) (when applicable) in estimating \(P(T_{ij}^{(a)} \geq t \mid \bm{V}_{ij} )\).
\item (marginal-o0c0) the proposed method with mis-specified \(K_c^{(a)}(t\mid \bm{V}_i)\) and \(P(T_{ij}^{(a)} \geq t \mid \bm{V}_{ij} )\) using marginal Cox models.
\item (frailty-o1c1) the proposed method with correct specification of \(K_c^{(a)}(t\mid \bm{V}_{ij})\) and \(P(T_{ij}^{(a)} \geq t \mid \bm{V}_{ij} )\) using frailty Cox models with Gamma frailty that match the data generating process.
\item (frailty-o1c0) the proposed method with mis-specified  \(K_c^{(a)}(t\mid \bm{V}_i)\) and correctly specified \(P(T_{ij}^{(a)} \geq t \mid \bm{V}_{ij} )\) using frailty Cox models with Gamma frailty.
\item (frailty-o0c1) the proposed method with correctly specified  \(K_c^{(a)}(t\mid \bm{V}_{ij})\) and mis-specified \(P(T_{ij}^{(a)} \geq t \mid \bm{V}_{ij} )\) using frailty Cox model with Gamma frailty.
\item (frailty-o0c0) the proposed method with mis-specified \(K_c^{(a)}(t\mid \bm{V}_i)\) and \(P(T_{ij}^{(a)} \geq t \mid \bm{V}_{ij} )\) using frailty Cox models with Gamma frailty.
\item (marginal-OR1) the outcome regression approach with correctly specified marginal Cox models, where the marginal Cox model was fitted adjusting for covariates within each treatment arm as discussed in \eqref{breslow_t} and the estimated survival probabilities 
\(\widehat{S}_{ij}^{(a)}(t \mid \bm{V}_{ij}) = \exp\left( -\int_{0}^{t} \widehat{\lambda}^{(a)}(u \mid \bm{V}_{ij}) \, du \right)\) were summarized at either the cluster or individual level by
\begin{align}
    \widehat{S}_C^{(a)}(t) = \frac{1}{M} \sum_{i=1}^{M} \frac{1}{N_i} \sum_{j=1}^{N_i} \widehat{S}_{ij}^{(a)}(t\mid\bm{V}_{ij}) \;, \quad \widehat{S}_I^{(a)}(t) = \frac{1}{\sum_{i=1}^{M}N_i} \sum_{i=1}^{M}\sum_{j=1}^{N_i} \widehat{S}_{ij}^{(a)}(t\mid\bm{V}_{ij}). \label{OR_est}
\end{align}
\item (marginal-OR0) the outcome regression approach with mis-specified marginal Cox models, where the mis-specification arises from omitting the interaction term $Z_{ij1}Z_{ij2}$ and the cluster size $N_i$, when applicable.
\item (frailty-OR1) the outcome regression with correctly specified frailty Cox models; the Laplace transformation was applied to estimate the marginal survival distribution as described in \eqref{frailty_t}, before the cluster- and individual-level estimators are calculated based on \eqref{OR_est}; 
\item (frailty-OR0) the outcome regression with mis-specified frailty Cox models, where the mis-specification arises from omitting the interaction term $Z_{ij1}Z_{ij2}$ and the cluster size $N_i$, when applicable.
\item (KM) the Kaplan-Meier estimator stratified by treatment group \(A_i\). In this approach, cluster-level estimators \( \widehat{S}_C^{(a)}(t) \) are obtained by weighting participants by the inverse of cluster size, \( 1/N_i \), while individual-level estimators \( \widehat{S}_I^{(a)}(t) \) use equal weights of 1 \citep{kahan2023estimands}. The estimators are computed separately for each treatment group (\( a = 0,1 \)).
\end{enumerate}
We conduct \(1000\) Monte Carlo iterations, and for reproducibility purposes, we use an iteration-indexed seeding rule for the random-number generator. We report the percentage bias (PBias), the empirical standard error from jackknife variance (AESE) described in Section \ref{jk_var} (cluster jackknife was also applied to outcome regression and Kaplan-Meier methods), the standard deviation from Monte Carlo simulations (MCSD), and empirical coverage probability of the nominal 95\% confidence interval (CP). The percentage bias (PBias) for a given estimator at time $t$ is defined as \(\left| \{E[\widehat{\Delta}(t)] - \Delta^{\text{true}}(t)\}/{\Delta^{\text{true}}(t)} \right| \times 100 \%\), and addresses whether the point estimator is targeting the right estimand on average. The MCSD metric reflects the empirical variability of the estimator across repeated experiments, whereas AESE reports the average standard error estimated from the cluster jackknife method. By comparing AESE with MCSD, one can assess the quality of uncertainty quantification and evaluate whether the estimated standard error is on average close to the simulation truth MCSD. The CP metric summarizes the proportion of intervals that contain the true value and evaluates the empirical validity of the estimated 95\% confidence intervals. Finally, although our primary simulation study includes three data generating processes with \(M = 50\) clusters and a 50\% censoring rate, we have extended our primary simulations to the following scenarios for robustness check. First, we examined a smaller sample size scenario with \(M = 26\) clusters under Scenario 3. Second, we considered two alternative censoring rates (25\% and 75\%) under Scenario 3 with \(M = 50\). The results under these additional scenarios are presented in Web Appendix D, and will be discussed in Section \ref{results1}.

\subsection{Simulation results}
\label{results1}

%\import{./resultsRev}{M50_C50_ICS_E0_C0.tex}

\begin{table}[htbp]
\centering
\caption{Simulation results for cluster-level survival probabilities \(S_C^{(a)}(t)\) and causal effects \(\Delta_C^{\text{SPCE}}\) on different scales under Scenario 1, estimated under 13 methods at time points \( t = \{0.1, 0.5, 1\} \). The number of clusters is \( M = 50 \), with censoring rate of approximately \( 50\% \). Reported metrics include PBias (percentage bias), MCSD (Monte Carlo standard deviation), AESE (average estimated standard error from the jackknife procedure), and CP (empirical coverage probability of the \( 95\% \) confidence interval).}
\label{M50_C50_ICS_E0_C0_Sc}
\resizebox{\textwidth}{!}{
\begin{tabular}{@{}l r 
  rrrr   rrrr   rrrr@{}}
\toprule
Method & \(t\) 
  & \multicolumn{4}{c}{$S_C^{(1)}(t)$} 
  & \multicolumn{4}{c}{$S_C^{(0)}(t)$} 
  & \multicolumn{4}{c}{$\Delta_C^{\mathrm{SPCE}}(t)$} \\
\cmidrule(lr){3-6} \cmidrule(lr){7-10} \cmidrule(lr){11-14}
 & & PBias & MCSD & AESE & CP 
     & PBias & MCSD & AESE & CP 
     & PBias & MCSD & AESE & CP \\
\midrule
\multicolumn{14}{@{}c}{\textbf{\textit{Doubly robust estimator (Marginal Cox)}}}\\
marginal-o1c1 & 0.1 & 0.112 & 0.027 & 0.028 & 0.932 
                      & 0.347 & 0.033 & 0.034 & 0.952 
                      & 2.506 & 0.028 & 0.030 & 0.960 \\
              & 0.5 & 0.205 & 0.042 & 0.043 & 0.949 
                      & 0.535 & 0.039 & 0.041 & 0.948 
                      & 1.735 & 0.038 & 0.039 & 0.950 \\
              & 1.0 & 0.196 & 0.046 & 0.047 & 0.953 
                      & 0.774 & 0.037 & 0.038 & 0.936 
                      & 1.567 & 0.040 & 0.042 & 0.956 \\
marginal-o1c0 & 0.1 & 0.183 & 0.027 & 0.028 & 0.928 
                      & 0.320 & 0.031 & 0.034 & 0.959 
                      & 2.802 & 0.028 & 0.030 & 0.964 \\
              & 0.5 & 0.153 & 0.042 & 0.043 & 0.945 
                      & 0.834 & 0.037 & 0.041 & 0.960 
                      & 2.196 & 0.039 & 0.039 & 0.947 \\
              & 1.0 & 0.054 & 0.046 & 0.047 & 0.952 
                      & 1.137 & 0.035 & 0.038 & 0.948 
                      & 1.736 & 0.041 & 0.041 & 0.949 \\
marginal-o0c1 & 0.1 & 0.094 & 0.027 & 0.028 & 0.934 
                      & 0.310 & 0.032 & 0.034 & 0.957 
                      & 2.201 & 0.028 & 0.030 & 0.961 \\
              & 0.5 & 0.187 & 0.042 & 0.043 & 0.950 
                      & 0.533 & 0.038 & 0.041 & 0.958 
                      & 1.676 & 0.038 & 0.040 & 0.951 \\
              & 1.0 & 0.180 & 0.046 & 0.047 & 0.954 
                      & 0.847 & 0.036 & 0.038 & 0.941 
                      & 1.631 & 0.040 & 0.042 & 0.957 \\
marginal-o0c0 & 0.1 & 0.240 & 0.027 & 0.028 & 0.937 
                      & 0.179 & 0.032 & 0.034 & 0.957 
                      & 2.426 & 0.028 & 0.030 & 0.965 \\
              & 0.5 & 0.772 & 0.042 & 0.043 & 0.949 
                      & 0.149 & 0.039 & 0.041 & 0.955 
                      & 2.061 & 0.038 & 0.040 & 0.951 \\
              & 1.0 & 1.183 & 0.046 & 0.047 & 0.958 
                      & 0.528 & 0.037 & 0.038 & 0.942 
                      & 2.109 & 0.040 & 0.042 & 0.955 \\
\midrule
\multicolumn{14}{@{}c}{\textbf{\textit{Doubly robust estimator (Frailty Cox)}}}\\
frailty-o1c1 & 0.1 & 0.256 & 0.027 & 0.028 & 0.946 
                      & 0.427 & 0.033 & 0.034 & 0.964 
                      & 3.816 & 0.027 & 0.029 & 0.962 \\
              & 0.5 & 0.541 & 0.041 & 0.043 & 0.956 
                      & 0.606 & 0.039 & 0.041 & 0.950 
                      & 2.913 & 0.037 & 0.040 & 0.955 \\
              & 1.0 & 0.710 & 0.045 & 0.047 & 0.963 
                      & 0.718 & 0.037 & 0.038 & 0.944 
                      & 2.728 & 0.039 & 0.043 & 0.956 \\
frailty-o1c0 & 0.1 & 0.265 & 0.026 & 0.028 & 0.948 
                      & 0.397 & 0.033 & 0.034 & 0.964 
                      & 3.716 & 0.027 & 0.030 & 0.964 \\
              & 0.5 & 0.572 & 0.041 & 0.043 & 0.959 
                      & 0.557 & 0.039 & 0.041 & 0.949 
                      & 2.908 & 0.036 & 0.040 & 0.960 \\
              & 1.0 & 0.676 & 0.045 & 0.047 & 0.968 
                      & 0.643 & 0.037 & 0.038 & 0.946 
                      & 2.539 & 0.038 & 0.043 & 0.959 \\
frailty-o0c1 & 0.1 & 0.226 & 0.027 & 0.028 & 0.949 
                      & 0.442 & 0.033 & 0.034 & 0.964 
                      & 3.709 & 0.027 & 0.029 & 0.961 \\
              & 0.5 & 0.522 & 0.041 & 0.043 & 0.963 
                      & 0.622 & 0.039 & 0.041 & 0.951 
                      & 2.887 & 0.036 & 0.040 & 0.960 \\
              & 1.0 & 0.666 & 0.044 & 0.047 & 0.965 
                      & 0.717 & 0.037 & 0.038 & 0.947 
                      & 2.621 & 0.039 & 0.043 & 0.963 \\
frailty-o0c0 & 0.1 & 0.408 & 0.026 & 0.028 & 0.949 
                      & 0.208 & 0.033 & 0.034 & 0.961 
                      & 3.615 & 0.027 & 0.029 & 0.961 \\
              & 0.5 & 1.208 & 0.040 & 0.043 & 0.962 
                      & 0.130 & 0.039 & 0.041 & 0.952 
                      & 3.439 & 0.036 & 0.040 & 0.957 \\
              & 1.0 & 1.833 & 0.044 & 0.047 & 0.967 
                      & 0.575 & 0.037 & 0.039 & 0.956 
                      & 3.610 & 0.039 & 0.043 & 0.964 \\
\midrule
\multicolumn{14}{@{}c}{\textbf{\textit{Outcome regression \& KM}}}\\
marginal-OR1 & 0.1 & 0.128 & 0.028 & 0.029 & 0.939 
                      & 0.753 & 0.032 & 0.035 & 0.954 
                      & 3.129 & 0.029 & 0.031 & 0.959 \\
              & 0.5 & 0.237 & 0.043 & 0.044 & 0.947 
                      & 1.220 & 0.037 & 0.041 & 0.960 
                      & 1.798 & 0.040 & 0.041 & 0.953 \\
              & 1.0 & 0.356 & 0.047 & 0.048 & 0.949 
                      & 1.328 & 0.035 & 0.039 & 0.953 
                      & 1.016 & 0.042 & 0.043 & 0.939 \\
marginal-OR0 & 0.1 & 0.455 & 0.029 & 0.030 & 0.949 
                      & 1.777 & 0.032 & 0.035 & 0.954 
                      & 6.438 & 0.030 & 0.032 & 0.954 \\
              & 0.5 & 0.471 & 0.043 & 0.045 & 0.953 
                      & 1.512 & 0.036 & 0.040 & 0.961 
                      & 1.684 & 0.040 & 0.041 & 0.951 \\
              & 1.0 & 0.059 & 0.046 & 0.048 & 0.953 
                      & 0.058 & 0.034 & 0.038 & 0.956 
                      & 0.062 & 0.042 & 0.043 & 0.943 \\
frailty-OR1 & 0.1 & 2.230 & 0.026 & 0.027 & 0.869 
                      & 1.888 & 0.032 & 0.034 & 0.928 
                      & 4.016 & 0.029 & 0.031 & 0.964 \\
              & 0.5 & 6.875 & 0.042 & 0.044 & 0.851 
                      & 6.430 & 0.041 & 0.042 & 0.918 
                      & 7.796 & 0.042 & 0.044 & 0.951 \\
              & 1.0 & 10.542 & 0.047 & 0.049 & 0.842 
                      & 10.236 & 0.040 & 0.041 & 0.918 
                      & 10.973 & 0.045 & 0.049 & 0.949 \\
frailty-OR0 & 0.1 & 1.313 & 0.027 & 0.028 & 0.919 
                      & 0.341 & 0.033 & 0.034 & 0.952 
                      & 6.382 & 0.029 & 0.032 & 0.961 \\
              & 0.5 & 5.924 & 0.043 & 0.044 & 0.875 
                      & 5.354 & 0.040 & 0.041 & 0.934 
                      & 7.102 & 0.041 & 0.044 & 0.952 \\
              & 1.0 & 10.042 & 0.048 & 0.049 & 0.848 
                      & 10.346 & 0.039 & 0.040 & 0.913 
                      & 9.612 & 0.045 & 0.048 & 0.950 \\
KM          & 0.1 & 1.236 & 0.031 & 0.032 & 0.891 
                      & 0.647 & 0.042 & 0.043 & 0.936 
                      & 4.306 & 0.052 & 0.055 & 0.956 \\
              & 0.5 & 5.752 & 0.050 & 0.051 & 0.870 
                      & 4.617 & 0.052 & 0.054 & 0.940 
                      & 8.100 & 0.072 & 0.075 & 0.944 \\
              & 1.0 & 10.358 & 0.055 & 0.057 & 0.840 
                      & 8.717 & 0.050 & 0.052 & 0.931 
                      & 12.677 & 0.074 & 0.077 & 0.938 \\
\bottomrule
\end{tabular}
}
\end{table}

Table \ref{M50_C50_ICS_E0_C0_Sc} presents the results for the cluster-level survival probability \(S_c^{(a)}(t)\) and \(\Delta_C^{\text{SPCE}}\) at time \(t=0.1,0.5, 1\) in Scenario 1. In this settings, cluster size \(N_i\) is not related to the event time nor censoring time, thus the estimands for \(S_C^{(a)}(t)\) and \(\Delta_C^{\text{SPCE}}(t)\) coincide with the \(S_I^{(a)}(t)\) and \(\Delta_I^{\text{SPCE}}(t)\) . When \( K_c^{(a)}(t, \bm{V}_{ij}) \) and \( P(T_{ij}^{(a)} \geq t \mid \bm{V}_{ij}) \) are correctly specified, the estimators are approximately unbiased with comparable Monte Carlo standard deviations under both the marginal and the frailty Cox models. The average jackknife standard error is close to the Monte Carlo standard deviation, leading to close to nominal coverage probabilities for the $95\%$ confidence intervals. When only one of \( K_c^{(a)}(t \mid \bm{V}_{ij}) \) and \( P(T_{ij}^{(a)} \geq t \mid \bm{V}_{ij}) \) is correctly specified, 
%either \( K_c^{(a)}(t \mid \bm{V}_{ij}) \) is mis-specified while \( P(T_{ij}^{(a)} \geq t \mid \bm{V}_{ij}) \) is correctly specified, or \( P(T_{ij}^{(a)} \geq t \mid \bm{V}_{ij}) \) is mis-specified while \( K_c^{(a)}(t \mid \bm{V}_{ij}) \) is correctly specified, 
the results remain similar to those obtained under the correctly specified models, regardless of whether marginal Cox or frailty Cox models are considered. However, when both models for \( K_c^{(a)}(t \mid \bm{V}_{ij}) \) and \( P(T_{ij}^{(a)} \geq t \mid \bm{V}_{ij}) \) are mis-specified, the doubly robust estimators show higher percentage bias, particularly at later time points. Additionally, the Monte Carlo standard deviations are larger than those in the previous settings, though the coverage probabilities remain around the nominal level. In contrast, the singly robust outcome regression methods are sensitive to model specification, showing poor coverage and higher percentage bias when the associated working outcome model is mis-specified, especially when the frailty Cox model is employed. Finally, the Kaplan-Meier estimator, included as a reference approach, also shows considerable percentage bias that intensifies with increasing $t$.

%\import{./resultsRev}{M50_C50_ICS_E1_C1.tex}
\begin{table}[ht!]
\centering
\caption{Simulation results for cluster-level survival probabilities and causal effects on different scales under Scenario 3, estimated under 13 methods at time points \( t = \{0.1, 0.5, 1\} \). The number of clusters is \( M = 50 \), with censoring rate of approximately \( 50\% \). Reported metrics include PBias (percentage bias), MCSD (Monte Carlo standard deviation), AESE (average estimated standard error from the jackknife procedure), and CP (empirical coverage probability of the \( 95\% \) confidence interval).}
\label{M50_C50_ICS_E1_C1_Sc}
\resizebox{\textwidth}{!}{
\begin{tabular}{@{}l r 
  rrrr   rrrr   rrrr@{}}
\toprule
Method & \(t\) 
  & \multicolumn{4}{c}{$S_C^{(1)}(t)$} 
  & \multicolumn{4}{c}{$S_C^{(0)}(t)$} 
  & \multicolumn{4}{c}{$\Delta_C^{\mathrm{SPCE}}(t)$} \\
\cmidrule(lr){3-6} \cmidrule(lr){7-10} \cmidrule(lr){11-14}
 & & PBias & MCSD & AESE & CP 
     & PBias & MCSD & AESE & CP 
     & PBias & MCSD & AESE & CP \\
\midrule
\multicolumn{14}{@{}c}{\textbf{ \textit{Doubly robust estimator (Marginal Cox)}}}\\
marginal-o1c1 & 0.1 & 0.053 & 0.009 & 0.010 & 0.940 
                      & 0.872 & 0.032 & 0.033 & 0.967 
                      & 2.656 & 0.033 & 0.035 & 0.969 \\
              & 0.5 & 0.287 & 0.022 & 0.023 & 0.947 
                      & 1.163 & 0.041 & 0.043 & 0.960 
                      & 1.797 & 0.047 & 0.050 & 0.962 \\
              & 1.0 & 0.450 & 0.032 & 0.032 & 0.951 
                      & 1.110 & 0.039 & 0.042 & 0.958 
                      & 1.501 & 0.052 & 0.054 & 0.960 \\
marginal-o1c0 & 0.1 & 0.078 & 0.009 & 0.010 & 0.937 
                      & 0.912 & 0.032 & 0.034 & 0.968 
                      & 2.862 & 0.033 & 0.035 & 0.969 \\
              & 0.5 & 0.278 & 0.022 & 0.024 & 0.945 
                      & 1.069 & 0.045 & 0.045 & 0.963 
                      & 1.720 & 0.049 & 0.052 & 0.969 \\
              & 1.0 & 0.415 & 0.030 & 0.032 & 0.940 
                      & 1.295 & 0.041 & 0.043 & 0.961 
                      & 1.599 & 0.052 & 0.056 & 0.961 \\
marginal-o0c1 & 0.1 & 0.193 & 0.009 & 0.010 & 0.924 
                      & 0.730 & 0.036 & 0.038 & 0.962 
                      & 2.789 & 0.036 & 0.039 & 0.965 \\
              & 0.5 & 0.773 & 0.022 & 0.024 & 0.933 
                      & 1.130 & 0.047 & 0.049 & 0.957 
                      & 2.756 & 0.051 & 0.054 & 0.949 \\
              & 1.0 & 1.072 & 0.033 & 0.033 & 0.942 
                      & 1.105 & 0.046 & 0.047 & 0.939 
                      & 2.540 & 0.055 & 0.057 & 0.948 \\
marginal-o0c0 & 0.1 & 0.323 & 0.009 & 0.010 & 0.903 
                      & 0.146 & 0.035 & 0.038 & 0.963 
                      & 1.641 & 0.036 & 0.039 & 0.969 \\
              & 0.5 & 1.163 & 0.022 & 0.024 & 0.918 
                      & 0.249 & 0.051 & 0.051 & 0.965 
                      & 2.157 & 0.053 & 0.056 & 0.958 \\
              & 1.0 & 1.581 & 0.031 & 0.033 & 0.922 
                      & 0.639 & 0.047 & 0.049 & 0.946 
                      & 2.216 & 0.055 & 0.059 & 0.954 \\
\midrule
\multicolumn{14}{@{}c}{\textbf{\textit{Doubly robust estimator (Frailty Cox)}}}\\
frailty-o1c1 & 0.1 & 0.072 & 0.009 & 0.010 & 0.937 
                      & 0.835 & 0.032 & 0.033 & 0.961 
                      & 2.623 & 0.033 & 0.035 & 0.973 \\
              & 0.5 & 0.454 & 0.020 & 0.023 & 0.953 
                      & 1.090 & 0.041 & 0.043 & 0.964 
                      & 2.063 & 0.046 & 0.049 & 0.971 \\
              & 1.0 & 0.791 & 0.027 & 0.030 & 0.958 
                      & 0.885 & 0.039 & 0.042 & 0.959 
                      & 1.919 & 0.048 & 0.053 & 0.963 \\
frailty-o1c0 & 0.1 & 0.085 & 0.009 & 0.009 & 0.939 
                      & 0.776 & 0.032 & 0.033 & 0.959 
                      & 2.508 & 0.033 & 0.035 & 0.971 \\
              & 0.5 & 0.493 & 0.020 & 0.022 & 0.948 
                      & 1.165 & 0.041 & 0.044 & 0.958 
                      & 2.222 & 0.046 & 0.050 & 0.967 \\
              & 1.0 & 0.845 & 0.027 & 0.030 & 0.955 
                      & 1.117 & 0.040 & 0.042 & 0.955 
                      & 2.167 & 0.048 & 0.053 & 0.959 \\
frailty-o0c1 & 0.1 & 0.118 & 0.009 & 0.010 & 0.942 
                      & 0.529 & 0.037 & 0.038 & 0.962 
                      & 1.939 & 0.037 & 0.039 & 0.960 \\
              & 0.5 & 0.628 & 0.021 & 0.024 & 0.951 
                      & 0.663 & 0.048 & 0.049 & 0.958 
                      & 1.973 & 0.051 & 0.054 & 0.963 \\
              & 1.0 & 1.038 & 0.028 & 0.032 & 0.954 
                      & 0.267 & 0.047 & 0.048 & 0.941 
                      & 1.918 & 0.053 & 0.057 & 0.966 \\
frailty-o0c0 & 0.1 & 0.304 & 0.009 & 0.010 & 0.920 
                      & 0.066 & 0.037 & 0.038 & 0.960 
                      & 1.344 & 0.037 & 0.039 & 0.960 \\
              & 0.5 & 1.208 & 0.022 & 0.023 & 0.921 
                      & 0.064 & 0.049 & 0.050 & 0.958 
                      & 2.400 & 0.051 & 0.054 & 0.960 \\
              & 1.0 & 1.793 & 0.029 & 0.032 & 0.925 
                      & 0.700 & 0.048 & 0.048 & 0.944 
                      & 2.529 & 0.054 & 0.058 & 0.965 \\
\midrule
\multicolumn{14}{@{}c}{\textbf{\textit{Outcome regression \& KM}}}\\
marginal-OR1 & 0.1 & 0.020 & 0.009 & 0.010 & 0.940 
                      & 1.599 & 0.032 & 0.033 & 0.961 
                      & 4.421 & 0.033 & 0.035 & 0.960 \\
              & 0.5 & 0.099 & 0.022 & 0.023 & 0.944 
                      & 2.960 & 0.040 & 0.043 & 0.956 
                      & 2.884 & 0.046 & 0.049 & 0.959 \\
              & 1.0 & 0.170 & 0.029 & 0.031 & 0.946 
                      & 3.696 & 0.039 & 0.041 & 0.950 
                      & 2.206 & 0.049 & 0.053 & 0.958 \\
marginal-OR0 & 0.1 & 0.716 & 0.009 & 0.010 & 0.832 
                      & 8.799 & 0.038 & 0.040 & 0.689 
                      & 27.481 & 0.039 & 0.041 & 0.638 \\
              & 0.5 & 2.466 & 0.022 & 0.024 & 0.831 
                      & 16.542 & 0.040 & 0.043 & 0.619 
                      & 22.274 & 0.046 & 0.049 & 0.548 \\
              & 1.0 & 3.600 & 0.031 & 0.033 & 0.852 
                      & 18.591 & 0.035 & 0.037 & 0.647 
                      & 18.551 & 0.046 & 0.049 & 0.616 \\
frailty-OR1 & 0.1 & 0.259 & 0.008 & 0.009 & 0.923 
                      & 2.011 & 0.032 & 0.033 & 0.932 
                      & 4.669 & 0.033 & 0.034 & 0.944 \\
              & 0.5 & 1.686 & 0.019 & 0.022 & 0.916 
                      & 7.009 & 0.042 & 0.045 & 0.923 
                      & 3.861 & 0.047 & 0.050 & 0.967 \\
              & 1.0 & 3.247 & 0.027 & 0.029 & 0.887 
                      & 11.556 & 0.043 & 0.045 & 0.923 
                      & 2.351 & 0.050 & 0.055 & 0.966 \\
frailty-OR0 & 0.1 & 0.916 & 0.009 & 0.009 & 0.794 
                      & 2.226 & 0.033 & 0.035 & 0.942 
                      & 9.756 & 0.034 & 0.036 & 0.916 \\
              & 0.5 & 4.159 & 0.021 & 0.023 & 0.671 
                      & 3.730 & 0.043 & 0.045 & 0.950 
                      & 4.607 & 0.046 & 0.050 & 0.949 \\
              & 1.0 & 7.383 & 0.029 & 0.031 & 0.589 
                      & 13.191 & 0.044 & 0.046 & 0.890 
                      & 3.470 & 0.052 & 0.055 & 0.956 \\
KM            & 0.1 & 0.764 & 0.009 & 0.009 & 0.807 
                      & 1.313 & 0.042 & 0.042 & 0.932 
                      & 0.779 & 0.043 & 0.043 & 0.942 \\
              & 0.5 & 2.952 & 0.022 & 0.023 & 0.781 
                      & 3.851 & 0.055 & 0.056 & 0.938 
                      & 2.015 & 0.060 & 0.061 & 0.941 \\
              & 1.0 & 4.441 & 0.031 & 0.033 & 0.794 
                      & 6.155 & 0.054 & 0.055 & 0.933 
                      & 3.285 & 0.062 & 0.064 & 0.940 \\
\bottomrule
\end{tabular}
}
\end{table}

Table \ref{M50_C50_ICS_E1_C1_Sc} presents the results for the cluster-level survival probability \(S_c^{(a)}(t)\) and \(\Delta_C^{\text{SPCE}}\) at time \(t=0.1,0.5, 1\) in Scenario 3. In this settings, cluster size \(N_i\) is related to both the event time and censoring time, thus the estimands \(S_C^{(a)}(t)\) and \(\Delta_C^{\text{SPCE}}(t)\) differ from \(S_I^{(a)}(t)\) and \(\Delta_I^{\text{SPCE}}(t)\). Similar to Scenario 1, when \( K_c^{(a)}(t, \bm{V}_{ij}) \) and \( P(T_{ij}^{(a)} \geq t \mid \bm{V}_{ij}) \) are correctly specified, or one of \( K_c^{(a)}(t, \bm{V}_{ij}) \) and \( P(T_{ij}^{(a)} \geq t \mid \bm{V}_{ij}) \) is mis-specified, our proposed estimators are approximately unbiased with comparable Monte Carlo standard deviations when the nuisance functions are estimated under both the marginal and the frailty Cox models. when both models for \( K_c^{(a)}(t \mid \bm{V}_{ij}) \) and \( P(T_{ij}^{(a)} \geq t \mid \bm{V}_{ij}) \) are mis-specified, the doubly robust estimators show larger percentage bias, particularly at later time points, and coverage rates deviate from the nominal levels at some time points. In contrast, the singly robust outcome regression methods are highly sensitive to model specification, showing poor coverage and relatively large percentage bias when the working outcome model is mis-specified. Additionally, the Kaplan-Meier estimator also fails to provide accurate estimation for the cluster-level estimand. These comparisons highlight the advantage of the doubly robust estimators over traditional survival methods applied to the CRT setting.

Additional results in Web Appendix D (Tables \ref{M50_C50_ICS_E0_C0_Si} to \ref{M50_C50_ICS_E0_C0_rmsti}) for the estimators \( S_I^{(a)}(t) \), \( \mu_C^{(a)}(t) \), \( \mu_I^{(a)}(t) \), \( \Delta_I^{\text{SPCE}}(t) \), \( \Delta_C^{\text{RMST}}(t) \), and \( \Delta_I^{\text{RMST}}(t) \) mirror the patterns observed in the cluster-level results under Scenario 1, under non-informative cluster size. In contrast, Tables \ref{M50_C50_ICS_E1_C1_Si} to \ref{M50_C50_ICS_E1_C1_rmsti} present those additional results under Scenario 3, and show that the proposed methods exhibit greater bias and undercoverage when both the censoring model \( K_c^{(a)}(t \mid \bm{V}_{ij}) \) and the survival model \( P(T_{ij}^{(a)} \geq t \mid \bm{V}_{ij}) \) are mis-specified. But typically when at least one of the specified models are correct, our estimators show minimal bias and close to nominal coverage. %These findings further support the double robustness of the proposed estimators. Similar performance degradation is observed when using mis-specified outcome regression models or the Kaplan-Meier estimator. 
A similar trend can be found under Scenario 2, as shown in Tables \ref{M50_C50_ICS_E1_C0_Sc} to \ref{M50_C50_ICS_E1_C0_rmsti}.

Finally, additional simulation results under varying censoring rates and sample sizes are provided in Web Appendix D, Tables \ref{M26_C50_ICS_E1_C1_Sc} to \ref{M50_C75_ICS_E1_C1_rmsti}. Specifically, we repeat Scenario 3 but now with censoring rates of \(25\%\) and \(75\%\), as well as a smaller number of clusters setting with \(M = 26\) clusters and a censoring rate of \(50\%\). Across all four estimands, \(\Delta_C^{\text{SPCE}}\), \(\Delta_I^{\text{SPCE}}\), \(\Delta_C^{\text{RMST}}\), and \(\Delta_I^{\text{RMST}}\), the results are consistent with those observed for \(M = 50\). Moreover, under a fixed sample size (\(M = 50\)), a higher censoring rate typically leads to increased bias when both the outcome model and the censoring model are mis-specified, as the effective sample size reduces with more incompletely observed event times.

\section{Illustrative Data Example}
\label{data_ex}

The Strategies to Reduce Injuries and Develop Confidence in Elders (STRIDE) is a pragmatic, CRT designed to evaluate the effectiveness of a multifactorial intervention aimed at preventing fall injuries among older adults at risk of fall \citep{Bhasin2018,Bhasin2020}. %The intervention consisted of individualized risk assessments and tailored care plans administered by trained nurses. 
A total of 5451 community-dwelling adults aged 70 years or older were enrolled from 86 primary care across 10 health care systems, with practices randomized equally into either the active intervention group (43 practices) or the usual care group (43 practices). The trial's primary outcome was time to first serious fall-related injury with maximum follow-up time 44 months. Of the total participants, 2,802 were assigned to the active intervention group and 2,649 to the usual care group. During the study, 3,002 participants reported no fall injuries, while 1,372 experienced one fall injury, 594 experienced two, 258 experienced three, and 225 experienced four or more fall injuries. Additionally, 232 participants died during the follow-up period. The individual-average event rates for first reported fall injuries were 25.6 and 28.6 per 100 person-years in the active intervention and usual care arms, respectively.
Cluster sizes varied widely across practices, ranging from 10 to 199 participants, with a mean cluster size of 63 and a coefficient of variation of 0.517; such non-trivial variation may reflect systematic differences in cluster-level characteristics (e.g., provider type or patient population) that are also associated with the outcome. This potential dependence between cluster size and outcome suggests the possibility of informative cluster size and hence motivates us to consider our methods.

For illustration, we consider the survival outcome as the time to the first self-reported fall injury, with right-censoring applied due to loss to follow-up, death, or study termination. This outcome has been previously analyzed \cite{blaha2022design} and \cite{chen2023clustered} using model-based methods. Among the participants, 2449 reported at least once fall injury, with 1238 events occurring in the usual care group and 1221 in the active intervention group. We consider the following four methods: (1) the proposed doubly robust method under marginal Cox models; (2) the proposed doubly robust method under frailty Cox models; (3) direct outcome regression with standardization based on a marginal Cox model as described in Section \ref{simulation1}; and (4) Kaplan-Meier estimator, with weight \(1/N_i\) for cluster-level estimators \(S_C^{(a)}(t)\), and equal weight for individual-level estimators \(S_I^{(a)}(t)\). In the proposed methods and outcome regression, we adjust for health care system as cluster-level fixed effects and additional individual-level baseline covariates including race, gender, age, and number of chronic diseases in the censoring and, additionally, cluster size as covariates in survival time models whenever applicable. %While fitting the Cox proportional hazards model, the hazard ratio (\(\text{HR} = 0.895\)) with a \(p\)-value of $0.006$. 
The target estimands included differences in cluster-level and individual-level survival functions, \(S_C^{(a)}(t)\) and \(S_I^{(a)}(t)\), as well as differences in cluster-level and individual-level RMST, \(\Delta_C^{\text{RMST}}(t)\) and \(\Delta_I^{\text{RMST}}(t)\), at years of time $\in\{1, 1.5, 2, 2.5, 3\}$. The estimators from the four methods are presented along with their \(95\%\) confidence intervals, with standard errors calculated using the cluster jackknife method introduced in Section \ref{jk_var}. %Additionally, \(p\)-values comparing the cluster-level and individual-level estimators are also reported.

\begin{table}[htbp]
\caption{Analysis of the STRIDE study, with the cluster-level $\Delta_C^{\text{RMST}}(t)$ and individual-level $\Delta_I^{\text{RMST}}(t)$ for restricted mean survival time along with their confidence intervals at specified time points using proposed method under marginal Cox model, proposed method under frailty Cox model, direct outcome regression method, and Kaplan-Meier method.
}
\label{stride_data_rmst}
\begin{center}
\begin{tabular}{ccll}
\toprule
 & \textbf{Year} & \multicolumn{1}{c}{$\boldsymbol{\Delta_C^{\text{RMST}}(t)}$ (Confidence Interval)} & \multicolumn{1}{c}{$\boldsymbol{\Delta_I^{\text{RMST}}(t)}$ (Confidence Interval)} 
\\ 
\midrule
\multirow{4}{*}{Doubly robust estimator (Marginal Cox)} 
& 1   & 0.015 (-0.001, 0.031) & 0.022 (0.009, 0.035) \\
& 1.5 & 0.036 (0.008, 0.063)  & 0.046 (0.023, 0.069) \\
& 2   & 0.060 (0.019, 0.101)  & 0.073 (0.038, 0.109) \\
& 2.5 & 0.075 (0.021, 0.128)  & 0.093 (0.044, 0.141) \\
& 3   & 0.082 (0.015, 0.148)  & 0.104 (0.042, 0.165) \\
\midrule
\multirow{4}{*}{Doubly robust estimator (Frailty Cox)}
& 1   & 0.015 (-0.001, 0.031) & 0.022 (0.009, 0.035) \\
& 1.5 & 0.036 (0.008, 0.063)  & 0.046 (0.023, 0.069) \\
& 2   & 0.060 (0.019, 0.101)  & 0.074 (0.038, 0.109) \\
& 2.5 & 0.075 (0.021, 0.128)  & 0.093 (0.044, 0.141) \\
& 3   & 0.082 (0.016, 0.148)  & 0.104 (0.042, 0.165) \\
\midrule
\multirow{4}{*}{Outcome regression (Marginal Cox)}
& 1   & 0.019 (0.006, 0.033)  & 0.023 (0.010, 0.036) \\
& 1.5 & 0.040 (0.016, 0.064)  & 0.047 (0.024, 0.071) \\
& 2   & 0.063 (0.026, 0.100)  & 0.075 (0.039, 0.111) \\
& 2.5 & 0.078 (0.027, 0.128)  & 0.093 (0.045, 0.142) \\
& 3   & 0.084 (0.018, 0.150)  & 0.105 (0.042, 0.167) \\
\midrule
\multirow{4}{*}{Kaplan-Meier estimator} 
& 1   & 0.015 (-0.012, 0.043) & 0.021 (0.008, 0.034) \\
& 1.5 & 0.038 (-0.018, 0.094) & 0.044 (0.022, 0.067) \\
& 2   & 0.061 (-0.021, 0.143) & 0.070 (0.036, 0.105) \\
& 2.5 & 0.078 (-0.027, 0.183) & 0.088 (0.042, 0.133) \\
& 3   & 0.090 (-0.038, 0.218) & 0.098 (0.039, 0.156) \\
\bottomrule
\end{tabular}
\end{center}
\end{table}

Table \ref{stride_data_rmst} summarizes the results by presenting the treatment effect on the time to first reported fall injury. The table presents cluster-level and individual-level effects on the RMST scale for specified time points. Across all methods and time points, except for the doubly robust estimators (at year 1) and the Kaplan Meier estimator for $\Delta_C^{\text{RMST}}(t)$, the confidence intervals for the treatment effect consistently exclude zero, indicating statistically significant treatment effect in reducing the risk of the first reported fall injury at the 5\% level. Under the marginal Cox models, the proposed method yields wider confidence intervals for both \(\Delta_C^{\text{RMST}}(t)\) and \(\Delta_I^{\text{RMST}}(t)\) compared to the OR method, which is consistent with findings from the simulation results when the outcome model is correctly specified. Under the frailty Cox models, the proposed method produces results nearly identical to those obtained under the marginal Cox models, similar to the simulation study. In general, our proposed methods show some differences between the cluster-level and individual-level treatment effect estimates on the RMST scale, with larger effect estimates on the individual level. On the other hand, the Kaplan-Meier method did not detect the treatment effect signal at the cluster level possibly due to potential bias and lack of efficiency. %Additionally, Kaplan-Meier produces similar estimates for cluster-level and individual-level estimands, suggesting a limitation in distinguishing between the two estimands in CRT.

\begin{figure}[htbp]
    \centering
    \includegraphics[scale=0.35]{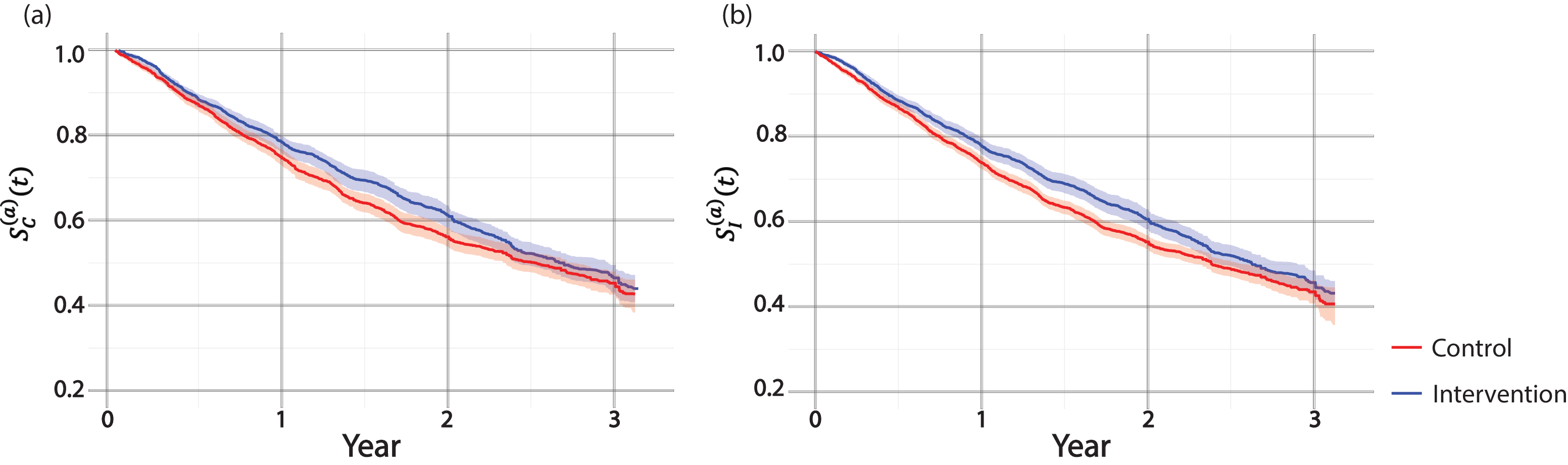}
    \caption{The cluster-level and individual-level survival curves for time to first occurrence of self-reported fall injuries estimated using the proposed doubly robust methods under marginal Cox model with the STRIDE  data. Panel (a) presents the cluster-level estimators for \(S_C^{(a)}(t)\) for \(a = 0, 1\), while Panel (b) displays the individual-level estimators for \(S_I^{(a)}(t)\) for \(a = 0, 1\).}
    \label{fig:stride_s}
\end{figure}

Figure \ref{fig:stride_s} displays the cluster-level and individual-level survival function estimates, \(\widehat{S}_C^{(a)}(t)\) and \(\widehat{S}_I^{(a)}(t)\), for \(a = 0, 1\). The survival functions correspond to the probability of experiencing self-reported fall injury over time, with separate curves for the active intervention and usual care groups. The active intervention group exhibits a relative benefit, as indicated by slightly higher survival probabilities compared to the usual care group. The \(95\%\) confidence bands for both groups did not overlap between 1 to 2 years in the cluster-level survival curves and between 0 to 2.2 years in the individual-level survival curves, indicating a non-negligible treatment effect between the two groups over these time intervals. In addition, the cluster-level and individual-level survival curves exhibit slightly different trends in the treatment effect, providing some potential evidence for informative cluster size.

\section{Discussion}
\label{discussion}

\cite{kahan2023estimands,kahan2023informative} have recently highlighted the importance of addressing informative cluster size in CRTs, and this article extends that discussion to survival data. When informative cluster size is present in survival data, the cluster-level and individual-level treatment effect estimands would carry different magnitudes; on the other hand, when the cluster size is not informative, these two estimands can still represent different units of inference conceptually \citep{hemming2023commentary} although they are numerically equivalent. Specifically, informative cluster size implies that cluster size is marginally correlated with the potential outcomes or their within-cluster contrasts, potentially leading to divergent estimates by traditional regression-based methods and estimation bias if ignored. For non-survival data, it has been shown that, when informative cluster size is present, the independence estimating equations can target i-ATE, whereas the inverse cluster size weighted generalized estimating equations or traditional cluster-level regression methods can target c-ATE \citep{wang2022two}; hence regression coefficients from model-based methods can sometimes target the treatment effect estimands defined with potential outcomes. For survival data, the presence of informative cluster size complicates estimation due to censoring and the time-dependent nature of the outcomes. Unlike the setting with non-survival data, it is often challenging to directly take the regression coefficients from the marginal Cox proportional hazards models and frailty Cox models to target the treatment effect estimands defined based on the survival function or RMST scale. Instead, our doubly robust methods serve as a principled approach to synthesize output from the outcome regression models (properly adjusting for covariate-dependent censoring and cluster size weights when applicable) to target the cluster-level and individual-level estimands.

With survival outcomes, the choice between cluster-level and individual-level treatment effect estimands can depend on the specific CRT at hand, and can depend on the type of intervention and scientific question \citep{kahan2023estimands}. For example, the cluster-level treatment effect could be of interest when evaluating a cluster-level intervention, whereas the individual-level treatment effect could often be of interest when evaluating an individual-level intervention (mimicking an individually-randomized trial that may be unfeasible or undesirable to conduct). But a study could be interested in both estimands which could provide complementary interpretations of the treatment effect. In our illustrative analysis of the STRIDE trial, an interesting finding is that the individual-level treatment effect estimates are generally larger in magnitude than their cluster-level counterparts. 
%With a multi-state outcome, we observe that only the 95\% confidence interval for the individual-level treatment effect estimand excludes the null at certain time points, suggesting that the statistical conclusion based on the conventional 5\% cutoff could change due to the unit of inference. 
This suggests some evidence of informative cluster size, and indirectly highlights the importance of estimand considerations in CRTs with heterogeneous cluster sizes. In the STRIDE trial, the active intervention being considered is a nurse-delivered, individualized fall prevention program that involves assessing and managing modifiable risk factors through personalized care plans, motivational interviewing, primary care provider collaboration, and ongoing follow-up with each patient \citep{Bhasin2018,Bhasin2020}. For this active intervention delivered to each patient, the individual-level treatment effect estimand could be of more direct interest, and actually carries a higher treatment effect size. Although this comparison between estimates suggests some evidence of informative cluster size, it is not a definitive assessment. To more formally assess the presence of informative cluster size, a statistical test assessing the difference between the cluster-level and individual-level estimators can be developed. However, if this difference is computed at one specific time point, the test may not fully capture the differences over the entire time span. More powerful testing methods that evaluate discrepancies between the cluster-level and individual-level estimators across the complete time horizon represent an area for our future research.

It is worth emphasizing that the estimands we exemplified are not exhaustive. In general, the survival function and RMST are among the most commonly used estimands for assessing treatment effects in clinical trials \citep{fay2024causal}.
%The restricted mean time in favor estimand has been introduced as a recent innovation to deal with multi-state outcomes and may be less familiar to applied statisticians. Nevertheless, we believe the connection between the restricted mean time in favor estimand and RMST is important and can potentially make the restricted mean time in favor time estimand more widely accessible; hence we provide an extension of this estimand and the associated estimation methods to CRTs. 
Beyond the estimands discussed in this article, other estimands for survival data can also be considered and merit further research in the context of CRTs. For example,  \cite{martinussen2020subtleties} introduced the \textit{causal hazard ratio}, which compares the instantaneous risk of the treatment versus usual care in a subpopulation containing participants who would have survived up to time \( t \) regardless of their treatment assignment---a continuous-time survivor average causal effect. This differs from the traditional hazard ratio, where the hazard function is conditioned on survival beyond \( t \), potentially comparing non-identical subpopulations \citep{fay2024causal}. %The causal hazard ratio addresses this selection bias, ensuring a more interpretable causal effect.\citep{axelrod2023sensitivity, fay2024causal} 
Another widely used estimand is the survival quantile, which represents the \( q \)-quantile of the survival time distribution \citep{hu2021estimating}. A special case of this is the median survival time, obtained when \( q = 0.5 \), which provides a complementary perspective to mean-based measures. 
Although these estimands were not the primary focus of this work, their extensions to CRTs under informative cluster sizes and the associated estimation methods remain of interest and will be developed in future studies. Furthermore, we focus on settings with a single time-to-event outcome and do not address competing risks in this work. When competing risks are present, alternative estimands must be defined to account for the fact that some events preclude the occurrence of the primary event of interest. Possible choices on causal estimands include the cumulative incidence function, estimands defined under hypothetical elimination of competing events \citep{young2020causal}, as well as principal stratum estimands for subpopulations \citep{nevo2022causal}. Extending the doubly robust estimation framework developed here to the more complex competing risks setting is also a fruitful direction for future research.

Our proposed doubly robust estimators are related to the two-stage framework originally described by \cite{hayes2017cluster} and more recently modernized through targeted minimum loss-based estimation (TMLE) by \cite{balzer2023two} and \cite{nugent2024blurring} for CRTs. In that framework, the causal estimand is defined as a (weighted) average of cluster-level potential outcomes. Stage 1 involves estimating cluster-specific outcomes using flexible methods (including machine learning) that can accommodate censoring, missingness, post-baseline covariates, and cluster-varying measurement patterns. Stage 2 then compares these cluster-level estimates across treatment arms, often using appropriate weights to address informative cluster size. This two-stage TMLE framework also naturally handles complex data features such as cluster-specific missingness mechanisms and post-randomization variables. In contrast, our approach constructs a single-stage, doubly robust estimator based on AIPWCC, which directly targets both cluster-level and individual-level estimands through (weighted) estimating equations. 
In their current forms, our estimators rely on identification assumptions such that right censoring depends only on fully observed baseline covariates, rather than post-baseline covariates (the potentially differential censoring process across clusters is thus reflected by heterogeneity in baseline cluster-/individual-level covariates and unmeasured frailty). While our method therefore does not yet address settings where post-treatment variables influence censoring or outcomes, in principle it could be extended in this direction, for example, by incorporating time-dependent survival models for the censoring process, as has been explored previously for independent data \citep{zhang2011estimating}. Developing such extensions in the CRT context along with clarification of the identification assumptions remains important future work. Despite these differences, both approaches are grounded in the estimand-centric principle and semiparametric efficiency theory. Two-stage TMLE achieves robustness and efficiency through substitution estimators that iteratively update initial regressions via targeted fluctuation steps, often leveraging ensemble machine learning (e.g., SuperLearner). In contrast, AIPWCC solves directly for the influence function through estimating equations, and can therefore be viewed as a one-step implementation within the broader class of targeted estimators. TMLE implementations for continuous-time survival outcomes have been developed for independent data (e.g.,\cite{cai2020one}), and to our knowledge, no extension to clustered continuous-time survival data has yet been implemented for our estimands. Expanding TMLE methods to allow flexible handling of informative censoring or outcome processes in CRTs, therefore represents a natural and important direction for future research.

Finally, we acknowledge that our exploration is confined to two commonly used models in practice: the marginal Cox model and the frailty Cox model. The marginal Cox model is computationally efficient and straightforward to implement, making it a popular choice for CRT \citep{wang2023improving}. However, it does not directly quantify within-cluster correlation by treating it as a nuisance component. In contrast, the frailty Cox model explicitly incorporates the within-cluster correlation measure, but can be computationally more demanding. 
In the empirical studies, the choice of the frailty distribution followed the true data generation scheme. A natural concern is whether mis-specification of the frailty distribution affects the accuracy of the proposed method, particularly in the context of marginalization.  \cite{gasparini2019impact} demonstrated through extensive simulations that the frailty proportional Cox model exhibits robustness to frailty distribution mis-specification. Nevertheless, exploring conditions under which frailty mis-specification may introduce bias or affect efficiency in our proposed methods, particularly in small-sample settings or in the presence of highly heterogeneous clusters, could be pursued in future work. Beyond the marginal Cox and frailty Cox models, which rely heavily on the proportional hazards assumption, alternative approaches exist for modeling censoring times and survival outcomes. For example, machine learning techniques offer flexible, non-parametric approaches with minimal assumptions. Methods such as random survival forests \citep{ishwaran2008random} and deep learning-based survival analysis \citep{hao2021deep} have shown significant promise, particularly in high-dimensional settings. These techniques can model covariate interactions in a data-adaptive fashion, often providing excellent predictive performance. As demonstrated in prior work with non-survival data \citep{wang2024model,wang2024handling}, the doubly robust estimation framework is a natural vehicle to integrate machine learning models to achieve robust and efficient influence of the c-ATE and i-ATE at the parametric rate, through a cross-fitting scheme. Thus, additional work is warranted to adapt recent debiased machine learning advancements to target our estimands, possibly extending the doubly robust machine learning method developed for non-clustered survival data to enhance model robustness and statistical efficiency \citep{westling2024inference}.

\section*{Acknowledgement}
Research in this article was supported by Patient-Centered Outcomes Research Institute Awards\textsuperscript{\textregistered} (PCORI\textsuperscript{\textregistered} Awards ME-2021C2-23685 and ME-2022C2-27676), the National Institute Of Allergy And Infectious Diseases of the National Institutes of Health under Award Number R00AI173395, and National Heart, Lung, and Blood Institute under Award Number 1R01HL178513-01. The statements presented in this article are solely the responsibility of the authors and do not necessarily represent the official views of the National Institutes of Health or PCORI\textsuperscript{\textregistered}, its Board of Governors, or the Methodology Committee. The STRIDE study was funded primarily by the Patient Centered Outcomes Research Institute (PCORI\textsuperscript{\textregistered}), with additional support from the National Institute on Aging (NIA) at NIH. Funding is provided and the award man-aged through a cooperative agreement (5U01AG048270) between the NIA and the Brigham and Women’s Hospital. The authors thank the Associate Editor and two anonymous reviewers for providing constructive comments that have improved the quality of this article.

\section*{Supplementary Material}

The supplementary material includes technical derivations and web tables referenced in the article.

\section*{Data Availability Statement}

An R package implementing our method is available at \url{https://github.com/fancy575/DRsurvCRT}, and a tutorial for using this package is provided in Web Appendix E. The STRIDE data can be obtained via \url{https://agingresearchbiobank.nia.nih.gov/studies/}, which is publicly available. The simulation code and the R code used for the illustrative example analysis can be found at \url{https://github.com/fancy575/surv_CRT}.

\clearpage
\newpage
\printbibliography

\newpage
\appendix

\section{Proof the Proposition 1}

In this section, we derive the following identity
\[
\mu_C^{(a)}(t) = \int_{0}^{t} S_C^{(a)}(u) \, du,
\]
where the cluster-level survival function is defined as
\[
S_C^{(a)}(u) = E\left[ \frac{1}{N_i} \sum_{j=1}^{N_i} \mathbb{I}(T_{ij}^{(a)} \geq u) \right].
\]
For any non-negative event time \(T_{ij}^{(a)}\), 
\[\min\{T_{ij}^{(a)},t \} = \int_{0}^{t}\mathbb{I}(T_{ij}^{(a)} \geq u ) du,  \]
almost surely.
Averaging over \(j\) within cluster \(i\) and taking expectation, by Fubini's theorem \citep{veraar2012stochastic} to interchange the order of expectation and integration, we can show that:
\begin{align*}
    \mu_C^{(a)}(t) & = E\left[ \frac{1}{N_i} \sum_{j=1}^{N_i} \min \{T_{ij}^{(a)},t \}  \right] \\
    & = E \left[\frac{1}{N_i} \sum_{j=1}^{N_i} \int_{0}^{t} \mathbb{I}\{T_{ij}^{(a)} \geq u \}du  \right] \\
    & = \int_{0}^{t} E \left[ \frac{1}{N_i} \sum_{j=1}^{N_i} \mathbb{I} (T_{ij}^{(a)} \geq u)   \right] du \\
    & = \int_{0}^{t} S_C^{(a)}(u) du.
\end{align*}

Therefore, the cluster-level estimand satisfies \(\mu_C^{(a)}(t)=\int_{0}^{t} S_C^{(a)}(u)\,du\). An entirely analogous proof gives the individual-level identity \(\mu_I^{(a)}(t)=\int_{0}^{t} S_I^{(a)}(u)\,du\), so we omit the details.

\section{Proof of the Proposition 2}

The following proof is for the estimators of the survival distribution, and the doubly robustness for RMST can be shown accordingly from Proposition 1, thus omitted here.
Define $\widetilde{N}_{ij}(u) = I(C_{ij} \leq u)$, then from \cite{bai2013doubly}, we have the following identity
\begin{align}
    \int_{0}^{t} \frac{d\widetilde{N}_{ij}(u) - \lambda_c(u\mid\bm{V}_{i})I(C_{ij} \geq u)}{K_c^{(1)}(u\mid\bm{V}_{i})}  = \int_{0}^{t} \frac{d\widetilde{M}_{ij}(t\mid\bm{V}_{i})}{K_c^{(1)}(t,\bm{V}_{i})} =1- \frac{I(C_{ij} \geq t)}{K_c^{(1)}(t\mid\bm{V}_{i})}. \label{cluster_dr_pre}
\end{align}

Assuming that $P(T_{ij}^{(1)} \geq t |\bm{V}_i) \stackrel{p}{\rightarrow} P^{*}(T_{ij}^{(1)} \geq t|\bm{V}_i)$, $K_c^{(1)}(t \mid\bm{V}_i) \stackrel{p}{\rightarrow} K_c^{(1),*}(t\mid\bm{V}_i)$, and under CRT, the treatment assignment probability \(\pi^{(a)}\) is known, then the estimators can be written as
\begin{align}
    \widehat{S}_C^{(1)}(t)  =& \frac{1}{M} \sum_{i=1}^{M} \frac{1}{N_i} \sum_{j=1}^{N_i} \Bigg[ \frac{A_i I (U_{ij} \geq t)}{\pi^{(1)}K_c^{(1)}(t\mid\bm{V}_{i})}  - \frac{A_i -\pi^{(1)}}{\pi^{(1)}} P(T_{ij}^{(1)} \geq t |\bm{V}_{i}) 
    + \frac{A_i}{\pi^{(1)}} \int_{0}^{t} \frac{dM_{c}^{(1)}(u\mid\bm{V}_{i})}{ K_c^{(1)}(u\mid\bm{V}_{i})} \left\{ \frac{P(T_{ij}^{(1)}\geq t |\bm{V}_{i})}{P(T_{ij}^{(1)} \geq u |\bm{V}_{i})} \right\}
    \Bigg] \nonumber\\
     = & \frac{1}{M} \sum_{i=1}^{M} \frac{1}{N_i} \sum_{N_i}^{j=1} \Bigg[
    \frac{A_i I(T_{ij}^{(1)} \geq t) I(C_{ij}^{(1)} \geq t)}{\pi^{(1)}K_c^{(1),*}(t\mid\bm{V}_{i})} - \frac{A_i -\pi(\bm{V}_{i})}{\pi^{(1)}} P(T_{ij}^{(1)} \geq t |\bm{V}_{i}) \nonumber\\
    & + \frac{A_i}{\pi^{(1)}} \int_{0}^{t} \frac{dM_{c}^{(1)}(u\mid\bm{V}_{i})}{ K_c^{(1),*}(u\mid\bm{V}_{i})} \left\{ \frac{P^{*}(T_{ij}^{(1)}\geq t |\bm{V}_{i})}{P^{*}(T_{ij}^{(1)} \geq u \mid\bm{V}_{i})} \right\}
    \Bigg] + o_p(1) \nonumber\\
     =& \frac{1}{M} \sum_{i=1}^{M} \frac{1}{N_i} \sum_{j=1}^{N_i} \Bigg[
    \frac{A_i I(T_{ij} \geq t)}{\pi^{(1)}}\left\{ 1 - \int_{0}^{t} \frac{d\widetilde{M}_{ij}(u\mid\bm{V}_{i})}{K_c^{(1),*}(u\mid\bm{V}_{i})} \right\} - \left\{ \frac{A_i - \pi^{(1)}}{\pi^{(1)}} \right\} P^{*}(T_{ij}^{(1)}\geq t |\bm{V}_{i}) \nonumber\\
    & + \frac{A_i}{\pi^{(1)}} \int_{0}^{t} \frac{dM_{c}(u\mid\bm{V}_{i})}{K_c^{(1),*}(u\mid\bm{V}_{i})} \left\{ \frac{P^{*}(T_{ij}^{(1)}\geq t |\bm{V}_{i})}{P^{*}(T_{ij}^{(1)} \geq u \mid\bm{V}_{i})} \right\}
    \Bigg] + o_p(1) \nonumber\\
     =& \frac{1}{M} \sum_{i=1}^{M} \frac{1}{N_i} \sum_{j=1}^{N_i} \Bigg[
    \left\{I(T_{ij} \geq t)  \right\} + \underbrace{ \left\{ \frac{A_i - \pi^{(1)}}{\pi^{(1)}} \right\} \left\{ I(T_{ij}^{(1)} \geq t) - P^{*}(T_{ij}^{(1)} \geq t\mid\bm{V}_{i}) \right\}}_{(8)} \nonumber\\
    & + \underbrace{\frac{A_i}{\pi^{(1)}} \int_{0}^{t} \frac{d\widetilde{M}_{ij}(u\mid\bm{V}_{i})}{K_c^{(1)}(u\mid\bm{V}_{i})} \left\{ \frac{I(T_{ij}^{(1)}\geq u)P^{*}(T_{ij}^{(1)} \geq t\mid\bm{V}_{i})}{P^{*}(T_{ij}^{(1)}\geq u |\bm{V}_{i})}  -I(T_{ij} \geq t) \right\} }_{(9)}\nonumber
    \Bigg] + o_p(1)
\end{align}
It can be shown that if either $P(T_{ij}^{(1)} \geq t\mid\bm{V}_{i})$ is correctly specified, $E[(8)] = 0 $. If $K_c^{(1)}(t\mid\bm{V}_i) = P(C_{ij} >t|A_i=1,\bm{V}_i)$ is correctly specified, then $\widetilde{M}_{ij}(t\mid\bm{V}_{i})$ is martingale increment, and then $E[(9)] = 0$. Then, we can show that $E[\widehat{S}_C^{(1)}(t)] = S_C^{(1)}(t)$ if either $P(T_{ij}^{(1)} \geq t\mid\bm{V}_{i})$ or $K_c^{(1)}(t\mid\bm{V}_i) = P(C_{ij} >t|A_i=1,\bm{V}_i)$ is correctly specified.
Following the similar steps, we can show that $E[\widehat{S}_I^{(1)}(t)] = S_I^{(1)}(t)$. The case \(a=0\) follows by similar steps; we omit the details here.

\section{Nonparametric structural equation model (NPSEM)} \label{NPSEM}

To formalize the hierarchical structure and dependence among variables in the cluster-randomized setting, here we define the data-generating process using a nonparametric structural equation model \citep[NPSEM,][]{pearl2009causality}, given by a series of equations:
\begin{align*}
N &= f_{N}(\varepsilon_{N}),\\
\bm{W} &= f_{\bm W}(N, \bm\varepsilon_{\bm W}),\\
\bm Z &= f_{\bm Z}(N, \bm W, \bm \varepsilon_{\bm Z}),\\
A &= f_{A}(\varepsilon_{A}),\\
\bm T &= f_{T}(N, \bm W, \bm Z, A, \bm \varepsilon_{\bm T}),\\
\bm C &= f_{C}(N, \bm W, \bm Z, A, \bm \varepsilon_{\bm C}),\\
\bm U &= \bm T \wedge^{\circ}  \bm C ,~~~\bm \Delta = \mathbb{I}\{ \bm T \leq^{\circ} \bm C\},
\end{align*}
where \(\wedge^{\circ} \) and  \(\le^{\circ}\) are component-wise versions of the operators $\wedge$ and $\le$. Here, each equation represents (in a sequential fashion) the data generator for cluster size, cluster-level covariates, individual-level covariates, treatment assignment, latent survival outcome, latent censoring time, as well as the observed survival time and censoring indicator. Each function \( f_{\cdot} \) in the NPSEM represents a mechanism for generating a random variable from underlying exogenous variation, denoted collectively as \( (\varepsilon_N, \bm\varepsilon_{\bm W}, \bm \varepsilon_{\bm Z}, \varepsilon_A, \bm \varepsilon_{\bm T}, \bm \varepsilon_{\bm C}) \). These exogenous terms are assumed to be mutually independent. The treatment assignment function \( f_A \) is determined by design, and hence does not depend on additional variables. The cluster size \( N \) may influence both the baseline covariates and the time-to-event/censoring outcomes \( \bm{T} \) and \( \bm{C} \), allowing for arbitrary forms of effect modification by cluster size. Within a cluster, dependence among individuals' event time and censoring time is induced by generating the entire vectors \(\bm T\) and \(\bm C\) at cluster-level through \(f_T\) and \(f_C\). This allows arbitrary within-cluster residual correlation in \(\bm T\) and \(\bm C\) beyond shared information in the measured covariates. The final observed data \( \bm U = \bm T \wedge^{\circ} \bm C \) and $\bm \Delta = \mathbb{I}\{\bm T \leq^{\circ} C\}$ are a deterministic function of the latent event and censoring times. 

\section{Additional simulation results}\label{additional_results}

In this section, we present additional simulation results under various scenarios, sample sizes, and censoring rates. Tables \ref{M50_C50_ICS_E0_C0_Si} to \ref{M50_C50_ICS_E0_C0_rmsti} report the individual-level survival probabilities, as well as the cluster-level and individual-level RMST estimates under Scenario 1. Tables \ref{M50_C50_ICS_E1_C1_Si} to \ref{M50_C50_ICS_E1_C1_rmsti} provide corresponding results for Scenario 3. Finally, Tables \ref{M50_C50_ICS_E1_C0_Sc} to \ref{M50_C50_ICS_E1_C0_rmsti} present results for Scenario 2.

%\import{./resultsRev/}{M50_C50_ICS_E0_C0_supp.tex}
\begin{table}[ht!]
\centering
\caption{Simulation results for individual-level survival probabilities \(S_I^{(a)}(t)\) and causal effects \(\Delta_I^{\text{SPCE}}(t)\) on different scales under Scenario 1, estimated under 13 methods at time points \( t = \{0.1, 0.5, 1\} \). The number of clusters is \( M = 50 \), with censoring rate of approximately \(50\%\). Reported metrics include PBias (percentage bias), MCSD (Monte Carlo standard deviation), AESE (average estimated standard error from the jackknife procedure), and CP (empirical coverage probability of the \(95\%\) confidence interval).}
\label{M50_C50_ICS_E0_C0_Si}
\resizebox{\textwidth}{!}{
\begin{tabular}{@{}l r 
  rrrr   rrrr   rrrr@{}}
\toprule
Method & \(t\) 
  & \multicolumn{4}{c}{$S_I^{(1)}(t)$} 
  & \multicolumn{4}{c}{$S_I^{(0)}(t)$} 
  & \multicolumn{4}{c}{$\Delta_I^{\mathrm{SPCE}}(t)$} \\
\cmidrule(lr){3-6} \cmidrule(lr){7-10} \cmidrule(lr){11-14}
 & & PBias & MCSD & AESE & CP 
     & PBias & MCSD & AESE & CP 
     & PBias & MCSD & AESE & CP \\
\midrule
\multicolumn{14}{@{}c}{\textbf{\textit{Doubly robust estimator (Marginal Cox)}}}\\
marginal-o1c1 & 0.1 & 0.221 & 0.029 & 0.030 & 0.932 
                      & 0.332 & 0.035 & 0.037 & 0.952 
                      & 3.104 & 0.030 & 0.032 & 0.967 \\
              & 0.5 & 0.419 & 0.045 & 0.047 & 0.952 
                      & 0.077 & 0.042 & 0.044 & 0.959 
                      & 1.431 & 0.040 & 0.042 & 0.961 \\
              & 1.0 & 0.618 & 0.049 & 0.051 & 0.952 
                      & 0.012 & 0.039 & 0.041 & 0.954 
                      & 1.495 & 0.043 & 0.044 & 0.957 \\
marginal-o1c0 & 0.1 & 0.301 & 0.029 & 0.030 & 0.928 
                      & 0.300 & 0.034 & 0.037 & 0.957 
                      & 3.436 & 0.030 & 0.032 & 0.961 \\
              & 0.5 & 0.475 & 0.046 & 0.047 & 0.952 
                      & 0.302 & 0.041 & 0.044 & 0.961 
                      & 2.060 & 0.041 & 0.042 & 0.955 \\
              & 1.0 & 0.596 & 0.049 & 0.051 & 0.955 
                      & 0.309 & 0.038 & 0.041 & 0.963 
                      & 1.855 & 0.044 & 0.044 & 0.943 \\
marginal-o0c1 & 0.1 & 0.201 & 0.029 & 0.030 & 0.933 
                      & 0.292 & 0.035 & 0.037 & 0.959 
                      & 2.773 & 0.030 & 0.032 & 0.970 \\
              & 0.5 & 0.395 & 0.045 & 0.047 & 0.954 
                      & 0.073 & 0.042 & 0.044 & 0.958 
                      & 1.350 & 0.040 & 0.042 & 0.963 \\
              & 1.0 & 0.587 & 0.049 & 0.051 & 0.955 
                      & 0.065 & 0.039 & 0.041 & 0.955 
                      & 1.495 & 0.043 & 0.044 & 0.958 \\
marginal-o0c0 & 0.1 & 0.367 & 0.029 & 0.030 & 0.933 
                      & 0.151 & 0.035 & 0.037 & 0.954 
                      & 3.064 & 0.030 & 0.032 & 0.967 \\
              & 0.5 & 1.032 & 0.046 & 0.047 & 0.946 
                      & 0.663 & 0.042 & 0.045 & 0.953 
                      & 1.785 & 0.041 & 0.042 & 0.963 \\
              & 1.0 & 1.694 & 0.050 & 0.051 & 0.945 
                      & 1.407 & 0.040 & 0.042 & 0.950 
                      & 2.094 & 0.043 & 0.045 & 0.958 \\
\midrule
\multicolumn{14}{@{}c}{\textbf{\textit{Doubly robust estimator (Frailty Cox)}}}\\
frailty-o1c1 & 0.1 & 0.364 & 0.029 & 0.030 & 0.940 
                      & 0.304 & 0.036 & 0.037 & 0.957 
                      & 3.849 & 0.029 & 0.031 & 0.964 \\
              & 0.5 & 0.867 & 0.044 & 0.047 & 0.960 
                      & 0.156 & 0.043 & 0.045 & 0.955 
                      & 2.317 & 0.039 & 0.043 & 0.955 \\
              & 1.0 & 1.271 & 0.048 & 0.051 & 0.960 
                      & 0.502 & 0.041 & 0.042 & 0.948 
                      & 2.342 & 0.042 & 0.046 & 0.962 \\
frailty-o1c0 & 0.1 & 0.373 & 0.029 & 0.030 & 0.942 
                      & 0.272 & 0.036 & 0.037 & 0.956 
                      & 3.734 & 0.029 & 0.031 & 0.960 \\
              & 0.5 & 0.913 & 0.044 & 0.047 & 0.963 
                      & 0.255 & 0.043 & 0.045 & 0.954 
                      & 2.257 & 0.039 & 0.043 & 0.961 \\
              & 1.0 & 1.271 & 0.048 & 0.051 & 0.961 
                      & 0.641 & 0.041 & 0.042 & 0.949 
                      & 2.148 & 0.041 & 0.046 & 0.960 \\
frailty-o0c1 & 0.1 & 0.317 & 0.029 & 0.030 & 0.943 
                      & 0.332 & 0.036 & 0.037 & 0.951 
                      & 3.700 & 0.030 & 0.031 & 0.956 \\
              & 0.5 & 0.830 & 0.044 & 0.047 & 0.963 
                      & 0.175 & 0.043 & 0.045 & 0.956 
                      & 2.168 & 0.039 & 0.042 & 0.956 \\
              & 1.0 & 1.223 & 0.048 & 0.051 & 0.960 
                      & 0.569 & 0.041 & 0.042 & 0.947 
                      & 2.134 & 0.041 & 0.046 & 0.966 \\
frailty-o0c0 & 0.1 & 0.513 & 0.029 & 0.030 & 0.940 
                      & 0.077 & 0.036 & 0.037 & 0.953 
                      & 3.586 & 0.029 & 0.031 & 0.958 \\
              & 0.5 & 1.574 & 0.044 & 0.047 & 0.956 
                      & 0.970 & 0.043 & 0.045 & 0.960 
                      & 2.805 & 0.039 & 0.043 & 0.957 \\
              & 1.0 & 2.467 & 0.048 & 0.051 & 0.954 
                      & 1.894 & 0.041 & 0.042 & 0.953 
                      & 3.265 & 0.041 & 0.046 & 0.961 \\
\midrule
\multicolumn{14}{@{}c}{\textbf{\textit{Outcome regression \& KM}}}\\
marginal-OR1 & 0.1 & 0.026 & 0.029 & 0.030 & 0.933 
                      & 0.672 & 0.034 & 0.037 & 0.960 
                      & 3.340 & 0.030 & 0.032 & 0.962 \\
              & 0.5 & 0.039 & 0.045 & 0.047 & 0.955 
                      & 0.519 & 0.040 & 0.044 & 0.962 
                      & 1.177 & 0.040 & 0.042 & 0.958 \\
              & 1.0 & 0.133 & 0.048 & 0.050 & 0.956 
                      & 0.241 & 0.037 & 0.041 & 0.967 
                      & 0.653 & 0.043 & 0.043 & 0.941 \\
marginal-OR0 & 0.1 & 0.352 & 0.030 & 0.031 & 0.942 
                      & 1.695 & 0.034 & 0.037 & 0.959 
                      & 6.647 & 0.030 & 0.032 & 0.963 \\
              & 0.5 & 0.199 & 0.045 & 0.047 & 0.951 
                      & 0.819 & 0.039 & 0.043 & 0.965 
                      & 1.066 & 0.040 & 0.041 & 0.954 \\
              & 1.0 & 0.430 & 0.048 & 0.050 & 0.955 
                      & 1.042 & 0.037 & 0.040 & 0.964 
                      & 0.423 & 0.042 & 0.043 & 0.946 \\
frailty-OR1  & 0.1 & 2.311 & 0.027 & 0.028 & 0.850 
                      & 1.961 & 0.035 & 0.036 & 0.930 
                      & 4.140 & 0.029 & 0.031 & 0.963 \\
              & 0.5 & 7.160 & 0.044 & 0.045 & 0.824 
                      & 7.184 & 0.044 & 0.045 & 0.900 
                      & 7.110 & 0.042 & 0.045 & 0.953 \\
              & 1.0 & 11.089 & 0.049 & 0.051 & 0.832 
                      & 11.470 & 0.043 & 0.043 & 0.903 
                      & 10.559 & 0.045 & 0.049 & 0.950 \\
frailty-OR0  & 0.1 & 1.386 & 0.028 & 0.029 & 0.904 
                      & 0.375 & 0.035 & 0.036 & 0.950 
                      & 6.656 & 0.030 & 0.032 & 0.962 \\
              & 0.5 & 6.169 & 0.044 & 0.045 & 0.870 
                      & 6.011 & 0.043 & 0.044 & 0.911 
                      & 6.491 & 0.041 & 0.044 & 0.950 \\
              & 1.0 & 10.528 & 0.050 & 0.051 & 0.836 
                      & 11.463 & 0.041 & 0.042 & 0.900 
                      & 9.226 & 0.045 & 0.048 & 0.949 \\
KM           & 0.1 & 1.361 & 0.034 & 0.035 & 0.873 
                      & 0.671 & 0.046 & 0.047 & 0.924 
                      & 4.958 & 0.057 & 0.059 & 0.951 \\
              & 0.5 & 6.069 & 0.054 & 0.056 & 0.860 
                      & 5.100 & 0.057 & 0.059 & 0.936 
                      & 8.045 & 0.078 & 0.082 & 0.946 \\
              & 1.0 & 10.910 & 0.060 & 0.062 & 0.846 
                      & 9.569 & 0.055 & 0.057 & 0.941 
                      & 12.776 & 0.081 & 0.085 & 0.933 \\
\bottomrule
\end{tabular}
}
\end{table}

\begin{table}[ht!]
\centering
\caption{Simulation results for cluster-level RMST \(\mu_C^{(a)}(t)\) and causal effects \(\Delta_C^{\text{RMST}}(t)\) on different scales under Scenario 1, estimated under 13 methods at time points \( t = \{0.1, 0.5, 1\} \). The number of clusters is \( M = 50 \), with censoring rate of approximately \(50\%\). Reported metrics include PBias (percentage bias), MCSD (Monte Carlo standard deviation), AESE (average estimated standard error from the jackknife procedure), and CP (empirical coverage probability of the \(95\%\) confidence interval).}
\label{M50_C50_ICS_E0_C0_rmstc}
\resizebox{\textwidth}{!}{
\begin{tabular}{@{}l r 
  rrrr   rrrr   rrrr@{}}
\toprule
Method & \(t\) 
  & \multicolumn{4}{c}{$\mu_C^{(1)}(t)$} 
  & \multicolumn{4}{c}{$\mu_C^{(0)}(t)$} 
  & \multicolumn{4}{c}{$\Delta_C^{\mathrm{RMST}}(t)$} \\
\cmidrule(lr){3-6} \cmidrule(lr){7-10} \cmidrule(lr){11-14}
 & & PBias & MCSD & AESE & CP 
     & PBias & MCSD & AESE & CP 
     & PBias & MCSD & AESE & CP \\
\midrule
\multicolumn{14}{@{}c}{\textbf{\textit{Doubly robust estimator (Marginal Cox)}}}\\
marginal-o1c1 & 0.1 & 0.081 & 0.002 & 0.002 & 0.929 
                      & 0.203 & 0.002 & 0.003 & 0.953 
                      & 2.574 & 0.002 & 0.002 & 0.962 \\
              & 0.5 & 0.155 & 0.016 & 0.017 & 0.945 
                      & 0.387 & 0.017 & 0.018 & 0.955 
                      & 2.101 & 0.015 & 0.016 & 0.950 \\
              & 1.0 & 0.176 & 0.038 & 0.039 & 0.946 
                      & 0.483 & 0.036 & 0.037 & 0.950 
                      & 1.842 & 0.034 & 0.035 & 0.955 \\
marginal-o1c0 & 0.1 & 0.142 & 0.002 & 0.002 & 0.929 
                      & 0.167 & 0.002 & 0.003 & 0.955 
                      & 2.853 & 0.002 & 0.002 & 0.966 \\
              & 0.5 & 0.153 & 0.016 & 0.017 & 0.941 
                      & 0.462 & 0.017 & 0.018 & 0.958 
                      & 2.363 & 0.016 & 0.016 & 0.953 \\
              & 1.0 & 0.142 & 0.038 & 0.039 & 0.944 
                      & 0.652 & 0.034 & 0.037 & 0.958 
                      & 2.148 & 0.034 & 0.035 & 0.951 \\
marginal-o0c1 & 0.1 & 0.069 & 0.002 & 0.002 & 0.932 
                      & 0.180 & 0.002 & 0.003 & 0.955 
                      & 2.256 & 0.002 & 0.002 & 0.960 \\
              & 0.5 & 0.140 & 0.016 & 0.017 & 0.946 
                      & 0.363 & 0.017 & 0.018 & 0.959 
                      & 1.944 & 0.015 & 0.016 & 0.956 \\
              & 1.0 & 0.157 & 0.038 & 0.039 & 0.951 
                      & 0.480 & 0.035 & 0.037 & 0.956 
                      & 1.768 & 0.034 & 0.035 & 0.954 \\
marginal-o0c0 & 0.1 & 0.138 & 0.002 & 0.002 & 0.929 
                      & 0.132 & 0.002 & 0.003 & 0.954 
                      & 2.519 & 0.002 & 0.002 & 0.963 \\
              & 0.5 & 0.425 & 0.016 & 0.017 & 0.946 
                      & 0.075 & 0.017 & 0.018 & 0.961 
                      & 2.222 & 0.015 & 0.016 & 0.956 \\
              & 1.0 & 0.658 & 0.038 & 0.039 & 0.955 
                      & 0.073 & 0.036 & 0.037 & 0.956 
                      & 2.137 & 0.034 & 0.035 & 0.953 \\
\midrule
\multicolumn{14}{@{}c}{\textbf{\textit{Doubly robust estimator (Frailty Cox)}}}\\
frailty-o1c1 & 0.1 & 0.176 & 0.002 & 0.002 & 0.941 
                      & 0.246 & 0.002 & 0.003 & 0.959 
                      & 3.879 & 0.002 & 0.002 & 0.963 \\
              & 0.5 & 0.339 & 0.016 & 0.017 & 0.951 
                      & 0.461 & 0.017 & 0.018 & 0.957 
                      & 3.210 & 0.015 & 0.016 & 0.959 \\
              & 1.0 & 0.436 & 0.037 & 0.039 & 0.955 
                      & 0.535 & 0.036 & 0.037 & 0.950 
                      & 2.889 & 0.033 & 0.036 & 0.959 \\
frailty-o1c0 & 0.1 & 0.180 & 0.002 & 0.002 & 0.945 
                      & 0.237 & 0.002 & 0.003 & 0.958 
                      & 3.847 & 0.002 & 0.002 & 0.964 \\
              & 0.5 & 0.363 & 0.016 & 0.017 & 0.954 
                      & 0.427 & 0.017 & 0.018 & 0.958 
                      & 3.197 & 0.015 & 0.016 & 0.961 \\
              & 1.0 & 0.453 & 0.037 & 0.039 & 0.960 
                      & 0.490 & 0.036 & 0.037 & 0.952 
                      & 2.839 & 0.032 & 0.036 & 0.963 \\
frailty-o0c1 & 0.1 & 0.157 & 0.002 & 0.002 & 0.947 
                      & 0.271 & 0.002 & 0.003 & 0.961 
                      & 3.915 & 0.002 & 0.002 & 0.964 \\
              & 0.5 & 0.315 & 0.016 & 0.017 & 0.959 
                      & 0.469 & 0.017 & 0.018 & 0.954 
                      & 3.127 & 0.015 & 0.016 & 0.960 \\
              & 1.0 & 0.411 & 0.037 & 0.039 & 0.964 
                      & 0.538 & 0.036 & 0.037 & 0.955 
                      & 2.812 & 0.032 & 0.035 & 0.962 \\
frailty-o0c0 & 0.1 & 0.236 & 0.002 & 0.002 & 0.943 
                      & 0.148 & 0.002 & 0.003 & 0.958 
                      & 3.613 & 0.002 & 0.002 & 0.964 \\
              & 0.5 & 0.674 & 0.016 & 0.017 & 0.960 
                      & 0.090 & 0.017 & 0.018 & 0.955 
                      & 3.419 & 0.015 & 0.016 & 0.958 \\
              & 1.0 & 1.012 & 0.036 & 0.039 & 0.960 
                      & 0.069 & 0.036 & 0.037 & 0.957 
                      & 3.396 & 0.032 & 0.036 & 0.961 \\
\midrule
\multicolumn{14}{@{}c}{\textbf{\textit{Outcome regression \& KM}}}\\
marginal-OR1 & 0.1 & 0.069 & 0.002 & 0.002 & 0.931 
                      & 0.497 & 0.002 & 0.003 & 0.954 
                      & 3.689 & 0.002 & 0.002 & 0.957 \\
              & 0.5 & 0.165 & 0.017 & 0.018 & 0.945 
                      & 0.877 & 0.017 & 0.018 & 0.960 
                      & 2.394 & 0.016 & 0.017 & 0.954 \\
              & 1.0 & 0.216 & 0.039 & 0.040 & 0.946 
                      & 1.027 & 0.034 & 0.038 & 0.962 
                      & 1.835 & 0.036 & 0.037 & 0.951 \\
marginal-OR0 & 0.1 & 0.256 & 0.002 & 0.002 & 0.941 
                      & 1.175 & 0.002 & 0.003 & 0.956 
                      & 7.824 & 0.002 & 0.002 & 0.958 \\
              & 0.5 & 0.476 & 0.017 & 0.018 & 0.950 
                      & 1.664 & 0.016 & 0.018 & 0.954 
                      & 3.788 & 0.016 & 0.017 & 0.955 \\
              & 1.0 & 0.390 & 0.039 & 0.040 & 0.949 
                      & 1.355 & 0.034 & 0.037 & 0.961 
                      & 2.050 & 0.036 & 0.037 & 0.949 \\
frailty-OR1  & 0.1 & 1.184 & 0.002 & 0.002 & 0.876 
                      & 0.958 & 0.002 & 0.002 & 0.927 
                      & 3.174 & 0.002 & 0.002 & 0.962 \\
              & 0.5 & 3.817 & 0.016 & 0.017 & 0.859 
                      & 3.263 & 0.018 & 0.018 & 0.916 
                      & 5.806 & 0.016 & 0.018 & 0.958 \\
              & 1.0 & 5.883 & 0.038 & 0.040 & 0.850 
                      & 5.124 & 0.037 & 0.039 & 0.914 
                      & 7.803 & 0.037 & 0.040 & 0.948 \\
frailty-OR0  & 0.1 & 0.593 & 0.002 & 0.002 & 0.924 
                      & 0.114 & 0.002 & 0.003 & 0.954 
                      & 6.810 & 0.002 & 0.002 & 0.959 \\
              & 0.5 & 2.895 & 0.016 & 0.017 & 0.898 
                      & 1.911 & 0.018 & 0.018 & 0.943 
                      & 6.429 & 0.016 & 0.018 & 0.954 \\
              & 1.0 & 5.042 & 0.039 & 0.040 & 0.882 
                      & 4.067 & 0.037 & 0.038 & 0.935 
                      & 7.506 & 0.037 & 0.040 & 0.950 \\
KM           & 0.1 & 0.587 & 0.002 & 0.002 & 0.895 
                      & 0.212 & 0.003 & 0.003 & 0.925 
                      & 3.889 & 0.004 & 0.004 & 0.961 \\
              & 0.5 & 2.785 & 0.019 & 0.020 & 0.887 
                      & 1.868 & 0.023 & 0.023 & 0.933 
                      & 6.077 & 0.030 & 0.031 & 0.951 \\
              & 1.0 & 5.003 & 0.045 & 0.047 & 0.873 
                      & 3.614 & 0.048 & 0.049 & 0.939 
                      & 8.517 & 0.065 & 0.068 & 0.949 \\
\bottomrule
\end{tabular}
}
\end{table}

\begin{table}[ht!]
\centering
\caption{Simulation results for individual-level RMST \(\mu_I^{(a)}(t)\) and causal effects \(\Delta_I^{\text{RMST}}(t)\) on different scales under Scenario 1, estimated under 13 methods at time points \( t = \{0.1, 0.5, 1\} \). The number of clusters is \( M = 50 \), with censoring rate of approximately \(50\%\). Reported metrics include PBias (percentage bias), MCSD (Monte Carlo standard deviation), AESE (average estimated standard error from the jackknife procedure), and CP (empirical coverage probability of the \(95\%\) confidence interval).}
\label{M50_C50_ICS_E0_C0_rmsti}
\resizebox{\textwidth}{!}{
\begin{tabular}{@{}l r 
  rrrr   rrrr   rrrr@{}}
\toprule
Method & \(t\) 
  & \multicolumn{4}{c}{$\mu_I^{(1)}(t)$} 
  & \multicolumn{4}{c}{$\mu_I^{(0)}(t)$} 
  & \multicolumn{4}{c}{$\Delta_I^{\mathrm{RMST}}(t)$} \\
\cmidrule(lr){3-6} \cmidrule(lr){7-10} \cmidrule(lr){11-14}
 & & PBias & MCSD & AESE & CP 
     & PBias & MCSD & AESE & CP 
     & PBias & MCSD & AESE & CP \\
\midrule
\multicolumn{14}{@{}c}{\textbf{\textit{Doubly robust estimator (Marginal Cox)}}}\\
marginal-o1c1 & 0.1 & 0.150 & 0.002 & 0.002 & 0.922 
                      & 0.243 & 0.003 & 0.003 & 0.949 
                      & 3.621 & 0.002 & 0.002 & 0.966 \\
              & 0.5 & 0.294 & 0.017 & 0.018 & 0.939 
                      & 0.225 & 0.018 & 0.020 & 0.956 
                      & 2.143 & 0.016 & 0.017 & 0.957 \\
              & 1.0 & 0.389 & 0.041 & 0.042 & 0.948 
                      & 0.157 & 0.038 & 0.040 & 0.957 
                      & 1.755 & 0.036 & 0.038 & 0.954 \\
marginal-o1c0 & 0.1 & 0.217 & 0.002 & 0.002 & 0.924 
                      & 0.215 & 0.003 & 0.003 & 0.949 
                      & 4.041 & 0.002 & 0.002 & 0.965 \\
              & 0.5 & 0.344 & 0.018 & 0.018 & 0.938 
                      & 0.282 & 0.018 & 0.020 & 0.959 
                      & 2.580 & 0.017 & 0.017 & 0.956 \\
              & 1.0 & 0.432 & 0.041 & 0.042 & 0.945 
                      & 0.280 & 0.037 & 0.040 & 0.961 
                      & 2.217 & 0.037 & 0.038 & 0.949 \\
marginal-o0c1 & 0.1 & 0.136 & 0.002 & 0.002 & 0.922 
                      & 0.221 & 0.003 & 0.003 & 0.956 
                      & 3.288 & 0.002 & 0.002 & 0.971 \\
              & 0.5 & 0.276 & 0.018 & 0.018 & 0.942 
                      & 0.199 & 0.018 & 0.020 & 0.957 
                      & 1.970 & 0.016 & 0.017 & 0.960 \\
              & 1.0 & 0.367 & 0.041 & 0.042 & 0.948 
                      & 0.149 & 0.038 & 0.040 & 0.958 
                      & 1.659 & 0.036 & 0.038 & 0.956 \\
marginal-o0c0 & 0.1 & 0.213 & 0.002 & 0.002 & 0.921 
                      & 0.168 & 0.003 & 0.003 & 0.953 
                      & 3.580 & 0.002 & 0.002 & 0.966 \\
              & 0.5 & 0.593 & 0.018 & 0.018 & 0.942 
                      & 0.112 & 0.019 & 0.020 & 0.956 
                      & 2.309 & 0.016 & 0.017 & 0.959 \\
              & 1.0 & 0.915 & 0.041 & 0.043 & 0.943 
                      & 0.445 & 0.039 & 0.041 & 0.957 
                      & 2.093 & 0.035 & 0.038 & 0.954 \\
\midrule
\multicolumn{14}{@{}c}{\textbf{\textit{Doubly robust estimator (Frailty Cox)}}}\\
frailty-o1c1 & 0.1 & 0.229 & 0.002 & 0.002 & 0.933 
                      & 0.243 & 0.003 & 0.003 & 0.949 
                      & 4.398 & 0.002 & 0.002 & 0.956 \\
              & 0.5 & 0.514 & 0.017 & 0.018 & 0.947 
                      & 0.140 & 0.019 & 0.020 & 0.959 
                      & 2.849 & 0.016 & 0.017 & 0.958 \\
              & 1.0 & 0.725 & 0.040 & 0.042 & 0.955 
                      & 0.034 & 0.039 & 0.041 & 0.957 
                      & 2.455 & 0.035 & 0.038 & 0.960 \\
frailty-o1c0 & 0.1 & 0.233 & 0.002 & 0.002 & 0.934 
                      & 0.234 & 0.003 & 0.003 & 0.950 
                      & 4.362 & 0.002 & 0.002 & 0.955 \\
              & 0.5 & 0.546 & 0.017 & 0.018 & 0.949 
                      & 0.090 & 0.019 & 0.020 & 0.959 
                      & 2.815 & 0.016 & 0.017 & 0.962 \\
              & 1.0 & 0.757 & 0.040 & 0.043 & 0.957 
                      & 0.109 & 0.040 & 0.041 & 0.954 
                      & 2.381 & 0.035 & 0.038 & 0.961 \\
frailty-o0c1 & 0.1 & 0.196 & 0.002 & 0.002 & 0.936 
                      & 0.278 & 0.003 & 0.003 & 0.948 
                      & 4.384 & 0.002 & 0.002 & 0.956 \\
              & 0.5 & 0.475 & 0.017 & 0.018 & 0.950 
                      & 0.149 & 0.019 & 0.020 & 0.958 
                      & 2.702 & 0.016 & 0.017 & 0.959 \\
              & 1.0 & 0.687 & 0.040 & 0.042 & 0.958 
                      & 0.044 & 0.040 & 0.041 & 0.958 
                      & 2.295 & 0.035 & 0.038 & 0.955 \\
frailty-o0c0 & 0.1 & 0.288 & 0.002 & 0.002 & 0.933 
                      & 0.141 & 0.003 & 0.003 & 0.953 
                      & 4.079 & 0.002 & 0.002 & 0.957 \\
              & 0.5 & 0.864 & 0.017 & 0.018 & 0.945 
                      & 0.259 & 0.019 & 0.020 & 0.962 
                      & 3.025 & 0.016 & 0.017 & 0.962 \\
              & 1.0 & 1.334 & 0.040 & 0.042 & 0.949 
                      & 0.686 & 0.039 & 0.041 & 0.965 
                      & 2.955 & 0.035 & 0.038 & 0.956 \\
\midrule
\multicolumn{14}{@{}c}{\textbf{\textit{Outcome regression \& KM}}}\\
marginal-OR1 & 0.1 & 0.010 & 0.002 & 0.002 & 0.926 
                      & 0.500 & 0.003 & 0.003 & 0.951 
                      & 4.322 & 0.002 & 0.002 & 0.966 \\
              & 0.5 & 0.009 & 0.018 & 0.018 & 0.944 
                      & 0.593 & 0.018 & 0.020 & 0.964 
                      & 2.074 & 0.016 & 0.017 & 0.962 \\
              & 1.0 & 0.033 & 0.041 & 0.042 & 0.954 
                      & 0.513 & 0.037 & 0.040 & 0.965 
                      & 1.401 & 0.036 & 0.037 & 0.950 \\
marginal-OR0 & 0.1 & 0.194 & 0.002 & 0.002 & 0.929 
                      & 1.176 & 0.003 & 0.003 & 0.957 
                      & 8.477 & 0.002 & 0.002 & 0.961 \\
              & 0.5 & 0.322 & 0.018 & 0.018 & 0.948 
                      & 1.383 & 0.018 & 0.019 & 0.962 
                      & 3.462 & 0.016 & 0.017 & 0.958 \\
              & 1.0 & 0.142 & 0.041 & 0.042 & 0.957 
                      & 0.845 & 0.036 & 0.040 & 0.963 
                      & 1.616 & 0.036 & 0.037 & 0.954 \\
frailty-OR1  & 0.1 & 1.228 & 0.002 & 0.002 & 0.861 
                      & 0.946 & 0.003 & 0.003 & 0.921 
                      & 3.713 & 0.002 & 0.002 & 0.959 \\
              & 0.5 & 3.960 & 0.017 & 0.017 & 0.852 
                      & 3.551 & 0.019 & 0.019 & 0.914 
                      & 5.418 & 0.017 & 0.018 & 0.958 \\
              & 1.0 & 6.136 & 0.040 & 0.041 & 0.826 
                      & 5.668 & 0.040 & 0.041 & 0.903 
                      & 7.307 & 0.038 & 0.040 & 0.954 \\
frailty-OR0  & 0.1 & 0.634 & 0.002 & 0.002 & 0.912 
                      & 0.145 & 0.003 & 0.003 & 0.950 
                      & 7.512 & 0.002 & 0.002 & 0.958 \\
              & 0.5 & 3.019 & 0.017 & 0.018 & 0.894 
                      & 2.142 & 0.019 & 0.019 & 0.937 
                      & 6.150 & 0.017 & 0.018 & 0.953 \\
              & 1.0 & 5.262 & 0.040 & 0.041 & 0.869 
                      & 4.533 & 0.040 & 0.040 & 0.915 
                      & 7.086 & 0.037 & 0.040 & 0.953 \\
KM           & 0.1 & 0.661 & 0.002 & 0.002 & 0.871 
                      & 0.176 & 0.003 & 0.003 & 0.925 
                      & 4.942 & 0.004 & 0.004 & 0.949 \\
              & 0.5 & 2.963 & 0.021 & 0.021 & 0.858 
                      & 2.036 & 0.025 & 0.026 & 0.931 
                      & 6.269 & 0.033 & 0.034 & 0.949 \\
              & 1.0 & 5.280 & 0.049 & 0.051 & 0.860 
                      & 3.966 & 0.053 & 0.054 & 0.934 
                      & 8.569 & 0.072 & 0.075 & 0.944 \\
\bottomrule
\end{tabular}
}
\end{table}

%\import{./resultsRev/}{M50_C50_ICS_E1_C1_supp.tex}
\begin{table}[ht!]
\centering
\caption{Simulation results for individual-level survival probabilities \(S_I^{(a)}(t)\) and causal effects \(\Delta_I^{\text{SPCE}}(t)\) on different scales under Scenario 3, estimated under 13 methods at time points \( t = \{0.1, 0.5, 1\} \). The number of clusters is \( M = 50 \), with censoring rate of approximately \(50\%\). Reported metrics include PBias (percentage bias), MCSD (Monte Carlo standard deviation), AESE (average estimated standard error from the jackknife procedure), and CP (empirical coverage probability of the \(95\%\) confidence interval).}
\label{M50_C50_ICS_E1_C1_Si}
\resizebox{\textwidth}{!}{
\begin{tabular}{@{}l r 
  rrrr   rrrr   rrrr@{}}
\toprule
Method & \(t\) 
  & \multicolumn{4}{c}{$S_I^{(1)}(t)$} 
  & \multicolumn{4}{c}{$S_I^{(0)}(t)$} 
  & \multicolumn{4}{c}{$\Delta_I^{\mathrm{SPCE}}(t)$} \\
\cmidrule(lr){3-6} \cmidrule(lr){7-10} \cmidrule(lr){11-14}
 & & PBias & MCSD & AESE & CP 
     & PBias & MCSD & AESE & CP 
     & PBias & MCSD & AESE & CP \\
\midrule
\multicolumn{14}{@{}c}{\textbf{\textit{Doubly robust estimator (Marginal Cox)}}}\\
marginal-o1c1 & 0.1 & 0.022 & 0.009 & 0.010 & 0.945 
                      & 0.686 & 0.035 & 0.036 & 0.951 
                      & 1.305 & 0.036 & 0.038 & 0.957 \\
              & 0.5 & 0.123 & 0.023 & 0.024 & 0.947 
                      & 0.901 & 0.036 & 0.038 & 0.957 
                      & 0.727 & 0.043 & 0.046 & 0.956 \\
              & 1.0 & 0.133 & 0.036 & 0.033 & 0.949 
                      & 1.079 & 0.030 & 0.032 & 0.961 
                      & 0.554 & 0.047 & 0.047 & 0.965 \\
marginal-o1c0 & 0.1 & 0.042 & 0.009 & 0.010 & 0.939 
                      & 0.772 & 0.036 & 0.037 & 0.953 
                      & 1.518 & 0.036 & 0.038 & 0.958 \\
              & 0.5 & 0.091 & 0.023 & 0.024 & 0.953 
                      & 0.781 & 0.043 & 0.041 & 0.959 
                      & 0.634 & 0.046 & 0.049 & 0.968 \\
              & 1.0 & 0.097 & 0.030 & 0.033 & 0.951 
                      & 1.370 & 0.032 & 0.033 & 0.956 
                      & 0.661 & 0.047 & 0.050 & 0.968 \\
marginal-o0c1 & 0.1 & 0.235 & 0.009 & 0.010 & 0.919 
                      & 0.155 & 0.038 & 0.040 & 0.948 
                      & 0.941 & 0.039 & 0.041 & 0.947 \\
              & 0.5 & 0.944 & 0.024 & 0.025 & 0.924 
                      & 0.346 & 0.041 & 0.042 & 0.956 
                      & 1.706 & 0.046 & 0.049 & 0.951 \\
              & 1.0 & 1.239 & 0.038 & 0.035 & 0.926 
                      & 0.488 & 0.034 & 0.035 & 0.954 
                      & 1.842 & 0.050 & 0.050 & 0.957 \\
marginal-o0c0 & 0.1 & 0.430 & 0.009 & 0.009 & 0.888 
                      & 0.833 & 0.038 & 0.041 & 0.947 
                      & 0.301 & 0.038 & 0.041 & 0.957 \\
              & 0.5 & 1.589 & 0.022 & 0.024 & 0.893 
                      & 1.906 & 0.049 & 0.045 & 0.960 
                      & 1.429 & 0.050 & 0.051 & 0.953 \\
              & 1.0 & 2.208 & 0.032 & 0.034 & 0.911 
                      & 2.308 & 0.037 & 0.038 & 0.962 
                      & 2.173 & 0.047 & 0.051 & 0.961 \\
\midrule
\multicolumn{14}{@{}c}{\textbf{\textit{Doubly robust estimator (Frailty Cox)}}}\\
frailty-o1c1 & 0.1 & 0.053 & 0.009 & 0.009 & 0.944 
                      & 0.723 & 0.034 & 0.036 & 0.962 
                      & 1.460 & 0.035 & 0.037 & 0.968 \\
              & 0.5 & 0.381 & 0.020 & 0.022 & 0.953 
                      & 0.922 & 0.036 & 0.038 & 0.957 
                      & 1.150 & 0.042 & 0.045 & 0.962 \\
              & 1.0 & 0.729 & 0.027 & 0.030 & 0.952 
                      & 0.927 & 0.030 & 0.032 & 0.951 
                      & 1.304 & 0.041 & 0.046 & 0.953 \\
frailty-o1c0 & 0.1 & 0.064 & 0.008 & 0.009 & 0.941 
                      & 0.603 & 0.034 & 0.036 & 0.960 
                      & 1.274 & 0.035 & 0.037 & 0.967 \\
              & 0.5 & 0.432 & 0.019 & 0.022 & 0.953 
                      & 0.879 & 0.036 & 0.039 & 0.956 
                      & 1.207 & 0.041 & 0.046 & 0.964 \\
              & 1.0 & 0.794 & 0.026 & 0.029 & 0.954 
                      & 0.967 & 0.030 & 0.032 & 0.952 
                      & 1.405 & 0.041 & 0.046 & 0.955 \\
frailty-o0c1 & 0.1 & 0.173 & 0.009 & 0.010 & 0.934 
                      & 0.145 & 0.040 & 0.042 & 0.954 
                      & 0.223 & 0.040 & 0.042 & 0.954 \\
              & 0.5 & 0.911 & 0.021 & 0.024 & 0.942 
                      & 0.789 & 0.043 & 0.047 & 0.964 
                      & 0.984 & 0.047 & 0.052 & 0.969 \\
              & 1.0 & 1.527 & 0.028 & 0.032 & 0.940 
                      & 1.790 & 0.037 & 0.041 & 0.968 
                      & 1.437 & 0.045 & 0.053 & 0.967 \\
frailty-o0c0 & 0.1 & 0.411 & 0.009 & 0.009 & 0.902 
                      & 0.899 & 0.039 & 0.041 & 0.942 
                      & 0.474 & 0.039 & 0.042 & 0.952 \\
              & 0.5 & 1.683 & 0.021 & 0.023 & 0.893 
                      & 1.979 & 0.043 & 0.047 & 0.963 
                      & 1.509 & 0.046 & 0.052 & 0.967 \\
              & 1.0 & 2.592 & 0.029 & 0.032 & 0.890 
                      & 3.560 & 0.037 & 0.041 & 0.966 
                      & 2.256 & 0.046 & 0.052 & 0.955 \\
\midrule
\multicolumn{14}{@{}c}{\textbf{\textit{Outcome regression \& KM}}}\\
marginal-OR1 & 0.1 & 0.021 & 0.010 & 0.010 & 0.942 
                      & 1.225 & 0.034 & 0.036 & 0.956 
                      & 2.163 & 0.035 & 0.037 & 0.953 \\
              & 0.5 & 0.105 & 0.022 & 0.024 & 0.949 
                      & 1.335 & 0.036 & 0.037 & 0.956 
                      & 0.622 & 0.042 & 0.045 & 0.958 \\
              & 1.0 & 0.189 & 0.030 & 0.031 & 0.949 
                      & 1.134 & 0.029 & 0.031 & 0.957 
                      & 0.138 & 0.043 & 0.046 & 0.965 \\
marginal-OR0 & 0.1 & 0.441 & 0.009 & 0.010 & 0.886 
                      & 3.302 & 0.040 & 0.042 & 0.935 
                      & 7.229 & 0.040 & 0.042 & 0.917 \\
              & 0.5 & 1.278 & 0.023 & 0.025 & 0.908 
                      & 2.532 & 0.037 & 0.039 & 0.951 
                      & 3.529 & 0.043 & 0.046 & 0.940 \\
              & 1.0 & 1.406 & 0.033 & 0.035 & 0.927 
                      & 2.457 & 0.030 & 0.032 & 0.964 
                      & 1.042 & 0.044 & 0.047 & 0.960 \\
frailty-OR1  & 0.1 & 0.255 & 0.008 & 0.009 & 0.922 
                      & 2.795 & 0.034 & 0.036 & 0.934 
                      & 4.351 & 0.035 & 0.037 & 0.936 \\
              & 0.5 & 1.617 & 0.019 & 0.021 & 0.908 
                      & 10.508 & 0.038 & 0.040 & 0.928 
                      & 3.634 & 0.043 & 0.046 & 0.966 \\
              & 1.0 & 3.060 & 0.026 & 0.029 & 0.878 
                      & 17.796 & 0.034 & 0.036 & 0.913 
                      & 2.051 & 0.043 & 0.048 & 0.967 \\
frailty-OR0  & 0.1 & 0.577 & 0.009 & 0.009 & 0.864 
                      & 4.684 & 0.034 & 0.036 & 0.889 
                      & 6.871 & 0.034 & 0.037 & 0.910 \\
              & 0.5 & 2.771 & 0.022 & 0.024 & 0.810 
                      & 25.105 & 0.040 & 0.042 & 0.595 
                      & 10.419 & 0.045 & 0.048 & 0.837 \\
              & 1.0 & 4.853 & 0.030 & 0.033 & 0.788 
                      & 49.504 & 0.041 & 0.042 & 0.401 
                      & 10.634 & 0.051 & 0.054 & 0.849 \\
KM           & 0.1 & 1.005 & 0.008 & 0.008 & 0.709 
                      & 2.877 & 0.045 & 0.044 & 0.902 
                      & 2.389 & 0.046 & 0.045 & 0.919 \\
              & 0.5 & 4.010 & 0.021 & 0.022 & 0.628 
                      & 6.916 & 0.049 & 0.049 & 0.927 
                      & 2.294 & 0.054 & 0.054 & 0.922 \\
              & 1.0 & 6.123 & 0.030 & 0.031 & 0.651 
                      & 10.269 & 0.042 & 0.042 & 0.941 
                      & 4.685 & 0.052 & 0.053 & 0.886 \\
\bottomrule
\end{tabular}
}
\end{table}

\begin{table}[ht!]
\centering
\caption{Simulation results for cluster-level RMST \(\mu_C^{(a)}(t)\) and causal effects \(\Delta_C^{\text{RMST}}(t)\) on different scales under Scenario 3, estimated under 13 methods at time points \( t = \{0.1, 0.5, 1\} \). The number of clusters is \( M = 50 \), with censoring rate of approximately \(50\%\). Reported metrics include PBias (percentage bias), MCSD (Monte Carlo standard deviation), AESE (average estimated standard error from the jackknife procedure), and CP (empirical coverage probability of the \(95\%\) confidence interval).}
\label{M50_C50_ICS_E1_C1_rmstc}
\resizebox{\textwidth}{!}{
\begin{tabular}{@{}l r rrrr rrrr rrrr@{}}
\toprule
Method & \(t\) & \multicolumn{4}{c}{$\mu_C^{(1)}(t)$} & \multicolumn{4}{c}{$\mu_C^{(0)}(t)$} & \multicolumn{4}{c}{$\Delta_C^{\mathrm{RMST}}(t)$} \\
\cmidrule(lr){3-6} \cmidrule(lr){7-10} \cmidrule(lr){11-14}
 & & PBias & MCSD & AESE & CP  & PBias & MCSD & AESE & CP  & PBias & MCSD & AESE & CP \\
\midrule
\multicolumn{14}{@{}c}{\textbf{\textit{Doubly robust estimator (Marginal Cox)}}}\\
marginal-o1c1  & 0.1 & 0.039 & 0.001 & 0.001 & 0.935 & 0.628 & 0.002 & 0.002 & 0.964 & 3.202 & 0.002 & 0.003 & 0.965 \\
               & 0.5 & 0.147 & 0.007 & 0.007 & 0.940 & 0.948 & 0.018 & 0.018 & 0.965 & 2.185 & 0.019 & 0.020 & 0.964 \\
               & 1.0 & 0.257 & 0.019 & 0.021 & 0.942 & 1.034 & 0.037 & 0.039 & 0.958 & 1.888 & 0.043 & 0.045 & 0.965 \\
marginal-o1c0  & 0.1 & 0.052 & 0.001 & 0.001 & 0.931 & 0.644 & 0.002 & 0.002 & 0.963 & 3.355 & 0.002 & 0.003 & 0.966 \\
               & 0.5 & 0.161 & 0.007 & 0.007 & 0.939 & 0.984 & 0.018 & 0.019 & 0.967 & 2.300 & 0.019 & 0.020 & 0.965 \\
               & 1.0 & 0.243 & 0.019 & 0.021 & 0.942 & 1.065 & 0.038 & 0.040 & 0.961 & 1.923 & 0.043 & 0.046 & 0.966 \\
marginal-o0c1  & 0.1 & 0.099 & 0.001 & 0.001 & 0.926 & 0.489 & 0.003 & 0.003 & 0.963 & 2.887 & 0.003 & 0.003 & 0.962 \\
               & 0.5 & 0.418 & 0.007 & 0.008 & 0.925 & 0.851 & 0.020 & 0.021 & 0.963 & 2.781 & 0.021 & 0.022 & 0.957 \\
               & 1.0 & 0.667 & 0.020 & 0.021 & 0.932 & 0.968 & 0.043 & 0.045 & 0.959 & 2.730 & 0.046 & 0.049 & 0.958 \\
marginal-o0c0  & 0.1 & 0.166 & 0.001 & 0.001 & 0.901 & 0.142 & 0.003 & 0.003 & 0.961 & 1.629 & 0.003 & 0.003 & 0.966 \\
               & 0.5 & 0.659 & 0.007 & 0.007 & 0.903 & 0.076 & 0.020 & 0.021 & 0.965 & 2.041 & 0.021 & 0.022 & 0.962 \\
               & 1.0 & 0.995 & 0.020 & 0.021 & 0.913 & 0.083 & 0.043 & 0.046 & 0.964 & 2.163 & 0.046 & 0.050 & 0.964 \\
\midrule
\multicolumn{14}{@{}c}{\textbf{\textit{Doubly robust estimator (Frailty Cox)}}}\\
frailty-o1c1  & 0.1 & 0.032 & 0.001 & 0.001 & 0.938 & 0.599 & 0.002 & 0.002 & 0.967 & 3.026 & 0.002 & 0.003 & 0.966 \\
               & 0.5 & 0.231 & 0.006 & 0.007 & 0.947 & 0.904 & 0.017 & 0.018 & 0.961 & 2.344 & 0.019 & 0.020 & 0.970 \\
               & 1.0 & 0.406 & 0.018 & 0.020 & 0.949 & 0.948 & 0.037 & 0.039 & 0.958 & 2.116 & 0.041 & 0.045 & 0.965 \\
frailty-o1c0  & 0.1 & 0.040 & 0.001 & 0.001 & 0.937 & 0.550 & 0.002 & 0.002 & 0.965 & 2.841 & 0.002 & 0.003 & 0.966 \\
               & 0.5 & 0.257 & 0.006 & 0.007 & 0.950 & 0.891 & 0.018 & 0.018 & 0.955 & 2.394 & 0.019 & 0.020 & 0.966 \\
               & 1.0 & 0.444 & 0.018 & 0.020 & 0.951 & 0.993 & 0.037 & 0.039 & 0.954 & 2.259 & 0.042 & 0.045 & 0.962 \\
frailty-o0c1  & 0.1 & 0.055 & 0.001 & 0.001 & 0.931 & 0.369 & 0.003 & 0.003 & 0.959 & 2.065 & 0.003 & 0.003 & 0.958 \\
               & 0.5 & 0.325 & 0.007 & 0.007 & 0.945 & 0.584 & 0.020 & 0.021 & 0.962 & 2.019 & 0.021 & 0.022 & 0.956 \\
               & 1.0 & 0.559 & 0.019 & 0.021 & 0.952 & 0.559 & 0.044 & 0.045 & 0.954 & 1.969 & 0.046 & 0.049 & 0.968 \\
frailty-o0c0  & 0.1 & 0.148 & 0.001 & 0.001 & 0.913 & 0.066 & 0.003 & 0.003 & 0.953 & 1.160 & 0.003 & 0.003 & 0.958 \\
               & 0.5 & 0.671 & 0.007 & 0.007 & 0.917 & 0.051 & 0.021 & 0.021 & 0.957 & 2.014 & 0.021 & 0.022 & 0.955 \\
               & 1.0 & 1.061 & 0.019 & 0.021 & 0.925 & 0.099 & 0.044 & 0.045 & 0.954 & 2.275 & 0.047 & 0.049 & 0.960 \\
\midrule
\multicolumn{14}{@{}c}{\textbf{\textit{Outcome regression \& KM}}}\\
marginal-OR1  & 0.1 & 0.001 & 0.001 & 0.001 & 0.934 & 1.077 & 0.002 & 0.002 & 0.961 & 5.116 & 0.002 & 0.003 & 0.957 \\
               & 0.5 & 0.049 & 0.007 & 0.007 & 0.939 & 2.003 & 0.017 & 0.018 & 0.957 & 3.592 & 0.019 & 0.020 & 0.959 \\
               & 1.0 & 0.098 & 0.019 & 0.021 & 0.940 & 2.514 & 0.037 & 0.039 & 0.957 & 2.953 & 0.042 & 0.045 & 0.960 \\
marginal-OR0  & 0.1 & 0.368 & 0.001 & 0.001 & 0.835 & 5.485 & 0.003 & 0.003 & 0.739 & 28.127 & 0.003 & 0.003 & 0.710 \\
               & 0.5 & 1.400 & 0.007 & 0.008 & 0.824 & 11.229 & 0.019 & 0.020 & 0.637 & 24.924 & 0.020 & 0.022 & 0.563 \\
               & 1.0 & 2.174 & 0.020 & 0.022 & 0.843 & 13.696 & 0.038 & 0.040 & 0.627 & 22.205 & 0.042 & 0.045 & 0.553 \\
frailty-OR1   & 0.1 & 0.110 & 0.000 & 0.001 & 0.922 & 1.049 & 0.002 & 0.002 & 0.932 & 4.344 & 0.002 & 0.002 & 0.940 \\
               & 0.5 & 0.798 & 0.006 & 0.007 & 0.910 & 3.517 & 0.018 & 0.019 & 0.931 & 4.266 & 0.019 & 0.020 & 0.955 \\
               & 1.0 & 1.567 & 0.017 & 0.020 & 0.904 & 5.670 & 0.039 & 0.041 & 0.928 & 3.612 & 0.043 & 0.046 & 0.967 \\
frailty-OR0   & 0.1 & 0.443 & 0.001 & 0.001 & 0.824 & 1.992 & 0.003 & 0.003 & 0.933 & 11.992 & 0.003 & 0.003 & 0.912 \\
               & 0.5 & 2.129 & 0.007 & 0.007 & 0.718 & 0.427 & 0.018 & 0.019 & 0.963 & 6.890 & 0.019 & 0.020 & 0.932 \\
               & 1.0 & 3.823 & 0.019 & 0.021 & 0.649 & 2.784 & 0.039 & 0.041 & 0.956 & 5.134 & 0.043 & 0.045 & 0.950 \\
KM            & 0.1 & 0.366 & 0.001 & 0.001 & 0.829 & 0.685 & 0.003 & 0.003 & 0.926 & 1.143 & 0.003 & 0.003 & 0.935 \\
               & 0.5 & 1.616 & 0.007 & 0.007 & 0.779 & 2.048 & 0.023 & 0.024 & 0.936 & 0.811 & 0.025 & 0.025 & 0.944 \\
               & 1.0 & 2.609 & 0.020 & 0.021 & 0.789 & 3.149 & 0.051 & 0.051 & 0.938 & 1.927 & 0.055 & 0.055 & 0.940 \\
\bottomrule
\end{tabular}
}
\end{table}

\begin{table}[ht!]
\centering
\caption{Simulation results for individual-level RMST \(\mu_I^{(a)}(t)\) and causal effects \(\Delta_I^{\text{RMST}}(t)\) on different scales under Scenario 3, estimated under 13 methods at time points \( t = \{0.1, 0.5, 1\} \). The number of clusters is \( M = 50 \), with censoring rate of approximately \(50\%\). Reported metrics include PBias (percentage bias), MCSD (Monte Carlo standard deviation), AESE (average estimated standard error from the jackknife procedure), and CP (empirical coverage probability of the \(95\%\) confidence interval).}
\label{M50_C50_ICS_E1_C1_rmsti}
\resizebox{\textwidth}{!}{
\begin{tabular}{@{}l r rrrr rrrr rrrr@{}}
\toprule
Method & \(t\) & \multicolumn{4}{c}{$\mu_I^{(1)}(t)$} & \multicolumn{4}{c}{$\mu_I^{(0)}(t)$} & \multicolumn{4}{c}{$\Delta_I^{\mathrm{RMST}}(t)$} \\
\cmidrule(lr){3-6} \cmidrule(lr){7-10} \cmidrule(lr){11-14}
 & & PBias & MCSD & AESE & CP  & PBias & MCSD & AESE & CP  & PBias & MCSD & AESE & CP \\
\midrule
\multicolumn{14}{@{}c}{\textbf{\textit{Doubly robust estimator (Marginal Cox)}}}\\
marginal-o1c1  & 0.1 & 0.024 & 0.001 & 0.001 & 0.933 & 0.507 & 0.003 & 0.003 & 0.952 & 1.702 & 0.003 & 0.003 & 0.954 \\
               & 0.5 & 0.064 & 0.007 & 0.008 & 0.940 & 0.722 & 0.018 & 0.018 & 0.961 & 0.983 & 0.019 & 0.020 & 0.959 \\
               & 1.0 & 0.111 & 0.020 & 0.021 & 0.944 & 0.818 & 0.034 & 0.035 & 0.954 & 0.815 & 0.039 & 0.042 & 0.960 \\
marginal-o1c0  & 0.1 & 0.035 & 0.001 & 0.001 & 0.931 & 0.546 & 0.003 & 0.003 & 0.951 & 1.870 & 0.003 & 0.003 & 0.957 \\
               & 0.5 & 0.070 & 0.007 & 0.008 & 0.940 & 0.803 & 0.018 & 0.019 & 0.960 & 1.098 & 0.019 & 0.020 & 0.959 \\
               & 1.0 & 0.080 & 0.020 & 0.022 & 0.943 & 0.874 & 0.035 & 0.036 & 0.953 & 0.828 & 0.040 & 0.044 & 0.964 \\
marginal-o0c1  & 0.1 & 0.117 & 0.001 & 0.001 & 0.921 & 0.096 & 0.003 & 0.003 & 0.944 & 0.788 & 0.003 & 0.003 & 0.946 \\
               & 0.5 & 0.512 & 0.008 & 0.008 & 0.915 & 0.207 & 0.019 & 0.020 & 0.958 & 1.354 & 0.020 & 0.021 & 0.954 \\
               & 1.0 & 0.809 & 0.021 & 0.022 & 0.916 & 0.285 & 0.038 & 0.039 & 0.956 & 1.639 & 0.042 & 0.044 & 0.952 \\
marginal-o0c0  & 0.1 & 0.213 & 0.001 & 0.001 & 0.899 & 0.486 & 0.003 & 0.003 & 0.944 & 0.648 & 0.003 & 0.003 & 0.952 \\
               & 0.5 & 0.891 & 0.007 & 0.007 & 0.886 & 1.035 & 0.020 & 0.021 & 0.957 & 0.732 & 0.020 & 0.022 & 0.963 \\
               & 1.0 & 1.373 & 0.020 & 0.022 & 0.892 & 1.373 & 0.039 & 0.040 & 0.959 & 1.391 & 0.042 & 0.045 & 0.957 \\
\midrule
\multicolumn{14}{@{}c}{\textbf{\textit{Doubly robust estimator (Frailty Cox)}}}\\
frailty-o1c1  & 0.1 & 0.027 & 0.001 & 0.001 & 0.936 & 0.530 & 0.003 & 0.003 & 0.961 & 1.790 & 0.003 & 0.003 & 0.963 \\
               & 0.5 & 0.188 & 0.006 & 0.007 & 0.950 & 0.767 & 0.017 & 0.018 & 0.964 & 1.307 & 0.019 & 0.020 & 0.964 \\
               & 1.0 & 0.355 & 0.018 & 0.020 & 0.948 & 0.825 & 0.033 & 0.035 & 0.963 & 1.249 & 0.038 & 0.042 & 0.963 \\
frailty-o1c0  & 0.1 & 0.035 & 0.000 & 0.001 & 0.941 & 0.442 & 0.003 & 0.003 & 0.959 & 1.543 & 0.003 & 0.003 & 0.961 \\
               & 0.5 & 0.217 & 0.006 & 0.007 & 0.951 & 0.669 & 0.017 & 0.018 & 0.965 & 1.255 & 0.018 & 0.020 & 0.962 \\
               & 1.0 & 0.398 & 0.017 & 0.020 & 0.952 & 0.758 & 0.033 & 0.035 & 0.963 & 1.274 & 0.038 & 0.042 & 0.957 \\
frailty-o0c1  & 0.1 & 0.080 & 0.001 & 0.001 & 0.934 & 0.059 & 0.003 & 0.003 & 0.950 & 0.146 & 0.003 & 0.003 & 0.952 \\
               & 0.5 & 0.465 & 0.007 & 0.007 & 0.940 & 0.285 & 0.020 & 0.022 & 0.956 & 0.675 & 0.021 & 0.022 & 0.961 \\
               & 1.0 & 0.819 & 0.019 & 0.021 & 0.941 & 0.598 & 0.040 & 0.043 & 0.962 & 0.988 & 0.043 & 0.047 & 0.968 \\
frailty-o0c0  & 0.1 & 0.201 & 0.001 & 0.001 & 0.900 & 0.550 & 0.003 & 0.003 & 0.934 & 0.904 & 0.003 & 0.003 & 0.946 \\
               & 0.5 & 0.914 & 0.007 & 0.007 & 0.893 & 1.132 & 0.020 & 0.022 & 0.953 & 0.660 & 0.021 & 0.022 & 0.967 \\
               & 1.0 & 1.495 & 0.019 & 0.021 & 0.894 & 1.649 & 0.040 & 0.043 & 0.962 & 1.379 & 0.043 & 0.047 & 0.965 \\
\midrule
\multicolumn{14}{@{}c}{\textbf{\textit{Outcome regression \& KM}}}\\
marginal-OR1  & 0.1 & 0.001 & 0.001 & 0.001 & 0.933 & 0.897 & 0.003 & 0.003 & 0.952 & 2.842 & 0.003 & 0.003 & 0.949 \\
               & 0.5 & 0.051 & 0.007 & 0.008 & 0.940 & 1.213 & 0.017 & 0.018 & 0.956 & 1.308 & 0.019 & 0.020 & 0.953 \\
               & 1.0 & 0.104 & 0.020 & 0.021 & 0.950 & 1.232 & 0.033 & 0.035 & 0.953 & 0.750 & 0.039 & 0.042 & 0.963 \\
marginal-OR0  & 0.1 & 0.228 & 0.001 & 0.001 & 0.886 & 1.967 & 0.003 & 0.003 & 0.936 & 7.163 & 0.003 & 0.003 & 0.934 \\
               & 0.5 & 0.797 & 0.007 & 0.008 & 0.896 & 3.137 & 0.019 & 0.020 & 0.938 & 5.400 & 0.020 & 0.021 & 0.916 \\
               & 1.0 & 1.068 & 0.021 & 0.023 & 0.912 & 2.205 & 0.035 & 0.037 & 0.947 & 3.547 & 0.040 & 0.043 & 0.944 \\
frailty-OR1   & 0.1 & 0.110 & 0.000 & 0.001 & 0.925 & 1.430 & 0.003 & 0.003 & 0.936 & 4.060 & 0.003 & 0.003 & 0.933 \\
               & 0.5 & 0.773 & 0.006 & 0.007 & 0.917 & 4.884 & 0.018 & 0.019 & 0.925 & 4.037 & 0.019 & 0.020 & 0.948 \\
               & 1.0 & 1.501 & 0.017 & 0.019 & 0.908 & 7.932 & 0.035 & 0.037 & 0.928 & 3.370 & 0.040 & 0.043 & 0.965 \\
frailty-OR0   & 0.1 & 0.267 & 0.001 & 0.001 & 0.882 & 2.069 & 0.003 & 0.003 & 0.905 & 5.430 & 0.003 & 0.003 & 0.922 \\
               & 0.5 & 1.411 & 0.007 & 0.007 & 0.826 & 10.216 & 0.017 & 0.019 & 0.764 & 8.894 & 0.019 & 0.020 & 0.870 \\
               & 1.0 & 2.537 & 0.020 & 0.021 & 0.807 & 18.954 & 0.037 & 0.039 & 0.626 & 9.900 & 0.042 & 0.045 & 0.839 \\
KM            & 0.1 & 0.481 & 0.000 & 0.000 & 0.751 & 1.649 & 0.003 & 0.003 & 0.896 & 3.211 & 0.003 & 0.003 & 0.908 \\
               & 0.5 & 2.172 & 0.007 & 0.007 & 0.637 & 3.892 & 0.023 & 0.023 & 0.918 & 0.159 & 0.024 & 0.024 & 0.938 \\
               & 1.0 & 3.559 & 0.020 & 0.020 & 0.630 & 5.441 & 0.045 & 0.045 & 0.926 & 2.133 & 0.050 & 0.050 & 0.922 \\
\bottomrule
\end{tabular}
}
\end{table}

%\import{./resultsRev/}{M50_C50_ICS_E1_C0.tex}
\begin{table}[ht!]
\centering
\caption{Simulation results for cluster-level survival probabilities \(S_C^{(a)}(t)\) and causal effects \(\Delta_C^{\text{SPCE}}(t)\) on different scales under Scenario 2, estimated under 13 methods at time points \( t = \{0.1, 0.5, 1\} \). The number of clusters is \( M = 50 \), with censoring rate of approximately \(50\%\). Reported metrics include PBias (percentage bias), MCSD (Monte Carlo standard deviation), AESE (average estimated standard error from the jackknife procedure), and CP (empirical coverage probability of the \(95\%\) confidence interval).}
\label{M50_C50_ICS_E1_C0_Sc}
\resizebox{\textwidth}{!}{
\begin{tabular}{@{}l r rrrr rrrr rrrr@{}}
\toprule
Method & \(t\) & \multicolumn{4}{c}{$S_C^{(1)}(t)$} & \multicolumn{4}{c}{$S_C^{(0)}(t)$} & \multicolumn{4}{c}{$\Delta_C^{\mathrm{SPCE}}(t)$} \\
\cmidrule(lr){3-6} \cmidrule(lr){7-10} \cmidrule(lr){11-14}
 & & PBias & MCSD & AESE & CP  & PBias & MCSD & AESE & CP  & PBias & MCSD & AESE & CP \\
\midrule
\multicolumn{14}{@{}c}{\textbf{\textit{Doubly robust estimator (Marginal Cox)}}}\\
marginal-o1c1  & 0.1 & 0.077 & 0.009 & 0.010 & 0.936 & 0.825 & 0.032 & 0.033 & 0.964 & 2.611 & 0.033 & 0.035 & 0.972 \\
               & 0.5 & 0.373 & 0.021 & 0.023 & 0.945 & 1.020 & 0.041 & 0.043 & 0.960 & 1.824 & 0.047 & 0.050 & 0.967 \\
               & 1.0 & 0.594 & 0.029 & 0.031 & 0.948 & 0.942 & 0.039 & 0.042 & 0.958 & 1.628 & 0.049 & 0.053 & 0.965 \\
marginal-o1c0  & 0.1 & 0.115 & 0.009 & 0.010 & 0.939 & 0.961 & 0.033 & 0.034 & 0.964 & 3.143 & 0.034 & 0.035 & 0.966 \\
               & 0.5 & 0.409 & 0.021 & 0.023 & 0.950 & 1.100 & 0.047 & 0.045 & 0.958 & 2.018 & 0.051 & 0.051 & 0.962 \\
               & 1.0 & 0.594 & 0.028 & 0.030 & 0.949 & 1.234 & 0.041 & 0.042 & 0.956 & 1.876 & 0.053 & 0.054 & 0.961 \\
marginal-o0c1  & 0.1 & 0.314 & 0.009 & 0.010 & 0.911 & 0.770 & 0.036 & 0.038 & 0.961 & 3.364 & 0.036 & 0.039 & 0.966 \\
               & 0.5 & 1.249 & 0.022 & 0.023 & 0.915 & 1.323 & 0.047 & 0.049 & 0.958 & 3.930 & 0.051 & 0.054 & 0.946 \\
               & 1.0 & 1.899 & 0.030 & 0.032 & 0.914 & 1.391 & 0.046 & 0.047 & 0.935 & 4.115 & 0.054 & 0.056 & 0.941 \\
marginal-o0c0  & 0.1 & 0.487 & 0.009 & 0.010 & 0.873 & 0.417 & 0.035 & 0.038 & 0.965 & 3.030 & 0.036 & 0.039 & 0.969 \\
               & 0.5 & 1.809 & 0.022 & 0.023 & 0.882 & 0.853 & 0.050 & 0.050 & 0.963 & 4.618 & 0.053 & 0.055 & 0.940 \\
               & 1.0 & 2.690 & 0.030 & 0.032 & 0.887 & 1.208 & 0.046 & 0.048 & 0.945 & 5.363 & 0.056 & 0.057 & 0.925 \\
\midrule
\multicolumn{14}{@{}c}{\textbf{\textit{Doubly robust estimator (Frailty Cox)}}}\\
frailty-o1c1  & 0.1 & 0.098 & 0.009 & 0.009 & 0.942 & 0.843 & 0.032 & 0.033 & 0.958 & 2.745 & 0.034 & 0.035 & 0.965 \\
               & 0.5 & 0.546 & 0.020 & 0.022 & 0.945 & 1.141 & 0.040 & 0.043 & 0.966 & 2.304 & 0.046 & 0.049 & 0.965 \\
               & 1.0 & 0.928 & 0.027 & 0.030 & 0.953 & 0.893 & 0.039 & 0.042 & 0.957 & 2.154 & 0.049 & 0.053 & 0.964 \\
frailty-o1c0  & 0.1 & 0.095 & 0.009 & 0.009 & 0.942 & 0.732 & 0.032 & 0.033 & 0.957 & 2.420 & 0.033 & 0.035 & 0.968 \\
               & 0.5 & 0.568 & 0.020 & 0.022 & 0.944 & 0.996 & 0.041 & 0.043 & 0.957 & 2.198 & 0.046 & 0.049 & 0.967 \\
               & 1.0 & 0.992 & 0.027 & 0.029 & 0.950 & 0.833 & 0.040 & 0.042 & 0.953 & 2.222 & 0.049 & 0.052 & 0.959 \\
frailty-o0c1  & 0.1 & 0.232 & 0.009 & 0.010 & 0.927 & 0.622 & 0.037 & 0.038 & 0.963 & 2.635 & 0.037 & 0.039 & 0.960 \\
               & 0.5 & 1.089 & 0.021 & 0.024 & 0.928 & 0.911 & 0.048 & 0.049 & 0.956 & 3.172 & 0.051 & 0.054 & 0.957 \\
               & 1.0 & 1.788 & 0.029 & 0.032 & 0.929 & 0.693 & 0.047 & 0.048 & 0.943 & 3.459 & 0.053 & 0.057 & 0.960 \\
frailty-o0c0  & 0.1 & 0.397 & 0.009 & 0.010 & 0.900 & 0.187 & 0.036 & 0.038 & 0.961 & 2.041 & 0.037 & 0.039 & 0.959 \\
               & 0.5 & 1.682 & 0.021 & 0.023 & 0.898 & 0.393 & 0.048 & 0.050 & 0.959 & 3.845 & 0.051 & 0.054 & 0.954 \\
               & 1.0 & 2.641 & 0.029 & 0.031 & 0.902 & 0.195 & 0.047 & 0.048 & 0.947 & 4.551 & 0.053 & 0.057 & 0.955 \\
\midrule
\multicolumn{14}{@{}c}{\textbf{\textit{Outcome regression \& KM}}}\\
marginal-OR1  & 0.1 & 0.015 & 0.009 & 0.010 & 0.939 & 1.497 & 0.032 & 0.033 & 0.968 & 4.268 & 0.033 & 0.035 & 0.961 \\
               & 0.5 & 0.015 & 0.022 & 0.023 & 0.940 & 2.875 & 0.040 & 0.043 & 0.961 & 2.966 & 0.047 & 0.049 & 0.958 \\
               & 1.0 & 0.016 & 0.029 & 0.030 & 0.941 & 3.655 & 0.039 & 0.041 & 0.955 & 2.436 & 0.050 & 0.053 & 0.956 \\
marginal-OR0  & 0.1 & 0.815 & 0.009 & 0.010 & 0.815 & 8.675 & 0.040 & 0.041 & 0.694 & 27.508 & 0.041 & 0.042 & 0.630 \\
               & 0.5 & 2.834 & 0.023 & 0.024 & 0.807 & 16.662 & 0.041 & 0.043 & 0.618 & 23.152 & 0.047 & 0.049 & 0.524 \\
               & 1.0 & 4.293 & 0.031 & 0.033 & 0.829 & 19.025 & 0.035 & 0.037 & 0.634 & 20.003 & 0.047 & 0.050 & 0.568 \\
frailty-OR1   & 0.1 & 0.261 & 0.009 & 0.009 & 0.915 & 2.031 & 0.032 & 0.033 & 0.937 & 4.714 & 0.033 & 0.035 & 0.946 \\
               & 0.5 & 1.690 & 0.020 & 0.021 & 0.903 & 7.291 & 0.043 & 0.045 & 0.912 & 4.147 & 0.047 & 0.050 & 0.959 \\
               & 1.0 & 3.229 & 0.027 & 0.029 & 0.871 & 12.024 & 0.043 & 0.045 & 0.916 & 2.696 & 0.051 & 0.055 & 0.966 \\
frailty-OR0   & 0.1 & 1.020 & 0.009 & 0.009 & 0.743 & 2.275 & 0.034 & 0.036 & 0.935 & 10.288 & 0.035 & 0.036 & 0.908 \\
               & 0.5 & 4.464 & 0.021 & 0.023 & 0.623 & 3.341 & 0.043 & 0.044 & 0.952 & 5.634 & 0.047 & 0.049 & 0.937 \\
               & 1.0 & 7.826 & 0.029 & 0.031 & 0.554 & 12.322 & 0.044 & 0.045 & 0.900 & 4.797 & 0.052 & 0.055 & 0.941 \\
KM            & 0.1 & 0.881 & 0.009 & 0.009 & 0.781 & 1.364 & 0.042 & 0.043 & 0.929 & 0.480 & 0.043 & 0.044 & 0.943 \\
               & 0.5 & 3.341 & 0.022 & 0.023 & 0.746 & 3.731 & 0.056 & 0.057 & 0.942 & 2.934 & 0.061 & 0.062 & 0.937 \\
               & 1.0 & 5.089 & 0.031 & 0.032 & 0.757 & 5.864 & 0.055 & 0.055 & 0.933 & 4.568 & 0.063 & 0.064 & 0.931 \\
\bottomrule
\end{tabular}
}
\end{table}

\begin{table}[ht!]
\centering
\caption{Simulation results for individual-level survival probabilities \(S_I^{(a)}(t)\) and causal effects \(\Delta_I^{\text{SPCE}}(t)\) on different scales under Scenario 2, estimated under 13 methods at time points \( t = \{0.1, 0.5, 1\} \). The number of clusters is \( M = 50 \), with censoring rate of approximately \(50\%\). Reported metrics include PBias (percentage bias), MCSD (Monte Carlo standard deviation), AESE (average estimated standard error from the jackknife procedure), and CP (empirical coverage probability of the \(95\%\) confidence interval).}
\label{M50_C50_ICS_E1_C0_Si}
\resizebox{\textwidth}{!}{
\begin{tabular}{@{}l r rrrr rrrr rrrr@{}}
\toprule
Method & \(t\) & \multicolumn{4}{c}{$S_I^{(1)}(t)$} 
             & \multicolumn{4}{c}{$S_I^{(0)}(t)$} 
             & \multicolumn{4}{c}{$\Delta_I^{\mathrm{SPCE}}(t)$} \\
\cmidrule(lr){3-6}\cmidrule(lr){7-10}\cmidrule(lr){11-14}
 &  & PBias & MCSD & AESE & CP 
     & PBias & MCSD & AESE & CP 
     & PBias & MCSD & AESE & CP \\
\midrule
\multicolumn{14}{@{}c}{\textbf{\textit{Doubly robust estimator (Marginal Cox)}}}\\
marginal-o1c1 & 0.1 & 0.048 & 0.009 & 0.010 & 0.938 
                      & 0.653 & 0.035 & 0.036 & 0.952 
                      & 1.318 & 0.036 & 0.038 & 0.958 \\
              & 0.5 & 0.235 & 0.022 & 0.023 & 0.940 
                      & 0.738 & 0.037 & 0.038 & 0.964 
                      & 0.809 & 0.043 & 0.045 & 0.959 \\
              & 1.0 & 0.337 & 0.029 & 0.031 & 0.952 
                      & 0.904 & 0.030 & 0.032 & 0.963 
                      & 0.768 & 0.043 & 0.046 & 0.964 \\
marginal-o1c0 & 0.1 & 0.094 & 0.009 & 0.010 & 0.933 
                      & 0.872 & 0.036 & 0.037 & 0.954 
                      & 1.844 & 0.037 & 0.038 & 0.960 \\
              & 0.5 & 0.280 & 0.021 & 0.023 & 0.948 
                      & 0.870 & 0.044 & 0.040 & 0.960 
                      & 1.010 & 0.049 & 0.047 & 0.965 \\
              & 1.0 & 0.374 & 0.028 & 0.030 & 0.949 
                      & 1.399 & 0.032 & 0.033 & 0.958 
                      & 1.027 & 0.045 & 0.046 & 0.961 \\
marginal-o0c1 & 0.1 & 0.380 & 0.009 & 0.010 & 0.904 
                      & 0.227 & 0.038 & 0.040 & 0.951 
                      & 1.482 & 0.038 & 0.041 & 0.951 \\
              & 0.5 & 1.488 & 0.023 & 0.024 & 0.902 
                      & 0.660 & 0.041 & 0.042 & 0.955 
                      & 2.757 & 0.045 & 0.048 & 0.938 \\
              & 1.0 & 2.175 & 0.032 & 0.034 & 0.903 
                      & 0.982 & 0.034 & 0.035 & 0.959 
                      & 3.270 & 0.045 & 0.049 & 0.943 \\
marginal-o0c0 & 0.1 & 0.595 & 0.009 & 0.009 & 0.864 
                      & 0.420 & 0.038 & 0.040 & 0.951 
                      & 0.911 & 0.038 & 0.041 & 0.954 \\
              & 0.5 & 2.176 & 0.022 & 0.024 & 0.849 
                      & 0.314 & 0.046 & 0.043 & 0.964 
                      & 3.298 & 0.048 & 0.049 & 0.942 \\
              & 1.0 & 3.187 & 0.031 & 0.033 & 0.869 
                      & 0.338 & 0.034 & 0.036 & 0.961 
                      & 4.443 & 0.047 & 0.049 & 0.926 \\
\midrule
\multicolumn{14}{@{}c}{\textbf{\textit{Doubly robust estimator (Frailty Cox)}}}\\
frailty-o1c1 & 0.1 & 0.086 & 0.009 & 0.009 & 0.942 
                      & 0.703 & 0.034 & 0.036 & 0.962 
                      & 1.516 & 0.036 & 0.037 & 0.966 \\
              & 0.5 & 0.483 & 0.020 & 0.022 & 0.952 
                      & 0.911 & 0.036 & 0.038 & 0.958 
                      & 1.306 & 0.042 & 0.045 & 0.956 \\
              & 1.0 & 0.842 & 0.026 & 0.029 & 0.948 
                      & 0.899 & 0.030 & 0.032 & 0.952 
                      & 1.447 & 0.042 & 0.045 & 0.949 \\
frailty-o1c0 & 0.1 & 0.079 & 0.008 & 0.009 & 0.942 
                      & 0.561 & 0.034 & 0.036 & 0.961 
                      & 1.240 & 0.035 & 0.037 & 0.966 \\
              & 0.5 & 0.509 & 0.019 & 0.022 & 0.955 
                      & 0.714 & 0.036 & 0.038 & 0.958 
                      & 1.232 & 0.042 & 0.045 & 0.956 \\
              & 1.0 & 0.919 & 0.026 & 0.029 & 0.946 
                      & 0.722 & 0.030 & 0.032 & 0.952 
                      & 1.488 & 0.041 & 0.045 & 0.948 \\
frailty-o0c1 & 0.1 & 0.294 & 0.009 & 0.010 & 0.926 
                      & 0.014 & 0.039 & 0.042 & 0.957 
                      & 0.801 & 0.040 & 0.042 & 0.959 \\
              & 0.5 & 1.312 & 0.021 & 0.024 & 0.922 
                      & 0.359 & 0.043 & 0.046 & 0.967 
                      & 1.875 & 0.046 & 0.052 & 0.962 \\
              & 1.0 & 2.139 & 0.029 & 0.032 & 0.921 
                      & 0.981 & 0.037 & 0.041 & 0.965 
                      & 2.540 & 0.046 & 0.052 & 0.950 \\
frailty-o0c0 & 0.1 & 0.510 & 0.009 & 0.009 & 0.881 
                      & 0.665 & 0.040 & 0.042 & 0.948 
                      & 0.229 & 0.040 & 0.042 & 0.955 \\
              & 0.5 & 2.092 & 0.021 & 0.023 & 0.852 
                      & 1.052 & 0.044 & 0.047 & 0.962 
                      & 2.706 & 0.047 & 0.052 & 0.958 \\
              & 1.0 & 3.289 & 0.029 & 0.031 & 0.859 
                      & 1.593 & 0.037 & 0.041 & 0.968 
                      & 3.878 & 0.046 & 0.052 & 0.945 \\
\midrule
\multicolumn{14}{@{}c}{\textbf{\textit{Outcome regression \& KM}}}\\
marginal-OR1 & 0.1 & 0.041 & 0.009 & 0.010 & 0.931 
                      & 1.093 & 0.035 & 0.036 & 0.960 
                      & 2.096 & 0.036 & 0.037 & 0.952 \\
              & 0.5 & 0.050 & 0.022 & 0.023 & 0.946 
                      & 1.185 & 0.036 & 0.037 & 0.965 
                      & 0.780 & 0.043 & 0.045 & 0.957 \\
              & 1.0 & 0.064 & 0.029 & 0.031 & 0.947 
                      & 1.004 & 0.029 & 0.031 & 0.962 
                      & 0.435 & 0.043 & 0.045 & 0.962 \\
marginal-OR0 & 0.1 & 0.600 & 0.009 & 0.010 & 0.858 
                      & 3.379 & 0.041 & 0.042 & 0.921 
                      & 7.815 & 0.041 & 0.043 & 0.914 \\
              & 0.5 & 1.836 & 0.023 & 0.025 & 0.877 
                      & 3.143 & 0.037 & 0.039 & 0.940 
                      & 4.777 & 0.043 & 0.046 & 0.915 \\
              & 1.0 & 2.386 & 0.032 & 0.035 & 0.897 
                      & 1.250 & 0.030 & 0.031 & 0.964 
                      & 2.780 & 0.044 & 0.047 & 0.944 \\
frailty-OR1  & 0.1 & 0.260 & 0.008 & 0.009 & 0.920 
                      & 2.839 & 0.035 & 0.036 & 0.936 
                      & 4.416 & 0.036 & 0.037 & 0.939 \\
              & 0.5 & 1.630 & 0.019 & 0.021 & 0.894 
                      & 10.906 & 0.039 & 0.041 & 0.914 
                      & 3.849 & 0.043 & 0.046 & 0.956 \\
              & 1.0 & 3.056 & 0.026 & 0.028 & 0.873 
                      & 18.426 & 0.035 & 0.036 & 0.895 
                      & 2.275 & 0.044 & 0.047 & 0.966 \\
frailty-OR0 & 0.1 & 0.681 & 0.009 & 0.009 & 0.837 
                      & 4.635 & 0.034 & 0.036 & 0.888 
                      & 6.491 & 0.035 & 0.037 & 0.911 \\
              & 0.5 & 3.072 & 0.021 & 0.023 & 0.789 
                      & 24.606 & 0.040 & 0.042 & 0.598 
                      & 9.646 & 0.045 & 0.048 & 0.852 \\
              & 1.0 & 5.286 & 0.030 & 0.032 & 0.759 
                      & 48.277 & 0.040 & 0.042 & 0.414 
                      & 9.626 & 0.051 & 0.054 & 0.864 \\
KM           & 0.1 & 1.104 & 0.008 & 0.008 & 0.675 
                      & 2.793 & 0.045 & 0.045 & 0.905 
                      & 1.959 & 0.046 & 0.046 & 0.922 \\
              & 0.5 & 4.273 & 0.021 & 0.022 & 0.597 
                      & 6.276 & 0.050 & 0.050 & 0.937 
                      & 3.091 & 0.055 & 0.055 & 0.912 \\
              & 1.0 & 6.566 & 0.030 & 0.031 & 0.620 
                      & 9.051 & 0.042 & 0.042 & 0.946 
                      & 5.704 & 0.063 & 0.064 & 0.931 \\
\bottomrule
\end{tabular}
}
\end{table}

\begin{table}[ht!]
\centering
\caption{Simulation results for cluster-level survival probabilities \(\mu_C^{(a)}(t)\) and causal effects \(\Delta_C^{\text{RMST}}(t)\) on different scales under Scenario 2, estimated under 13 methods at time points \( t = \{0.1, 0.5, 1\} \). The number of clusters is \( M = 50 \), with censoring rate of approximately \(50\%\). Reported metrics include PBias (percentage bias), MCSD (Monte Carlo standard deviation), AESE (average estimated standard error from the jackknife procedure), and CP (empirical coverage probability of the \(95\%\) confidence interval).}
\label{M50_C50_ICS_E1_C0_rmstc}
\resizebox{\textwidth}{!}{
\begin{tabular}{@{}l r rrrr rrrr rrrr@{}}
\toprule
Method & \(t\) & \multicolumn{4}{c}{$\mu_C^{(1)}(t)$} & \multicolumn{4}{c}{$\mu_C^{(0)}(t)$} & \multicolumn{4}{c}{$\Delta_C^{\mathrm{RMST}}(t)$} \\
\cmidrule(lr){3-6}\cmidrule(lr){7-10}\cmidrule(lr){11-14}
 &  & PBias & MCSD & AESE & CP & PBias & MCSD & AESE & CP & PBias & MCSD & AESE & CP \\
\midrule
\multicolumn{14}{@{}c}{\textbf{\textit{Doubly robust estimator (Marginal Cox)}}}\\
marginal-o1c1 & 0.1 & 0.043 & 0.001 & 0.001 & 0.935 & 0.604 & 0.002 & 0.002 & 0.959 & 3.112 & 0.002 & 0.003 & 0.966 \\
              & 0.5 & 0.190 & 0.007 & 0.007 & 0.939 & 0.879 & 0.018 & 0.019 & 0.963 & 2.182 & 0.019 & 0.020 & 0.963 \\
              & 1.0 & 0.330 & 0.019 & 0.020 & 0.939 & 0.938 & 0.037 & 0.039 & 0.960 & 1.931 & 0.042 & 0.045 & 0.964 \\
marginal-o1c0 & 0.1 & 0.066 & 0.001 & 0.001 & 0.932 & 0.686 & 0.002 & 0.003 & 0.960 & 3.629 & 0.003 & 0.003 & 0.961 \\
              & 0.5 & 0.227 & 0.007 & 0.007 & 0.941 & 1.029 & 0.019 & 0.019 & 0.960 & 2.580 & 0.020 & 0.021 & 0.959 \\
              & 1.0 & 0.351 & 0.019 & 0.020 & 0.942 & 1.139 & 0.039 & 0.040 & 0.959 & 2.262 & 0.044 & 0.046 & 0.960 \\
marginal-o0c1 & 0.1 & 0.157 & 0.001 & 0.001 & 0.904 & 0.506 & 0.003 & 0.003 & 0.966 & 3.303 & 0.003 & 0.003 & 0.963 \\
              & 0.5 & 0.671 & 0.007 & 0.007 & 0.915 & 0.959 & 0.020 & 0.021 & 0.960 & 3.707 & 0.021 & 0.022 & 0.961 \\
              & 1.0 & 1.101 & 0.020 & 0.021 & 0.916 & 1.136 & 0.043 & 0.045 & 0.956 & 3.924 & 0.046 & 0.049 & 0.948 \\
marginal-o0c0 & 0.1 & 0.246 & 0.001 & 0.001 & 0.876 & 0.259 & 0.003 & 0.003 & 0.963 & 2.643 & 0.003 & 0.003 & 0.967 \\
              & 0.5 & 0.998 & 0.007 & 0.007 & 0.873 & 0.592 & 0.020 & 0.021 & 0.963 & 3.967 & 0.021 & 0.022 & 0.963 \\
              & 1.0 & 1.591 & 0.020 & 0.021 & 0.872 & 0.827 & 0.042 & 0.045 & 0.965 & 4.672 & 0.046 & 0.049 & 0.944 \\
\midrule
\multicolumn{14}{@{}c}{\textbf{\textit{Doubly robust estimator (Frailty Cox)}}}\\
frailty-o1c1 & 0.1 & 0.041 & 0.001 & 0.001 & 0.936 & 0.603 & 0.002 & 0.002 & 0.961 & 3.094 & 0.002 & 0.003 & 0.960 \\
              & 0.5 & 0.268 & 0.006 & 0.007 & 0.942 & 0.939 & 0.017 & 0.018 & 0.961 & 2.516 & 0.019 & 0.020 & 0.965 \\
              & 1.0 & 0.469 & 0.018 & 0.020 & 0.948 & 0.985 & 0.037 & 0.039 & 0.959 & 2.303 & 0.042 & 0.045 & 0.966 \\
frailty-o1c0 & 0.1 & 0.043 & 0.001 & 0.001 & 0.937 & 0.531 & 0.002 & 0.002 & 0.960 & 2.767 & 0.002 & 0.003 & 0.965 \\
              & 0.5 & 0.289 & 0.006 & 0.007 & 0.949 & 0.814 & 0.018 & 0.018 & 0.957 & 2.343 & 0.019 & 0.020 & 0.967 \\
              & 1.0 & 0.515 & 0.018 & 0.020 & 0.946 & 0.866 & 0.038 & 0.039 & 0.955 & 2.257 & 0.042 & 0.045 & 0.964 \\
frailty-o0c1 & 0.1 & 0.108 & 0.001 & 0.001 & 0.926 & 0.415 & 0.003 & 0.003 & 0.962 & 2.593 & 0.003 & 0.003 & 0.961 \\
              & 0.5 & 0.565 & 0.007 & 0.007 & 0.923 & 0.715 & 0.020 & 0.021 & 0.962 & 2.948 & 0.021 & 0.022 & 0.959 \\
              & 1.0 & 0.961 & 0.019 & 0.021 & 0.928 & 0.766 & 0.044 & 0.045 & 0.954 & 3.140 & 0.046 & 0.049 & 0.957 \\
frailty-o0c0 & 0.1 & 0.189 & 0.001 & 0.001 & 0.900 & 0.116 & 0.003 & 0.003 & 0.954 & 1.632 & 0.003 & 0.003 & 0.959 \\
              & 0.5 & 0.901 & 0.007 & 0.007 & 0.905 & 0.280 & 0.020 & 0.021 & 0.962 & 3.102 & 0.021 & 0.022 & 0.964 \\
              & 1.0 & 1.494 & 0.019 & 0.021 & 0.900 & 0.305 & 0.044 & 0.045 & 0.955 & 3.765 & 0.047 & 0.049 & 0.957 \\
\midrule
\multicolumn{14}{@{}c}{\textbf{\textit{Outcome regression \& KM}}}\\
marginal-OR1 & 0.1 & 0.027 & 0.001 & 0.001 & 0.930 & 1.014 & 0.002 & 0.002 & 0.965 & 4.967 & 0.002 & 0.003 & 0.960 \\
              & 0.5 & 0.003 & 0.007 & 0.007 & 0.942 & 1.930 & 0.017 & 0.018 & 0.960 & 3.603 & 0.019 & 0.020 & 0.957 \\
              & 1.0 & 0.016 & 0.020 & 0.021 & 0.940 & 2.446 & 0.037 & 0.039 & 0.960 & 3.053 & 0.043 & 0.045 & 0.957 \\
marginal-OR0 & 0.1 & 0.417 & 0.001 & 0.001 & 0.809 & 5.365 & 0.003 & 0.003 & 0.741 & 27.840 & 0.003 & 0.003 & 0.699 \\
              & 0.5 & 1.586 & 0.007 & 0.008 & 0.799 & 11.174 & 0.020 & 0.021 & 0.650 & 25.354 & 0.021 & 0.022 & 0.558 \\
              & 1.0 & 2.513 & 0.021 & 0.022 & 0.809 & 13.768 & 0.039 & 0.040 & 0.628 & 23.065 & 0.043 & 0.046 & 0.535 \\
frailty-OR1  & 0.1 & 0.110 & 0.000 & 0.001 & 0.905 & 1.044 & 0.002 & 0.002 & 0.935 & 4.320 & 0.002 & 0.002 & 0.945 \\
              & 0.5 & 0.790 & 0.006 & 0.007 & 0.903 & 3.623 & 0.018 & 0.019 & 0.923 & 4.488 & 0.019 & 0.020 & 0.952 \\
              & 1.0 & 1.562 & 0.018 & 0.019 & 0.884 & 5.876 & 0.039 & 0.041 & 0.916 & 3.883 & 0.043 & 0.046 & 0.961 \\
frailty-OR0 & 0.1 & 0.497 & 0.001 & 0.001 & 0.784 & 1.982 & 0.003 & 0.003 & 0.928 & 12.255 & 0.003 & 0.003 & 0.906 \\
              & 0.5 & 2.301 & 0.007 & 0.007 & 0.671 & 0.571 & 0.018 & 0.019 & 0.963 & 7.650 & 0.019 & 0.020 & 0.919 \\
              & 1.0 & 4.092 & 0.019 & 0.021 & 0.609 & 2.467 & 0.040 & 0.041 & 0.952 & 6.144 & 0.043 & 0.045 & 0.932 \\
KM           & 0.1 & 0.427 & 0.001 & 0.001 & 0.803 & 0.751 & 0.003 & 0.003 & 0.920 & 1.111 & 0.003 & 0.003 & 0.933 \\
              & 0.5 & 1.833 & 0.007 & 0.007 & 0.742 & 2.046 & 0.024 & 0.024 & 0.935 & 1.436 & 0.025 & 0.025 & 0.947 \\
              & 1.0 & 2.965 & 0.020 & 0.021 & 0.747 & 3.069 & 0.051 & 0.052 & 0.939 & 2.834 & 0.055 & 0.056 & 0.941 \\
\bottomrule
\end{tabular}
}
\end{table}

\begin{table}[ht!]
\centering
\caption{Simulation results for individual-level survival probabilities \(\mu_I^{(a)}(t)\) and causal effects \(\Delta_I^{\text{RMST}}(t)\) on different scales under Scenario 2, estimated under 13 methods at time points \( t = \{0.1, 0.5, 1\} \). The number of clusters is \( M = 50 \), with censoring rate of approximately \(50\%\). Reported metrics include PBias (percentage bias), MCSD (Monte Carlo standard deviation), AESE (average estimated standard error from the jackknife procedure), and CP (empirical coverage probability of the \(95\%\) confidence interval).}
\label{M50_C50_ICS_E1_C0_rmsti}
\resizebox{\textwidth}{!}{
\begin{tabular}{@{}l r rrrr rrrr rrrr@{}}
\toprule
Method & \(t\) & \multicolumn{4}{c}{$\mu_I^{(1)}(t)$} 
             & \multicolumn{4}{c}{$\mu_I^{(0)}(t)$} 
             & \multicolumn{4}{c}{$\Delta_I^{\mathrm{RMST}}(t)$} \\
\cmidrule(lr){3-6}\cmidrule(lr){7-10}\cmidrule(lr){11-14}
 &  & PBias & MCSD & AESE & CP 
     & PBias & MCSD & AESE & CP 
     & PBias & MCSD & AESE & CP \\
\midrule
\multicolumn{14}{@{}c}{\textbf{\textit{Doubly robust estimator (Marginal Cox)}}}\\
marginal-o1c1 & 0.1 & 0.029 & 0.001 & 0.001 & 0.932 
                      & 0.494 & 0.003 & 0.003 & 0.950 
                      & 1.682 & 0.003 & 0.003 & 0.952 \\
              & 0.5 & 0.119 & 0.007 & 0.007 & 0.940 
                      & 0.666 & 0.018 & 0.018 & 0.963 
                      & 1.037 & 0.019 & 0.020 & 0.962 \\
              & 1.0 & 0.202 & 0.019 & 0.021 & 0.940 
                      & 0.727 & 0.034 & 0.035 & 0.961 
                      & 0.905 & 0.039 & 0.042 & 0.959 \\
marginal-o1c0 & 0.1 & 0.057 & 0.001 & 0.001 & 0.928 
                      & 0.620 & 0.003 & 0.003 & 0.949 
                      & 2.195 & 0.003 & 0.003 & 0.952 \\
              & 0.5 & 0.163 & 0.007 & 0.007 & 0.938 
                      & 0.891 & 0.019 & 0.019 & 0.959 
                      & 1.416 & 0.020 & 0.020 & 0.955 \\
              & 1.0 & 0.237 & 0.019 & 0.020 & 0.945 
                      & 1.028 & 0.035 & 0.036 & 0.956 
                      & 1.229 & 0.041 & 0.042 & 0.959 \\
marginal-o0c1 & 0.1 & 0.186 & 0.001 & 0.001 & 0.899 
                      & 0.122 & 0.003 & 0.003 & 0.949 
                      & 1.161 & 0.003 & 0.003 & 0.950 \\
              & 0.5 & 0.806 & 0.007 & 0.007 & 0.891 
                      & 0.374 & 0.019 & 0.020 & 0.958 
                      & 2.187 & 0.020 & 0.021 & 0.948 \\
              & 1.0 & 1.302 & 0.020 & 0.022 & 0.896 
                      & 0.540 & 0.037 & 0.039 & 0.959 
                      & 2.699 & 0.042 & 0.044 & 0.942 \\
marginal-o0c0 & 0.1 & 0.295 & 0.001 & 0.001 & 0.867 
                      & 0.305 & 0.003 & 0.003 & 0.946 
                      & 0.264 & 0.003 & 0.003 & 0.951 \\
              & 0.5 & 1.207 & 0.007 & 0.007 & 0.855 
                      & 0.325 & 0.020 & 0.021 & 0.965 
                      & 2.247 & 0.020 & 0.022 & 0.952 \\
              & 1.0 & 1.913 & 0.020 & 0.022 & 0.863 
                      & 0.138 & 0.037 & 0.039 & 0.967 
                      & 3.278 & 0.041 & 0.044 & 0.948 \\
\midrule
\multicolumn{14}{@{}c}{\textbf{\textit{Doubly robust estimator (Frailty Cox)}}}\\
frailty-o1c1 & 0.1 & 0.041 & 0.001 & 0.001 & 0.937 
                      & 0.513 & 0.003 & 0.003 & 0.960 
                      & 1.780 & 0.003 & 0.003 & 0.965 \\
              & 0.5 & 0.235 & 0.006 & 0.007 & 0.958 
                      & 0.762 & 0.017 & 0.018 & 0.962 
                      & 1.402 & 0.019 & 0.020 & 0.961 \\
              & 1.0 & 0.422 & 0.018 & 0.020 & 0.956 
                      & 0.819 & 0.033 & 0.035 & 0.963 
                      & 1.362 & 0.039 & 0.041 & 0.958 \\
frailty-o1c0 & 0.1 & 0.039 & 0.000 & 0.001 & 0.935 
                      & 0.426 & 0.003 & 0.003 & 0.959 
                      & 1.511 & 0.003 & 0.003 & 0.959 \\
              & 0.5 & 0.252 & 0.006 & 0.007 & 0.954 
                      & 0.601 & 0.017 & 0.018 & 0.963 
                      & 1.250 & 0.019 & 0.020 & 0.961 \\
              & 1.0 & 0.465 & 0.017 & 0.020 & 0.951 
                      & 0.652 & 0.033 & 0.035 & 0.961 
                      & 1.312 & 0.038 & 0.041 & 0.955 \\
frailty-o0c1 & 0.1 & 0.138 & 0.001 & 0.001 & 0.924 
                      & 0.001 & 0.003 & 0.003 & 0.954 
                      & 0.573 & 0.003 & 0.003 & 0.956 \\
              & 0.5 & 0.685 & 0.007 & 0.007 & 0.921 
                      & 0.083 & 0.020 & 0.022 & 0.956 
                      & 1.389 & 0.021 & 0.022 & 0.964 \\
              & 1.0 & 1.168 & 0.019 & 0.021 & 0.922 
                      & 0.261 & 0.040 & 0.043 & 0.967 
                      & 1.856 & 0.043 & 0.047 & 0.963 \\
frailty-o0c0 & 0.1 & 0.247 & 0.001 & 0.001 & 0.884 
                      & 0.449 & 0.003 & 0.003 & 0.936 
                      & 0.393 & 0.003 & 0.003 & 0.950 \\
              & 0.5 & 0.901 & 0.007 & 0.007 & 0.905 
                      & 0.280 & 0.020 & 0.021 & 0.962 
                      & 3.102 & 0.021 & 0.022 & 0.964 \\
              & 1.0 & 1.494 & 0.019 & 0.021 & 0.900 
                      & 0.305 & 0.044 & 0.045 & 0.955 
                      & 3.765 & 0.047 & 0.049 & 0.957 \\
\midrule
\multicolumn{14}{@{}c}{\textbf{\textit{Outcome regression \& KM}}}\\
marginal-OR1 & 0.1 & 0.041 & 0.001 & 0.001 & 0.924 
                      & 0.822 & 0.003 & 0.003 & 0.956 
                      & 2.769 & 0.003 & 0.003 & 0.949 \\
              & 0.5 & 0.044 & 0.007 & 0.007 & 0.940 
                      & 1.108 & 0.017 & 0.018 & 0.962 
                      & 1.392 & 0.019 & 0.020 & 0.957 \\
              & 1.0 & 0.041 & 0.020 & 0.021 & 0.943 
                      & 1.117 & 0.033 & 0.035 & 0.961 
                      & 0.919 & 0.039 & 0.041 & 0.956 \\
marginal-OR0 & 0.1 & 0.309 & 0.001 & 0.001 & 0.863 
                      & 1.968 & 0.003 & 0.003 & 0.932 
                      & 7.506 & 0.003 & 0.003 & 0.930 \\
              & 0.5 & 1.092 & 0.007 & 0.008 & 0.869 
                      & 3.351 & 0.019 & 0.020 & 0.923 
                      & 6.292 & 0.020 & 0.021 & 0.906 \\
              & 1.0 & 1.579 & 0.021 & 0.023 & 0.886 
                      & 2.650 & 0.035 & 0.037 & 0.939 
                      & 4.783 & 0.040 & 0.043 & 0.921 \\
frailty-OR1  & 0.1 & 0.112 & 0.000 & 0.001 & 0.921 
                      & 1.436 & 0.003 & 0.003 & 0.934 
                      & 4.074 & 0.003 & 0.003 & 0.932 \\
              & 0.5 & 0.790 & 0.006 & 0.007 & 0.905 
                      & 3.623 & 0.018 & 0.019 & 0.922 
                      & 4.220 & 0.019 & 0.020 & 0.945 \\
              & 1.0 & 1.504 & 0.018 & 0.019 & 0.897 
                      & 8.201 & 0.036 & 0.037 & 0.918 
                      & 3.569 & 0.040 & 0.043 & 0.963 \\
frailty-OR0 & 0.1 & 0.321 & 0.001 & 0.001 & 0.857 
                      & 2.090 & 0.003 & 0.003 & 0.904 
                      & 5.269 & 0.003 & 0.003 & 0.925 \\
              & 0.5 & 2.301 & 0.007 & 0.007 & 0.671 
                      & 0.571 & 0.018 & 0.019 & 0.963 
                      & 7.650 & 0.019 & 0.020 & 0.919 \\
              & 1.0 & 4.092 & 0.019 & 0.021 & 0.609 
                      & 2.467 & 0.040 & 0.041 & 0.952 
                      & 6.144 & 0.043 & 0.045 & 0.932 \\
KM           & 0.1 & 0.536 & 0.000 & 0.000 & 0.722 
                      & 1.667 & 0.003 & 0.003 & 0.892 
                      & 3.038 & 0.004 & 0.004 & 0.912 \\
              & 0.5 & 2.331 & 0.007 & 0.007 & 0.597 
                      & 3.660 & 0.023 & 0.023 & 0.919 
                      & 0.776 & 0.025 & 0.024 & 0.937 \\
              & 1.0 & 3.806 & 0.020 & 0.020 & 0.601 
                      & 4.979 & 0.046 & 0.046 & 0.931 
                      & 2.919 & 0.051 & 0.050 & 0.914 \\
\bottomrule
\end{tabular}
}
\end{table}

To evaluate performance under varying censoring rates, we examine Scenario 3 at censoring rates of \(25\%\) and \(75\%\), as well as a smaller sample size setting with \(M = 26\) clusters and a censoring rate of \(50\%\). The performance of all four estimands,\(\Delta_C^{\text{SPCE}}\), \(\Delta_I^{\text{SPCE}}\), \(\Delta_C^{\text{RMST}}\), and \(\Delta_I^{\text{RMST}}\) show similar trends to those observed with \(M = 50\), but with increased standard errors. Under the same sample size (\(M = 50\)), a higher censoring rate leads to greater bias when both the outcome model \(P(T_{ij}^{(a)} \geq t \mid \bm{V}_{ij})\) and the censoring model \(K_c^{(a)}(t \mid \bm{V}_{ij})\) are mis-specified. It also shows that performance is more sensitive to mis-specification of the censoring model than to reductions in sample size.

%\import{./resultsRev/}{M26_C50_ICS_E1_C1.tex}
\begin{table}[ht!]
\centering
\caption{Simulation results for cluster-level survival probabilities \(S_C^{(a)}(t)\) and causal effects \(\Delta_C^{\text{SPCE}}(t)\) on different scales under Scenario 3, estimated under 13 methods at time points \( t = \{0.1, 0.5, 1\} \). The number of clusters is \( M = 26 \), with censoring rate of approximately \(50\%\). Reported metrics include PBias (percentage bias), MCSD (Monte Carlo standard deviation), AESE (average estimated standard error from the jackknife procedure), and CP (empirical coverage probability of the \(95\%\) confidence interval).}
\label{M26_C50_ICS_E1_C1_Sc}
\resizebox{\textwidth}{!}{
\begin{tabular}{@{}l r rrrr rrrr rrrr@{}}
\toprule
Method & \(t\) & \multicolumn{4}{c}{$S_C^{(1)}(t)$} & \multicolumn{4}{c}{$S_C^{(0)}(t)$} & \multicolumn{4}{c}{$\Delta_C^{\mathrm{SPCE}}(t)$} \\
\cmidrule(lr){3-6} \cmidrule(lr){7-10} \cmidrule(lr){11-14}
 & & PBias & MCSD & AESE & CP  & PBias & MCSD & AESE & CP  & PBias & MCSD & AESE & CP \\
\midrule
\multicolumn{14}{@{}c}{\textbf{\textit{Doubly robust estimator (Marginal Cox)}}}\\
marginal-o1c1  & 0.1 & 0.021 & 0.015 & 0.018 & 0.929 & 1.337 & 0.051 & 0.053 & 0.968 & 3.681 & 0.053 & 0.056 & 0.966 \\
               & 0.5 & 0.164 & 0.033 & 0.038 & 0.934 & 1.661 & 0.060 & 0.065 & 0.966 & 2.065 & 0.069 & 0.077 & 0.955 \\
               & 1.0 & 0.451 & 0.045 & 0.050 & 0.937 & 1.321 & 0.057 & 0.062 & 0.953 & 1.713 & 0.075 & 0.082 & 0.962 \\
marginal-o1c0  & 0.1 & 0.012 & 0.014 & 0.018 & 0.927 & 1.440 & 0.055 & 0.054 & 0.971 & 4.110 & 0.058 & 0.057 & 0.966 \\
               & 0.5 & 0.184 & 0.033 & 0.038 & 0.940 & 1.689 & 0.070 & 0.068 & 0.968 & 2.294 & 0.078 & 0.080 & 0.957 \\
               & 1.0 & 0.311 & 0.045 & 0.051 & 0.933 & 1.375 & 0.064 & 0.064 & 0.953 & 1.555 & 0.079 & 0.085 & 0.958 \\
marginal-o0c1  & 0.1 & 0.093 & 0.015 & 0.018 & 0.911 & 1.043 & 0.055 & 0.060 & 0.963 & 3.287 & 0.057 & 0.063 & 0.962 \\
               & 0.5 & 0.589 & 0.035 & 0.038 & 0.933 & 1.701 & 0.067 & 0.074 & 0.952 & 2.975 & 0.074 & 0.083 & 0.953 \\
               & 1.0 & 1.026 & 0.046 & 0.050 & 0.927 & 1.599 & 0.064 & 0.070 & 0.941 & 2.795 & 0.077 & 0.085 & 0.951 \\
marginal-o0c0  & 0.1 & 0.265 & 0.015 & 0.017 & 0.896 & 0.403 & 0.056 & 0.062 & 0.965 & 2.169 & 0.058 & 0.064 & 0.963 \\
               & 0.5 & 1.140 & 0.034 & 0.038 & 0.914 & 0.495 & 0.072 & 0.077 & 0.954 & 2.915 & 0.079 & 0.086 & 0.955 \\
               & 1.0 & 1.675 & 0.046 & 0.051 & 0.915 & 0.185 & 0.068 & 0.072 & 0.942 & 2.720 & 0.080 & 0.088 & 0.946 \\
\midrule
\multicolumn{14}{@{}c}{\textbf{\textit{Doubly robust estimator (Frailty Cox)}}}\\
frailty-o1c1  & 0.1 & 0.061 & 0.014 & 0.015 & 0.927 & 1.131 & 0.046 & 0.049 & 0.952 & 3.412 & 0.049 & 0.052 & 0.946 \\
               & 0.5 & 0.423 & 0.031 & 0.035 & 0.938 & 1.539 & 0.058 & 0.062 & 0.954 & 2.468 & 0.066 & 0.073 & 0.950 \\
               & 1.0 & 0.809 & 0.040 & 0.046 & 0.946 & 1.429 & 0.056 & 0.060 & 0.944 & 2.316 & 0.069 & 0.078 & 0.955 \\
frailty-o1c0  & 0.1 & 0.095 & 0.014 & 0.015 & 0.923 & 1.192 & 0.045 & 0.049 & 0.957 & 3.717 & 0.047 & 0.052 & 0.951 \\
               & 0.5 & 0.485 & 0.030 & 0.034 & 0.927 & 1.794 & 0.057 & 0.063 & 0.959 & 2.860 & 0.064 & 0.073 & 0.954 \\
               & 1.0 & 0.844 & 0.040 & 0.045 & 0.943 & 1.944 & 0.055 & 0.061 & 0.950 & 2.723 & 0.068 & 0.078 & 0.956 \\
frailty-o0c1  & 0.1 & 0.140 & 0.014 & 0.015 & 0.910 & 0.885 & 0.053 & 0.057 & 0.952 & 3.025 & 0.054 & 0.059 & 0.949 \\
               & 0.5 & 0.671 & 0.032 & 0.036 & 0.935 & 1.328 & 0.067 & 0.073 & 0.954 & 2.754 & 0.072 & 0.080 & 0.957 \\
               & 1.0 & 1.128 & 0.042 & 0.048 & 0.940 & 1.197 & 0.064 & 0.070 & 0.939 & 2.694 & 0.074 & 0.085 & 0.950 \\
frailty-o0c0  & 0.1 & 0.310 & 0.014 & 0.015 & 0.890 & 0.457 & 0.053 & 0.057 & 0.945 & 2.468 & 0.054 & 0.058 & 0.945 \\
               & 0.5 & 1.285 & 0.032 & 0.035 & 0.916 & 0.698 & 0.067 & 0.074 & 0.955 & 3.351 & 0.072 & 0.081 & 0.955 \\
               & 1.0 & 1.957 & 0.042 & 0.047 & 0.915 & 0.318 & 0.065 & 0.071 & 0.942 & 3.489 & 0.074 & 0.085 & 0.948 \\
\midrule
\multicolumn{14}{@{}c}{\textbf{\textit{Outcome regression \& KM}}}\\
marginal-OR1   & 0.1 & 0.066 & 0.014 & 0.018 & 0.921 & 1.858 & 0.050 & 0.053 & 0.967 & 4.975 & 0.053 & 0.056 & 0.960 \\
               & 0.5 & 0.080 & 0.033 & 0.038 & 0.930 & 3.000 & 0.059 & 0.065 & 0.958 & 2.964 & 0.069 & 0.077 & 0.957 \\
               & 1.0 & 0.023 & 0.044 & 0.049 & 0.928 & 3.348 & 0.056 & 0.062 & 0.951 & 2.217 & 0.073 & 0.081 & 0.957 \\
marginal-OR0   & 0.1 & 0.539 & 0.015 & 0.017 & 0.876 & 8.320 & 0.058 & 0.063 & 0.863 & 25.457 & 0.060 & 0.066 & 0.850 \\
               & 0.5 & 2.110 & 0.035 & 0.039 & 0.874 & 15.183 & 0.059 & 0.067 & 0.815 & 20.131 & 0.068 & 0.078 & 0.803 \\
               & 1.0 & 3.245 & 0.047 & 0.051 & 0.888 & 16.673 & 0.051 & 0.058 & 0.809 & 16.664 & 0.068 & 0.078 & 0.822 \\
frailty-OR1    & 0.1 & 0.199 & 0.013 & 0.015 & 0.906 & 1.029 & 0.045 & 0.048 & 0.946 & 2.135 & 0.047 & 0.051 & 0.950 \\
               & 0.5 & 1.314 & 0.029 & 0.034 & 0.913 & 4.641 & 0.058 & 0.064 & 0.956 & 2.152 & 0.066 & 0.074 & 0.962 \\
               & 1.0 & 2.589 & 0.039 & 0.046 & 0.918 & 7.948 & 0.059 & 0.064 & 0.953 & 1.021 & 0.072 & 0.080 & 0.965 \\
frailty-OR0    & 0.1 & 0.798 & 0.013 & 0.015 & 0.836 & 2.467 & 0.049 & 0.054 & 0.946 & 9.980 & 0.050 & 0.056 & 0.948 \\
               & 0.5 & 3.686 & 0.031 & 0.035 & 0.796 & 2.450 & 0.060 & 0.068 & 0.953 & 4.973 & 0.067 & 0.076 & 0.955 \\
               & 1.0 & 6.583 & 0.043 & 0.047 & 0.783 & 10.600 & 0.061 & 0.068 & 0.945 & 3.877 & 0.074 & 0.083 & 0.956 \\
KM              & 0.1 & 0.764 & 0.013 & 0.013 & 0.813 & 1.260 & 0.056 & 0.060 & 0.933 & 0.632 & 0.058 & 0.061 & 0.940 \\
               & 0.5 & 3.028 & 0.033 & 0.033 & 0.802 & 3.748 & 0.074 & 0.079 & 0.949 & 2.278 & 0.081 & 0.086 & 0.955 \\
               & 1.0 & 4.605 & 0.046 & 0.046 & 0.827 & 6.165 & 0.072 & 0.077 & 0.940 & 3.555 & 0.086 & 0.090 & 0.941 \\
\bottomrule
\end{tabular}
}
\end{table}

\begin{table}[ht!]
\centering
\caption{Simulation results for individual-level survival probabilities \(S_I^{(a)}(t)\) and causal effects \(\Delta_I^{\text{SPCE}}(t)\) on different scales under Scenario 3, estimated under 13 methods at time points \( t = \{0.1, 0.5, 1\} \). The number of clusters is \( M = 26 \), with censoring rate of approximately \(50\%\). Reported metrics include PBias (percentage bias), MCSD (Monte Carlo standard deviation), AESE (average estimated standard error from the jackknife procedure), and CP (empirical coverage probability of the \(95\%\) confidence interval).}
\label{M26_C50_ICS_E1_C1_Si}
\resizebox{\textwidth}{!}{
\begin{tabular}{@{}l r rrrr rrrr rrrr@{}}
\toprule
Method & \(t\) & \multicolumn{4}{c}{$S_I^{(1)}(t)$} & \multicolumn{4}{c}{$S_I^{(0)}(t)$} & \multicolumn{4}{c}{$\Delta_I^{\mathrm{SPCE}}(t)$} \\
\cmidrule(lr){3-6} \cmidrule(lr){7-10} \cmidrule(lr){11-14}
 & & PBias & MCSD & AESE & CP  & PBias & MCSD & AESE & CP  & PBias & MCSD & AESE & CP \\
\midrule
\multicolumn{14}{@{}c}{\textbf{\textit{Doubly robust estimator (Marginal Cox)}}}\\
marginal-o1c1  & 0.1 & 0.063 & 0.014 & 0.018 & 0.936 & 1.168 & 0.053 & 0.058 & 0.947 & 1.941 & 0.056 & 0.061 & 0.953 \\
               & 0.5 & 0.059 & 0.034 & 0.039 & 0.938 & 1.079 & 0.053 & 0.058 & 0.949 & 0.544 & 0.064 & 0.072 & 0.956 \\
               & 1.0 & 0.081 & 0.045 & 0.051 & 0.945 & 0.594 & 0.044 & 0.048 & 0.942 & 0.348 & 0.065 & 0.073 & 0.961 \\
marginal-o1c0  & 0.1 & 0.016 & 0.014 & 0.017 & 0.933 & 1.283 & 0.057 & 0.059 & 0.946 & 2.293 & 0.059 & 0.062 & 0.954 \\
               & 0.5 & 0.036 & 0.033 & 0.039 & 0.934 & 1.009 & 0.065 & 0.062 & 0.948 & 0.688 & 0.074 & 0.075 & 0.954 \\
               & 1.0 & 0.096 & 0.046 & 0.052 & 0.943 & 0.607 & 0.053 & 0.050 & 0.941 & 0.216 & 0.072 & 0.076 & 0.962 \\
marginal-o0c1  & 0.1 & 0.085 & 0.014 & 0.017 & 0.923 & 0.052 & 0.058 & 0.063 & 0.948 & 0.334 & 0.059 & 0.066 & 0.956 \\
               & 0.5 & 0.571 & 0.035 & 0.040 & 0.933 & 0.170 & 0.059 & 0.065 & 0.955 & 0.808 & 0.068 & 0.077 & 0.961 \\
               & 1.0 & 0.930 & 0.046 & 0.053 & 0.941 & 0.685 & 0.049 & 0.055 & 0.948 & 1.014 & 0.067 & 0.076 & 0.955 \\
marginal-o0c0  & 0.1 & 0.334 & 0.014 & 0.017 & 0.913 & 0.943 & 0.061 & 0.066 & 0.947 & 0.750 & 0.061 & 0.068 & 0.952 \\
               & 0.5 & 1.372 & 0.033 & 0.039 & 0.917 & 2.091 & 0.069 & 0.069 & 0.956 & 1.026 & 0.075 & 0.080 & 0.951 \\
               & 1.0 & 1.995 & 0.047 & 0.054 & 0.924 & 3.316 & 0.057 & 0.057 & 0.951 & 1.575 & 0.071 & 0.080 & 0.951 \\
\midrule
\multicolumn{14}{@{}c}{\textbf{\textit{Doubly robust estimator (Frailty Cox)}}}\\
frailty-o1c1  & 0.1 & 0.026 & 0.013 & 0.015 & 0.936 & 0.798 & 0.050 & 0.054 & 0.941 & 1.519 & 0.053 & 0.057 & 0.952 \\
               & 0.5 & 0.270 & 0.031 & 0.036 & 0.945 & 0.577 & 0.051 & 0.056 & 0.951 & 0.770 & 0.061 & 0.069 & 0.957 \\
               & 1.0 & 0.602 & 0.040 & 0.047 & 0.944 & 0.122 & 0.043 & 0.047 & 0.948 & 0.853 & 0.060 & 0.070 & 0.957 \\
frailty-o1c0  & 0.1 & 0.070 & 0.013 & 0.015 & 0.930 & 0.854 & 0.049 & 0.054 & 0.947 & 1.744 & 0.051 & 0.057 & 0.956 \\
               & 0.5 & 0.348 & 0.030 & 0.034 & 0.939 & 0.903 & 0.050 & 0.057 & 0.954 & 1.086 & 0.060 & 0.069 & 0.964 \\
               & 1.0 & 0.655 & 0.039 & 0.045 & 0.937 & 0.798 & 0.042 & 0.047 & 0.953 & 1.159 & 0.059 & 0.069 & 0.959 \\
frailty-o0c1  & 0.1 & 0.140 & 0.013 & 0.015 & 0.915 & 0.372 & 0.058 & 0.063 & 0.943 & 0.281 & 0.059 & 0.064 & 0.937 \\
               & 0.5 & 0.748 & 0.032 & 0.037 & 0.929 & 1.491 & 0.062 & 0.071 & 0.956 & 0.309 & 0.068 & 0.079 & 0.964 \\
               & 1.0 & 1.284 & 0.043 & 0.050 & 0.929 & 2.864 & 0.053 & 0.063 & 0.958 & 0.736 & 0.066 & 0.081 & 0.971 \\
frailty-o0c0  & 0.1 & 0.364 & 0.013 & 0.015 & 0.886 & 1.054 & 0.058 & 0.063 & 0.930 & 0.886 & 0.059 & 0.064 & 0.937 \\
               & 0.5 & 1.531 & 0.031 & 0.035 & 0.899 & 2.486 & 0.063 & 0.071 & 0.956 & 0.968 & 0.068 & 0.079 & 0.961 \\
               & 1.0 & 2.381 & 0.042 & 0.047 & 0.911 & 4.400 & 0.054 & 0.063 & 0.961 & 1.681 & 0.066 & 0.079 & 0.961 \\
\midrule
\multicolumn{14}{@{}c}{\textbf{\textit{Outcome regression \& KM}}}\\
marginal-OR1   & 0.1 & 0.068 & 0.014 & 0.018 & 0.937 & 1.549 & 0.053 & 0.057 & 0.951 & 2.619 & 0.055 & 0.061 & 0.954 \\
               & 0.5 & 0.126 & 0.033 & 0.039 & 0.936 & 1.256 & 0.053 & 0.057 & 0.948 & 0.541 & 0.063 & 0.071 & 0.955 \\
               & 1.0 & 0.118 & 0.044 & 0.050 & 0.942 & 0.333 & 0.044 & 0.047 & 0.941 & 0.043 & 0.063 & 0.072 & 0.964 \\
marginal-OR0   & 0.1 & 0.314 & 0.014 & 0.017 & 0.906 & 2.898 & 0.060 & 0.066 & 0.950 & 6.140 & 0.061 & 0.068 & 0.953 \\
               & 0.5 & 1.026 & 0.035 & 0.040 & 0.927 & 1.081 & 0.055 & 0.062 & 0.949 & 2.270 & 0.065 & 0.074 & 0.943 \\
               & 1.0 & 1.189 & 0.048 & 0.054 & 0.929 & 4.727 & 0.045 & 0.050 & 0.957 & 0.038 & 0.065 & 0.074 & 0.955 \\
frailty-OR1    & 0.1 & 0.177 & 0.013 & 0.014 & 0.926 & 1.792 & 0.048 & 0.054 & 0.941 & 2.752 & 0.051 & 0.056 & 0.943 \\
               & 0.5 & 1.201 & 0.029 & 0.034 & 0.925 & 8.055 & 0.052 & 0.058 & 0.957 & 2.847 & 0.061 & 0.069 & 0.965 \\
               & 1.0 & 2.343 & 0.038 & 0.045 & 0.926 & 13.927 & 0.046 & 0.051 & 0.964 & 1.675 & 0.061 & 0.071 & 0.970 \\
frailty-OR0    & 0.1 & 0.480 & 0.013 & 0.015 & 0.868 & 4.157 & 0.050 & 0.057 & 0.928 & 6.186 & 0.051 & 0.058 & 0.939 \\
               & 0.5 & 2.353 & 0.032 & 0.036 & 0.865 & 22.743 & 0.056 & 0.064 & 0.857 & 9.690 & 0.065 & 0.074 & 0.937 \\
               & 1.0 & 4.137 & 0.044 & 0.049 & 0.866 & 44.745 & 0.056 & 0.063 & 0.812 & 9.948 & 0.072 & 0.081 & 0.936 \\
KM              & 0.1 & 0.967 & 0.012 & 0.012 & 0.770 & 3.125 & 0.060 & 0.064 & 0.911 & 2.948 & 0.062 & 0.065 & 0.922 \\
               & 0.5 & 3.909 & 0.031 & 0.031 & 0.725 & 7.874 & 0.067 & 0.071 & 0.946 & 1.567 & 0.074 & 0.078 & 0.947 \\
               & 1.0 & 6.024 & 0.044 & 0.044 & 0.757 & 12.144 & 0.057 & 0.060 & 0.950 & 3.901 & 0.072 & 0.076 & 0.929 \\
\bottomrule
\end{tabular}
}
\end{table}

\begin{table}[ht!]
\centering
\caption{Simulation results for cluster-level RMST \(\mu_C^{(a)}(t)\) and causal effects \(\Delta_C^{\text{RMST}}(t)\) on different scales under Scenario 3, estimated under 13 methods at time points \( t = \{0.1, 0.5, 1\} \). The number of clusters is \( M = 26 \), with censoring rate of approximately \(50\%\). Reported metrics include PBias (percentage bias), MCSD (Monte Carlo standard deviation), AESE (average estimated standard error from the jackknife procedure), and CP (empirical coverage probability of the \(95\%\) confidence interval).}
\label{M26_C50_ICS_E1_C1_rmstc}
\resizebox{\textwidth}{!}{
\begin{tabular}{@{}l r rrrr rrrr rrrr@{}}
\toprule
Method & \(t\) & \multicolumn{4}{c}{$\mu_C^{(1)}(t)$} & \multicolumn{4}{c}{$\mu_C^{(0)}(t)$} & \multicolumn{4}{c}{$\Delta_C^{\mathrm{RMST}}(t)$} \\
\cmidrule(lr){3-6} \cmidrule(lr){7-10} \cmidrule(lr){11-14}
 & & PBias & MCSD & AESE & CP  & PBias & MCSD & AESE & CP  & PBias & MCSD & AESE & CP \\
\midrule
\multicolumn{14}{@{}c}{\textbf{\textit{Doubly robust estimator (Marginal Cox)}}}\\
marginal-o1c1  & 0.1 & 0.039 & 0.001 & 0.001 & 0.935 & 0.957 & 0.004 & 0.004 & 0.963 & 4.314 & 0.004 & 0.004 & 0.964 \\
               & 0.5 & 0.033 & 0.011 & 0.013 & 0.942 & 1.420 & 0.027 & 0.029 & 0.967 & 2.740 & 0.029 & 0.032 & 0.963 \\
               & 1.0 & 0.146 & 0.030 & 0.035 & 0.930 & 1.479 & 0.055 & 0.060 & 0.959 & 2.203 & 0.064 & 0.070 & 0.957 \\
marginal-o1c0  & 0.1 & 0.016 & 0.001 & 0.001 & 0.935 & 1.034 & 0.005 & 0.004 & 0.965 & 4.826 & 0.005 & 0.004 & 0.966 \\
               & 0.5 & 0.062 & 0.011 & 0.013 & 0.941 & 1.486 & 0.029 & 0.029 & 0.967 & 3.021 & 0.032 & 0.033 & 0.961 \\
               & 1.0 & 0.130 & 0.030 & 0.035 & 0.931 & 1.535 & 0.061 & 0.062 & 0.961 & 2.350 & 0.069 & 0.072 & 0.956 \\
marginal-o0c1  & 0.1 & 0.010 & 0.001 & 0.001 & 0.917 & 0.692 & 0.004 & 0.005 & 0.963 & 3.343 & 0.005 & 0.005 & 0.964 \\
               & 0.5 & 0.272 & 0.011 & 0.013 & 0.929 & 1.245 & 0.029 & 0.033 & 0.956 & 3.099 & 0.031 & 0.035 & 0.954 \\
               & 1.0 & 0.508 & 0.031 & 0.034 & 0.939 & 1.433 & 0.061 & 0.068 & 0.950 & 2.959 & 0.068 & 0.076 & 0.950 \\
marginal-o0c0  & 0.1 & 0.101 & 0.001 & 0.001 & 0.900 & 0.335 & 0.005 & 0.005 & 0.961 & 2.182 & 0.005 & 0.005 & 0.963 \\
               & 0.5 & 0.591 & 0.011 & 0.013 & 0.908 & 0.409 & 0.031 & 0.034 & 0.962 & 2.521 & 0.032 & 0.036 & 0.956 \\
               & 1.0 & 0.968 & 0.031 & 0.035 & 0.915 & 0.379 & 0.064 & 0.070 & 0.958 & 2.728 & 0.069 & 0.077 & 0.949 \\
\midrule
\multicolumn{14}{@{}c}{\textbf{\textit{Doubly robust estimator (Frailty Cox)}}}\\
frailty-o1c1  & 0.1 & 0.006 & 0.001 & 0.001 & 0.929 & 0.819 & 0.003 & 0.004 & 0.955 & 3.922 & 0.004 & 0.004 & 0.948 \\
               & 0.5 & 0.186 & 0.010 & 0.011 & 0.931 & 1.243 & 0.025 & 0.027 & 0.953 & 2.846 & 0.027 & 0.029 & 0.945 \\
               & 1.0 & 0.376 & 0.027 & 0.031 & 0.941 & 1.353 & 0.053 & 0.057 & 0.949 & 2.559 & 0.060 & 0.066 & 0.955 \\
frailty-o1c0  & 0.1 & 0.029 & 0.001 & 0.001 & 0.927 & 0.847 & 0.003 & 0.004 & 0.963 & 4.188 & 0.003 & 0.004 & 0.955 \\
               & 0.5 & 0.231 & 0.010 & 0.011 & 0.927 & 1.367 & 0.025 & 0.027 & 0.957 & 3.208 & 0.027 & 0.030 & 0.954 \\
               & 1.0 & 0.427 & 0.027 & 0.031 & 0.933 & 1.572 & 0.052 & 0.057 & 0.954 & 2.950 & 0.059 & 0.066 & 0.960 \\
frailty-o0c1  & 0.1 & 0.044 & 0.001 & 0.001 & 0.917 & 0.616 & 0.004 & 0.004 & 0.954 & 3.175 & 0.004 & 0.004 & 0.949 \\
               & 0.5 & 0.331 & 0.010 & 0.011 & 0.930 & 1.025 & 0.029 & 0.031 & 0.953 & 2.856 & 0.030 & 0.033 & 0.959 \\
               & 1.0 & 0.590 & 0.028 & 0.032 & 0.938 & 1.135 & 0.061 & 0.067 & 0.948 & 2.767 & 0.065 & 0.073 & 0.955 \\
frailty-o0c0  & 0.1 & 0.127 & 0.001 & 0.001 & 0.896 & 0.333 & 0.004 & 0.004 & 0.950 & 2.310 & 0.004 & 0.004 & 0.944 \\
               & 0.5 & 0.676 & 0.010 & 0.011 & 0.909 & 0.559 & 0.029 & 0.031 & 0.955 & 2.976 & 0.030 & 0.033 & 0.960 \\
               & 1.0 & 1.121 & 0.028 & 0.031 & 0.914 & 0.568 & 0.061 & 0.067 & 0.953 & 3.253 & 0.065 & 0.073 & 0.955 \\
\midrule
\multicolumn{14}{@{}c}{\textbf{\textit{Outcome regression \& KM}}}\\
marginal-OR1   & 0.1 & 0.053 & 0.001 & 0.001 & 0.937 & 1.289 & 0.004 & 0.004 & 0.966 & 5.812 & 0.004 & 0.004 & 0.968 \\
               & 0.5 & 0.080 & 0.011 & 0.013 & 0.937 & 2.187 & 0.027 & 0.029 & 0.959 & 3.844 & 0.029 & 0.032 & 0.959 \\
               & 1.0 & 0.083 & 0.030 & 0.034 & 0.930 & 2.592 & 0.055 & 0.060 & 0.958 & 3.084 & 0.063 & 0.070 & 0.955 \\
marginal-OR0   & 0.1 & 0.239 & 0.001 & 0.001 & 0.877 & 5.228 & 0.005 & 0.005 & 0.886 & 26.166 & 0.005 & 0.005 & 0.886 \\
               & 0.5 & 1.139 & 0.011 & 0.013 & 0.877 & 10.444 & 0.029 & 0.032 & 0.840 & 22.713 & 0.031 & 0.035 & 0.818 \\
               & 1.0 & 1.854 & 0.032 & 0.035 & 0.885 & 12.604 & 0.056 & 0.063 & 0.823 & 20.104 & 0.063 & 0.072 & 0.811 \\
frailty-OR1    & 0.1 & 0.069 & 0.001 & 0.001 & 0.915 & 0.404 & 0.003 & 0.004 & 0.951 & 1.523 & 0.003 & 0.004 & 0.945 \\
               & 0.5 & 0.598 & 0.009 & 0.011 & 0.916 & 2.107 & 0.025 & 0.027 & 0.960 & 2.213 & 0.027 & 0.030 & 0.960 \\
               & 1.0 & 1.211 & 0.026 & 0.031 & 0.918 & 3.652 & 0.053 & 0.059 & 0.954 & 1.869 & 0.061 & 0.068 & 0.964 \\
frailty-OR0    & 0.1 & 0.368 & 0.001 & 0.001 & 0.844 & 2.138 & 0.004 & 0.004 & 0.947 & 12.251 & 0.004 & 0.004 & 0.946 \\
               & 0.5 & 1.861 & 0.010 & 0.011 & 0.805 & 0.986 & 0.026 & 0.029 & 0.956 & 7.163 & 0.028 & 0.031 & 0.952 \\
               & 1.0 & 3.376 & 0.029 & 0.032 & 0.793 & 1.707 & 0.055 & 0.063 & 0.955 & 5.481 & 0.061 & 0.070 & 0.956 \\
KM              & 0.1 & 0.352 & 0.001 & 0.001 & 0.819 & 0.692 & 0.004 & 0.004 & 0.932 & 1.259 & 0.004 & 0.004 & 0.935 \\
               & 0.5 & 1.631 & 0.010 & 0.010 & 0.793 & 2.013 & 0.031 & 0.034 & 0.942 & 0.921 & 0.033 & 0.035 & 0.951 \\
               & 1.0 & 2.665 & 0.030 & 0.030 & 0.814 & 3.108 & 0.067 & 0.072 & 0.949 & 2.106 & 0.074 & 0.079 & 0.948 \\
\bottomrule
\end{tabular}
}
\end{table}

\begin{table}[ht!]
\centering
\caption{Simulation results for individual-level RMST \(\mu_I^{(a)}(t)\) and causal effects \(\Delta_I^{\text{RMST}}(t)\) on different scales under Scenario 3, estimated under 13 methods at time points \( t = \{0.1, 0.5, 1\} \). The number of clusters is \( M = 26 \), with censoring rate of approximately \(50\%\). Reported metrics include PBias (percentage bias), MCSD (Monte Carlo standard deviation), AESE (average estimated standard error from the jackknife procedure), and CP (empirical coverage probability of the \(95\%\) confidence interval).}
\label{M26_C50_ICS_E1_C1_rmsti}
\resizebox{\textwidth}{!}{
\begin{tabular}{@{}l r rrrr rrrr rrrr@{}}
\toprule
Method & \(t\) & \multicolumn{4}{c}{$\mu_I^{(1)}(t)$} & \multicolumn{4}{c}{$\mu_I^{(0)}(t)$} & \multicolumn{4}{c}{$\Delta_I^{\mathrm{RMST}}(t)$} \\
\cmidrule(lr){3-6} \cmidrule(lr){7-10} \cmidrule(lr){11-14}
 & & PBias & MCSD & AESE & CP  & PBias & MCSD & AESE & CP  & PBias & MCSD & AESE & CP \\
\midrule
\multicolumn{14}{@{}c}{\textbf{\textit{Doubly robust estimator (Marginal Cox)}}}\\
marginal-o1c1  & 0.1 & 0.053 & 0.001 & 0.001 & 0.932 & 0.853 & 0.005 & 0.005 & 0.950 & 2.482 & 0.005 & 0.005 & 0.952 \\
               & 0.5 & 0.074 & 0.011 & 0.013 & 0.936 & 1.094 & 0.026 & 0.029 & 0.950 & 1.122 & 0.029 & 0.032 & 0.955 \\
               & 1.0 & 0.046 & 0.030 & 0.035 & 0.943 & 1.037 & 0.050 & 0.054 & 0.946 & 0.707 & 0.059 & 0.066 & 0.953 \\
marginal-o1c0  & 0.1 & 0.018 & 0.001 & 0.001 & 0.932 & 0.933 & 0.005 & 0.005 & 0.949 & 2.883 & 0.005 & 0.005 & 0.952 \\
               & 0.5 & 0.033 & 0.010 & 0.013 & 0.934 & 1.139 & 0.029 & 0.030 & 0.952 & 1.339 & 0.031 & 0.033 & 0.955 \\
               & 1.0 & 0.061 & 0.030 & 0.035 & 0.934 & 1.049 & 0.057 & 0.056 & 0.945 & 0.808 & 0.066 & 0.069 & 0.954 \\
marginal-o0c1  & 0.1 & 0.007 & 0.001 & 0.001 & 0.926 & 0.040 & 0.005 & 0.005 & 0.945 & 0.162 & 0.005 & 0.005 & 0.953 \\
               & 0.5 & 0.267 & 0.011 & 0.013 & 0.930 & 0.019 & 0.029 & 0.032 & 0.955 & 0.560 & 0.030 & 0.034 & 0.959 \\
               & 1.0 & 0.489 & 0.031 & 0.036 & 0.939 & 0.120 & 0.055 & 0.061 & 0.955 & 0.771 & 0.062 & 0.071 & 0.962 \\
marginal-o0c0  & 0.1 & 0.140 & 0.001 & 0.001 & 0.911 & 0.515 & 0.005 & 0.005 & 0.942 & 1.033 & 0.005 & 0.005 & 0.949 \\
               & 0.5 & 0.729 & 0.011 & 0.013 & 0.908 & 1.262 & 0.031 & 0.033 & 0.949 & 0.172 & 0.032 & 0.035 & 0.956 \\
               & 1.0 & 1.184 & 0.030 & 0.036 & 0.925 & 1.642 & 0.060 & 0.063 & 0.960 & 0.895 & 0.065 & 0.073 & 0.955 \\
\midrule
\multicolumn{14}{@{}c}{\textbf{\textit{Doubly robust estimator (Frailty Cox)}}}\\
frailty-o1c1  & 0.1 & 0.001 & 0.001 & 0.001 & 0.927 & 0.635 & 0.004 & 0.004 & 0.936 & 2.003 & 0.004 & 0.004 & 0.945 \\
               & 0.5 & 0.103 & 0.010 & 0.011 & 0.936 & 0.724 & 0.025 & 0.027 & 0.951 & 1.072 & 0.027 & 0.030 & 0.954 \\
               & 1.0 & 0.247 & 0.027 & 0.032 & 0.942 & 0.621 & 0.048 & 0.052 & 0.953 & 0.905 & 0.056 & 0.063 & 0.954 \\
frailty-o1c0  & 0.1 & 0.023 & 0.001 & 0.001 & 0.916 & 0.652 & 0.004 & 0.004 & 0.940 & 2.157 & 0.004 & 0.004 & 0.949 \\
               & 0.5 & 0.160 & 0.010 & 0.011 & 0.934 & 0.855 & 0.024 & 0.027 & 0.958 & 1.347 & 0.027 & 0.030 & 0.957 \\
               & 1.0 & 0.310 & 0.027 & 0.031 & 0.937 & 0.874 & 0.047 & 0.052 & 0.957 & 1.207 & 0.055 & 0.063 & 0.959 \\
frailty-o0c1  & 0.1 & 0.047 & 0.001 & 0.001 & 0.915 & 0.169 & 0.004 & 0.005 & 0.934 & 0.337 & 0.005 & 0.005 & 0.939 \\
               & 0.5 & 0.358 & 0.010 & 0.012 & 0.924 & 0.651 & 0.029 & 0.033 & 0.950 & 0.016 & 0.030 & 0.034 & 0.952 \\
               & 1.0 & 0.661 & 0.029 & 0.033 & 0.928 & 1.124 & 0.057 & 0.065 & 0.958 & 0.311 & 0.063 & 0.073 & 0.960 \\
frailty-o0c0  & 0.1 & 0.159 & 0.001 & 0.001 & 0.890 & 0.620 & 0.004 & 0.005 & 0.926 & 1.296 & 0.005 & 0.005 & 0.930 \\
               & 0.5 & 0.804 & 0.010 & 0.011 & 0.897 & 1.372 & 0.030 & 0.033 & 0.943 & 0.139 & 0.031 & 0.034 & 0.949 \\
               & 1.0 & 1.350 & 0.028 & 0.032 & 0.904 & 2.010 & 0.058 & 0.065 & 0.954 & 0.850 & 0.063 & 0.072 & 0.956 \\
\midrule
\multicolumn{14}{@{}c}{\textbf{\textit{Outcome regression \& KM}}}\\
marginal-OR1   & 0.1 & 0.051 & 0.001 & 0.001 & 0.937 & 1.153 & 0.005 & 0.005 & 0.952 & 3.430 & 0.005 & 0.005 & 0.954 \\
               & 0.5 & 0.100 & 0.011 & 0.013 & 0.940 & 1.418 & 0.026 & 0.028 & 0.952 & 1.442 & 0.028 & 0.032 & 0.958 \\
               & 1.0 & 0.125 & 0.030 & 0.035 & 0.938 & 1.251 & 0.049 & 0.054 & 0.945 & 0.727 & 0.058 & 0.066 & 0.954 \\
marginal-OR0   & 0.1 & 0.129 & 0.001 & 0.001 & 0.904 & 1.777 & 0.005 & 0.005 & 0.949 & 6.154 & 0.005 & 0.005 & 0.948 \\
               & 0.5 & 0.603 & 0.011 & 0.013 & 0.917 & 2.404 & 0.028 & 0.031 & 0.949 & 4.122 & 0.030 & 0.034 & 0.940 \\
               & 1.0 & 0.841 & 0.032 & 0.036 & 0.931 & 1.118 & 0.052 & 0.059 & 0.945 & 2.326 & 0.061 & 0.069 & 0.948 \\
frailty-OR1    & 0.1 & 0.059 & 0.001 & 0.001 & 0.919 & 0.763 & 0.004 & 0.004 & 0.941 & 2.168 & 0.004 & 0.004 & 0.948 \\
               & 0.5 & 0.545 & 0.009 & 0.011 & 0.931 & 3.480 & 0.024 & 0.027 & 0.945 & 2.890 & 0.027 & 0.030 & 0.960 \\
               & 1.0 & 1.106 & 0.026 & 0.031 & 0.929 & 5.929 & 0.048 & 0.054 & 0.957 & 2.549 & 0.056 & 0.064 & 0.968 \\
frailty-OR0    & 0.1 & 0.205 & 0.001 & 0.001 & 0.876 & 1.762 & 0.004 & 0.004 & 0.931 & 4.714 & 0.004 & 0.005 & 0.938 \\
               & 0.5 & 1.177 & 0.010 & 0.011 & 0.860 & 9.189 & 0.025 & 0.029 & 0.905 & 8.200 & 0.027 & 0.031 & 0.939 \\
               & 1.0 & 2.139 & 0.029 & 0.033 & 0.873 & 17.098 & 0.052 & 0.060 & 0.866 & 9.194 & 0.060 & 0.069 & 0.935 \\
KM              & 0.1 & 0.449 & 0.001 & 0.001 & 0.796 & 1.832 & 0.005 & 0.005 & 0.898 & 3.922 & 0.005 & 0.005 & 0.910 \\
               & 0.5 & 2.100 & 0.010 & 0.010 & 0.730 & 4.384 & 0.031 & 0.033 & 0.929 & 0.574 & 0.033 & 0.035 & 0.942 \\
               & 1.0 & 3.469 & 0.028 & 0.029 & 0.738 & 6.232 & 0.062 & 0.066 & 0.947 & 1.376 & 0.068 & 0.072 & 0.946 \\
\bottomrule
\end{tabular}
}
\end{table}

%\import{./resultsRev/}{M50_C25_ICS_E1_C1.tex}
% Table 1: Cluster‐level survival (censoring ≈25%)
\begin{table}[ht!]
\centering
\caption{Simulation results for cluster-level survival probabilities \(S_C^{(a)}(t)\) and causal effects \(\Delta_C^{\text{SPCE}}(t)\) on different scales under Scenario 3, estimated under 13 methods at time points \( t = \{0.1, 0.5, 1\} \). The number of clusters is \( M = 50 \), with censoring rate of approximately \(25\%\). Reported metrics include PBias (percentage bias), MCSD (Monte Carlo standard deviation), AESE (average estimated standard error from the jackknife procedure), and CP (empirical coverage probability of the \(95\%\) confidence interval).}
\label{M50_C25_ICS_E1_C1_Sc}
\resizebox{\textwidth}{!}{
\begin{tabular}{@{}l r rrrr rrrr rrrr@{}}
\toprule
Method & \(t\) & \multicolumn{4}{c}{$S_C^{(1)}(t)$} & \multicolumn{4}{c}{$S_C^{(0)}(t)$} & \multicolumn{4}{c}{$\Delta_C^{\mathrm{SPCE}}(t)$} \\
\cmidrule(lr){3-6} \cmidrule(lr){7-10} \cmidrule(lr){11-14}
 & & PBias & MCSD & AESE & CP  & PBias & MCSD & AESE & CP  & PBias & MCSD & AESE & CP \\
\midrule
\multicolumn{14}{@{}c}{\textbf{\textit{Doubly robust estimator (Marginal Cox)}}}\\
marginal-o1c1  & 0.1 & 0.107 & 0.009 & 0.009 & 0.929 & 0.839 & 0.032 & 0.033 & 0.964 & 2.766 & 0.034 & 0.035 & 0.964 \\
               & 0.5 & 0.499 & 0.021 & 0.022 & 0.939 & 1.054 & 0.041 & 0.043 & 0.961 & 2.117 & 0.047 & 0.049 & 0.956 \\
               & 1.0 & 0.833 & 0.028 & 0.029 & 0.938 & 0.959 & 0.040 & 0.042 & 0.956 & 2.041 & 0.050 & 0.052 & 0.955 \\
marginal-o1c0  & 0.1 & 0.115 & 0.009 & 0.009 & 0.929 & 0.861 & 0.032 & 0.033 & 0.962 & 2.860 & 0.033 & 0.035 & 0.964 \\
               & 0.5 & 0.528 & 0.021 & 0.022 & 0.938 & 1.102 & 0.040 & 0.043 & 0.963 & 2.226 & 0.046 & 0.049 & 0.959 \\
               & 1.0 & 0.876 & 0.028 & 0.029 & 0.937 & 0.998 & 0.039 & 0.042 & 0.960 & 2.138 & 0.049 & 0.052 & 0.958 \\
marginal-o0c1  & 0.1 & 0.129 & 0.009 & 0.009 & 0.937 & 0.965 & 0.035 & 0.038 & 0.970 & 3.207 & 0.036 & 0.039 & 0.973 \\
               & 0.5 & 0.561 & 0.021 & 0.022 & 0.950 & 1.747 & 0.046 & 0.049 & 0.964 & 2.966 & 0.049 & 0.053 & 0.961 \\
               & 1.0 & 0.917 & 0.028 & 0.029 & 0.945 & 1.944 & 0.044 & 0.047 & 0.949 & 2.843 & 0.051 & 0.054 & 0.953 \\
marginal-o0c0  & 0.1 & 0.107 & 0.009 & 0.009 & 0.924 & 0.809 & 0.035 & 0.038 & 0.964 & 2.685 & 0.035 & 0.039 & 0.970 \\
               & 0.5 & 0.502 & 0.021 & 0.022 & 0.939 & 1.270 & 0.046 & 0.049 & 0.959 & 2.350 & 0.049 & 0.053 & 0.953 \\
               & 1.0 & 0.841 & 0.029 & 0.029 & 0.938 & 1.389 & 0.045 & 0.047 & 0.943 & 2.344 & 0.051 & 0.054 & 0.949 \\
\midrule
\multicolumn{14}{@{}c}{\textbf{\textit{Doubly robust estimator(Frailty Cox)}}}\\
frailty-o1c1  & 0.1 & 0.117 & 0.009 & 0.009 & 0.936 & 0.758 & 0.032 & 0.033 & 0.961 & 2.579 & 0.033 & 0.035 & 0.969 \\
               & 0.5 & 0.591 & 0.020 & 0.021 & 0.943 & 0.966 & 0.041 & 0.043 & 0.958 & 2.214 & 0.045 & 0.049 & 0.967 \\
               & 1.0 & 1.020 & 0.026 & 0.028 & 0.943 & 0.737 & 0.040 & 0.042 & 0.953 & 2.203 & 0.048 & 0.052 & 0.961 \\
frailty-o1c0  & 0.1 & 0.132 & 0.009 & 0.009 & 0.935 & 0.789 & 0.032 & 0.033 & 0.959 & 2.722 & 0.034 & 0.035 & 0.967 \\
               & 0.5 & 0.621 & 0.020 & 0.021 & 0.943 & 1.026 & 0.041 & 0.043 & 0.962 & 2.337 & 0.046 & 0.049 & 0.968 \\
               & 1.0 & 1.059 & 0.026 & 0.028 & 0.944 & 0.813 & 0.039 & 0.042 & 0.955 & 2.321 & 0.048 & 0.052 & 0.963 \\
frailty-o0c1  & 0.1 & 0.124 & 0.009 & 0.009 & 0.939 & 0.666 & 0.036 & 0.038 & 0.960 & 2.346 & 0.037 & 0.038 & 0.964 \\
               & 0.5 & 0.610 & 0.020 & 0.022 & 0.943 & 0.947 & 0.047 & 0.049 & 0.964 & 2.232 & 0.050 & 0.052 & 0.962 \\
               & 1.0 & 1.053 & 0.027 & 0.029 & 0.946 & 0.811 & 0.046 & 0.048 & 0.950 & 2.308 & 0.051 & 0.055 & 0.968 \\
frailty-o0c0  & 0.1 & 0.113 & 0.009 & 0.009 & 0.938 & 0.670 & 0.037 & 0.038 & 0.958 & 2.315 & 0.037 & 0.038 & 0.964 \\
               & 0.5 & 0.609 & 0.020 & 0.022 & 0.939 & 0.978 & 0.048 & 0.049 & 0.953 & 2.262 & 0.050 & 0.052 & 0.960 \\
               & 1.0 & 1.058 & 0.027 & 0.029 & 0.940 & 0.889 & 0.047 & 0.048 & 0.941 & 2.369 & 0.051 & 0.055 & 0.965 \\
\midrule
\multicolumn{14}{@{}c}{\textbf{\textit{Outcome regression \& KM}}}\\
marginal-OR1   & 0.1 & 0.114 & 0.009 & 0.010 & 0.946 & 1.609 & 0.032 & 0.033 & 0.963 & 4.089 & 0.034 & 0.035 & 0.958 \\
               & 0.5 & 0.323 & 0.022 & 0.023 & 0.947 & 2.918 & 0.041 & 0.043 & 0.953 & 2.381 & 0.047 & 0.049 & 0.963 \\
               & 1.0 & 0.315 & 0.029 & 0.030 & 0.948 & 3.692 & 0.039 & 0.041 & 0.950 & 1.960 & 0.050 & 0.052 & 0.957 \\
marginal-OR0   & 0.1 & 0.363 & 0.009 & 0.009 & 0.896 & 9.934 & 0.039 & 0.040 & 0.607 & 29.323 & 0.040 & 0.041 & 0.592 \\
               & 0.5 & 1.121 & 0.022 & 0.023 & 0.920 & 17.400 & 0.040 & 0.042 & 0.585 & 20.421 & 0.045 & 0.047 & 0.576 \\
               & 1.0 & 1.890 & 0.029 & 0.030 & 0.920 & 19.471 & 0.035 & 0.036 & 0.619 & 16.280 & 0.045 & 0.046 & 0.655 \\
frailty-OR1    & 0.1 & 0.403 & 0.008 & 0.009 & 0.902 & 2.378 & 0.032 & 0.033 & 0.926 & 5.154 & 0.033 & 0.035 & 0.932 \\
               & 0.5 & 2.461 & 0.019 & 0.021 & 0.836 & 8.008 & 0.043 & 0.045 & 0.914 & 3.318 & 0.047 & 0.051 & 0.968 \\
               & 1.0 & 4.701 & 0.026 & 0.028 & 0.768 & 13.094 & 0.043 & 0.046 & 0.913 & 0.954 & 0.051 & 0.055 & 0.970 \\
frailty-OR0    & 0.1 & 0.649 & 0.009 & 0.009 & 0.842 & 3.033 & 0.034 & 0.035 & 0.922 & 11.003 & 0.034 & 0.036 & 0.906 \\
               & 0.5 & 3.264 & 0.022 & 0.023 & 0.757 & 3.272 & 0.043 & 0.045 & 0.954 & 3.255 & 0.047 & 0.049 & 0.941 \\
               & 1.0 & 6.354 & 0.029 & 0.031 & 0.656 & 12.577 & 0.044 & 0.046 & 0.900 & 2.161 & 0.053 & 0.054 & 0.952 \\
KM             & 0.1 & 0.212 & 0.009 & 0.009 & 0.916 & 1.086 & 0.043 & 0.043 & 0.941 & 3.861 & 0.044 & 0.044 & 0.944 \\
               & 0.5 & 0.772 & 0.020 & 0.021 & 0.924 & 1.486 & 0.055 & 0.056 & 0.941 & 3.124 & 0.059 & 0.060 & 0.939 \\
               & 1.0 & 1.211 & 0.027 & 0.028 & 0.926 & 1.375 & 0.053 & 0.053 & 0.937 & 2.954 & 0.060 & 0.060 & 0.940 \\
\bottomrule
\end{tabular}
}
\end{table}

% Table 2: Individual‐level survival (censoring ≈25%)
\begin{table}[ht!]
\centering
\caption{Simulation results for individual-level survival probabilities \(S_I^{(a)}(t)\) and causal effects \(\Delta_I^{\text{SPCE}}(t)\) on different scales under Scenario 3, estimated under 13 methods at time points \(t=\{0.1,0.5,1\}\). The number of clusters is \(M=50\), with censoring rate of approximately \(25\%\). Reported metrics include PBias, MCSD, AESE, and CP (empirical coverage probability of the 95\% CI).}
\label{M50_C25_ICS_E1_C1_Si}
\resizebox{\textwidth}{!}{
\begin{tabular}{@{}l r rrrr rrrr rrrr@{}}
\toprule
Method & \(t\) & \multicolumn{4}{c}{$S_I^{(1)}(t)$} & \multicolumn{4}{c}{$S_I^{(0)}(t)$} & \multicolumn{4}{c}{$\Delta_I^{\mathrm{SPCE}}(t)$} \\
\cmidrule(lr){3-6}\cmidrule(lr){7-10}\cmidrule(lr){11-14}
 & & PBias & MCSD & AESE & CP  & PBias & MCSD & AESE & CP  & PBias & MCSD & AESE & CP \\
\midrule
\multicolumn{14}{@{}c}{\textbf{\textit{Doubly robust estimator (Marginal Cox)}}}\\
marginal-o1c1 & 0.1 & 0.103 & 0.009 & 0.009 & 0.925 & 0.669 & 0.035 & 0.036 & 0.954 & 1.503 & 0.036 & 0.037 & 0.952 \\
              & 0.5 & 0.471 & 0.020 & 0.021 & 0.929 & 0.783 & 0.037 & 0.038 & 0.957 & 1.212 & 0.042 & 0.044 & 0.954 \\
              & 1.0 & 0.790 & 0.027 & 0.028 & 0.934 & 0.884 & 0.030 & 0.031 & 0.956 & 1.371 & 0.041 & 0.043 & 0.956 \\
marginal-o1c0 & 0.1 & 0.107 & 0.009 & 0.009 & 0.926 & 0.657 & 0.035 & 0.036 & 0.954 & 1.494 & 0.035 & 0.037 & 0.952 \\
              & 0.5 & 0.479 & 0.020 & 0.021 & 0.927 & 0.747 & 0.036 & 0.038 & 0.959 & 1.202 & 0.042 & 0.044 & 0.957 \\
              & 1.0 & 0.797 & 0.027 & 0.028 & 0.934 & 0.816 & 0.030 & 0.031 & 0.957 & 1.357 & 0.041 & 0.043 & 0.956 \\
marginal-o0c1 & 0.1 & 0.078 & 0.009 & 0.009 & 0.927 & 0.451 & 0.037 & 0.039 & 0.958 & 1.037 & 0.037 & 0.040 & 0.958 \\
              & 0.5 & 0.411 & 0.021 & 0.022 & 0.945 & 1.013 & 0.039 & 0.042 & 0.958 & 1.251 & 0.042 & 0.046 & 0.961 \\
              & 1.0 & 0.683 & 0.028 & 0.029 & 0.943 & 1.517 & 0.033 & 0.035 & 0.964 & 1.447 & 0.041 & 0.044 & 0.959 \\
marginal-o0c0 & 0.1 & 0.054 & 0.009 & 0.009 & 0.928 & 0.200 & 0.037 & 0.039 & 0.956 & 0.515 & 0.037 & 0.040 & 0.953 \\
              & 0.5 & 0.323 & 0.021 & 0.022 & 0.929 & 0.185 & 0.040 & 0.042 & 0.958 & 0.623 & 0.043 & 0.046 & 0.954 \\
              & 1.0 & 0.580 & 0.028 & 0.029 & 0.935 & 0.350 & 0.034 & 0.035 & 0.964 & 0.903 & 0.042 & 0.044 & 0.965 \\
\midrule
\multicolumn{14}{@{}c}{\textbf{\textit{Doubly robust estimator (Frailty Cox)}}}\\
frailty-o1c1 & 0.1 & 0.116 & 0.008 & 0.009 & 0.935 & 0.643 & 0.034 & 0.036 & 0.964 & 1.491 & 0.035 & 0.037 & 0.961 \\
              & 0.5 & 0.538 & 0.018 & 0.020 & 0.948 & 0.806 & 0.036 & 0.038 & 0.959 & 1.333 & 0.041 & 0.044 & 0.958 \\
              & 1.0 & 0.921 & 0.025 & 0.027 & 0.943 & 0.869 & 0.030 & 0.032 & 0.946 & 1.542 & 0.040 & 0.044 & 0.951 \\
frailty-o1c0 & 0.1 & 0.134 & 0.008 & 0.009 & 0.936 & 0.653 & 0.034 & 0.036 & 0.962 & 1.561 & 0.036 & 0.037 & 0.961 \\
              & 0.5 & 0.571 & 0.018 & 0.020 & 0.948 & 0.832 & 0.036 & 0.038 & 0.961 & 1.399 & 0.041 & 0.044 & 0.959 \\
              & 1.0 & 0.955 & 0.025 & 0.027 & 0.945 & 0.946 & 0.030 & 0.032 & 0.952 & 1.614 & 0.040 & 0.044 & 0.950 \\
frailty-o0c1 & 0.1 & 0.085 & 0.009 & 0.009 & 0.946 & 0.094 & 0.039 & 0.041 & 0.953 & 0.409 & 0.039 & 0.041 & 0.956 \\
              & 0.5 & 0.448 & 0.019 & 0.022 & 0.953 & 0.192 & 0.043 & 0.046 & 0.966 & 0.599 & 0.045 & 0.050 & 0.968 \\
              & 1.0 & 0.790 & 0.026 & 0.028 & 0.954 & 0.355 & 0.037 & 0.041 & 0.967 & 0.941 & 0.043 & 0.049 & 0.971 \\
frailty-o0c0 & 0.1 & 0.069 & 0.008 & 0.009 & 0.947 & 0.092 & 0.039 & 0.041 & 0.955 & 0.360 & 0.039 & 0.041 & 0.961 \\
              & 0.5 & 0.426 & 0.020 & 0.022 & 0.952 & 0.170 & 0.043 & 0.046 & 0.967 & 0.577 & 0.044 & 0.050 & 0.969 \\
              & 1.0 & 0.775 & 0.026 & 0.029 & 0.949 & 0.309 & 0.037 & 0.041 & 0.963 & 0.937 & 0.043 & 0.049 & 0.968 \\
\midrule
\multicolumn{14}{@{}c}{\textbf{\textit{Outcome regression \& KM}}}\\
marginal-OR1 & 0.1 & 0.008 & 0.009 & 0.009 & 0.938 & 1.204 & 0.035 & 0.036 & 0.959 & 2.945 & 0.003 & 0.003 & 0.947 \\
              & 0.5 & 0.004 & 0.007 & 0.007 & 0.945 & 1.122 & 0.036 & 0.037 & 0.963 & 1.332 & 0.019 & 0.020 & 0.953 \\
              & 1.0 & 0.014 & 0.019 & 0.019 & 0.943 & 1.111 & 0.033 & 0.035 & 0.961 & 0.866 & 0.039 & 0.040 & 0.959 \\
marginal-OR0 & 0.1 & 0.007 & 0.009 & 0.009 & 0.936 & 4.794 & 0.040 & 0.041 & 0.893 & 8.714 & 0.041 & 0.041 & 0.884 \\
              & 0.5 & 0.309 & 0.021 & 0.023 & 0.955 & 3.904 & 0.037 & 0.038 & 0.931 & 1.814 & 0.042 & 0.043 & 0.943 \\
              & 1.0 & 0.658 & 0.028 & 0.030 & 0.952 & 0.883 & 0.030 & 0.030 & 0.954 & 1.193 & 0.041 & 0.042 & 0.957 \\
frailty-OR1  & 0.1 & 0.184 & 0.000 & 0.000 & 0.902 & 1.689 & 0.003 & 0.003 & 0.925 & 4.572 & 0.003 & 0.003 & 0.931 \\
              & 0.5 & 1.163 & 0.006 & 0.006 & 0.851 & 5.578 & 0.018 & 0.019 & 0.916 & 4.004 & 0.019 & 0.020 & 0.948 \\
              & 1.0 & 2.233 & 0.016 & 0.018 & 0.802 & 8.937 & 0.036 & 0.038 & 0.914 & 2.845 & 0.040 & 0.043 & 0.967 \\
frailty-OR0  & 0.1 & 0.075 & 0.001 & 0.001 & 0.931 & 1.224 & 0.003 & 0.003 & 0.933 & 3.555 & 0.003 & 0.003 & 0.940 \\
              & 0.5 & 0.667 & 0.007 & 0.007 & 0.922 & 9.205 & 0.018 & 0.018 & 0.790 & 9.325 & 0.019 & 0.019 & 0.852 \\
              & 1.0 & 1.481 & 0.019 & 0.020 & 0.899 & 17.914 & 0.037 & 0.038 & 0.630 & 10.969 & 0.041 & 0.043 & 0.795 \\
KM           & 0.1 & 0.101 & 0.000 & 0.000 & 0.912 & 0.511 & 0.003 & 0.003 & 0.932 & 2.035 & 0.004 & 0.003 & 0.938 \\
              & 0.5 & 0.393 & 0.006 & 0.007 & 0.921 & 0.790 & 0.022 & 0.022 & 0.942 & 1.776 & 0.023 & 0.023 & 0.936 \\
              & 1.0 & 0.633 & 0.018 & 0.018 & 0.925 & 0.901 & 0.043 & 0.043 & 0.941 & 1.794 & 0.047 & 0.047 & 0.930 \\
\bottomrule
\end{tabular}
}
\end{table}

% Table 3: Cluster‐level RMST (censoring ≈25%)
\begin{table}[ht!]
\centering
\caption{Simulation results for cluster-level RMST \(\mu_C^{(a)}(t)\) and causal effects \(\Delta_C^{\text{RMST}}(t)\) on different scales under Scenario 3, estimated under 13 methods at time points \( t = \{0.1, 0.5, 1\} \). The number of clusters is \( M = 50 \), with censoring rate of approximately \(25\%\). Reported metrics include PBias (percentage bias), MCSD (Monte Carlo standard deviation), AESE (average estimated standard error from the jackknife procedure), and CP (empirical coverage probability of the \(95\%\) confidence interval).}
\label{M50_C25_ICS_E1_C1_rmstc}
\resizebox{\textwidth}{!}{
\begin{tabular}{@{}l r rrrr rrrr rrrr@{}}
\toprule
Method & \(t\) & \multicolumn{4}{c}{$\mu_C^{(1)}(t)$} & \multicolumn{4}{c}{$\mu_C^{(0)}(t)$} & \multicolumn{4}{c}{$\Delta_C^{\mathrm{RMST}}(t)$} \\
\cmidrule(lr){3-6}\cmidrule(lr){7-10}\cmidrule(lr){11-14}
 & & PBias & MCSD & AESE & CP  & PBias & MCSD & AESE & CP  & PBias & MCSD & AESE & CP \\
\midrule
\multicolumn{14}{@{}c}{\textbf{\textit{Doubly robust estimator (Marginal Cox)}}}\\
marginal-o1c1 & 0.1 & 0.062 & 0.001 & 0.001 & 0.928 & 0.603 & 0.002 & 0.002 & 0.961 & 3.216 & 0.002 & 0.003 & 0.965 \\
              & 0.5 & 0.253 & 0.007 & 0.007 & 0.934 & 0.897 & 0.018 & 0.019 & 0.965 & 2.397 & 0.019 & 0.020 & 0.957 \\
              & 1.0 & 0.442 & 0.019 & 0.020 & 0.932 & 0.955 & 0.038 & 0.039 & 0.959 & 2.207 & 0.043 & 0.045 & 0.958 \\
marginal-o1c0 & 0.1 & 0.065 & 0.001 & 0.001 & 0.928 & 0.618 & 0.002 & 0.002 & 0.962 & 3.306 & 0.002 & 0.003 & 0.967 \\
              & 0.5 & 0.270 & 0.007 & 0.007 & 0.931 & 0.923 & 0.017 & 0.019 & 0.964 & 2.492 & 0.019 & 0.020 & 0.959 \\
              & 1.0 & 0.469 & 0.019 & 0.020 & 0.930 & 0.990 & 0.037 & 0.039 & 0.961 & 2.310 & 0.042 & 0.045 & 0.960 \\
marginal-o0c1 & 0.1 & 0.073 & 0.001 & 0.001 & 0.930 & 0.657 & 0.003 & 0.003 & 0.968 & 3.531 & 0.003 & 0.003 & 0.971 \\
              & 0.5 & 0.295 & 0.007 & 0.007 & 0.943 & 1.233 & 0.020 & 0.021 & 0.967 & 3.139 & 0.020 & 0.022 & 0.969 \\
              & 1.0 & 0.499 & 0.019 & 0.020 & 0.947 & 1.475 & 0.042 & 0.045 & 0.966 & 2.990 & 0.044 & 0.048 & 0.964 \\
marginal-o0c0 & 0.1 & 0.061 & 0.001 & 0.001 & 0.927 & 0.552 & 0.003 & 0.003 & 0.964 & 2.970 & 0.003 & 0.003 & 0.969 \\
              & 0.5 & 0.254 & 0.007 & 0.007 & 0.931 & 0.942 & 0.020 & 0.021 & 0.961 & 2.482 & 0.020 & 0.022 & 0.961 \\
              & 1.0 & 0.446 & 0.019 & 0.020 & 0.930 & 1.107 & 0.042 & 0.045 & 0.958 & 2.406 & 0.045 & 0.048 & 0.956 \\
\midrule
\multicolumn{14}{@{}c}{\textbf{\textit{Doubly robust estimator (Frailty Cox)}}}\\
frailty-o1c1 & 0.1 & 0.056 & 0.001 & 0.001 & 0.930 & 0.555 & 0.002 & 0.002 & 0.962 & 2.957 & 0.002 & 0.003 & 0.963 \\
              & 0.5 & 0.311 & 0.006 & 0.007 & 0.942 & 0.809 & 0.018 & 0.018 & 0.959 & 2.396 & 0.019 & 0.020 & 0.971 \\
              & 1.0 & 0.534 & 0.018 & 0.019 & 0.943 & 0.840 & 0.037 & 0.039 & 0.956 & 2.269 & 0.041 & 0.045 & 0.967 \\
frailty-o1c0 & 0.1 & 0.063 & 0.001 & 0.001 & 0.929 & 0.570 & 0.002 & 0.002 & 0.960 & 3.065 & 0.002 & 0.003 & 0.962 \\
              & 0.5 & 0.324 & 0.006 & 0.007 & 0.939 & 0.856 & 0.018 & 0.018 & 0.959 & 2.522 & 0.019 & 0.020 & 0.970 \\
              & 1.0 & 0.557 & 0.018 & 0.019 & 0.942 & 0.897 & 0.037 & 0.039 & 0.956 & 2.391 & 0.042 & 0.045 & 0.967 \\
frailty-o0c1 & 0.1 & 0.058 & 0.001 & 0.001 & 0.931 & 0.467 & 0.003 & 0.003 & 0.957 & 2.549 & 0.003 & 0.003 & 0.963 \\
              & 0.5 & 0.315 & 0.007 & 0.007 & 0.939 & 0.747 & 0.020 & 0.021 & 0.966 & 2.293 & 0.021 & 0.022 & 0.961 \\
              & 1.0 & 0.548 & 0.018 & 0.020 & 0.946 & 0.811 & 0.043 & 0.045 & 0.958 & 2.263 & 0.045 & 0.048 & 0.962 \\
frailty-o0c0 & 0.1 & 0.053 & 0.001 & 0.001 & 0.933 & 0.472 & 0.003 & 0.003 & 0.956 & 2.542 & 0.003 & 0.003 & 0.963 \\
              & 0.5 & 0.314 & 0.007 & 0.007 & 0.936 & 0.759 & 0.020 & 0.021 & 0.960 & 2.311 & 0.021 & 0.022 & 0.958 \\
              & 1.0 & 0.551 & 0.018 & 0.020 & 0.939 & 0.839 & 0.044 & 0.045 & 0.949 & 2.305 & 0.045 & 0.048 & 0.957 \\
\midrule
\multicolumn{14}{@{}c}{\textbf{\textit{Outcome regression \& KM}}}\\
marginal-OR1 & 0.1 & 0.047 & 0.001 & 0.001 & 0.948 & 1.095 & 0.002 & 0.002 & 0.967 & 4.927 & 0.002 & 0.003 & 0.959 \\
              & 0.5 & 0.209 & 0.007 & 0.007 & 0.953 & 2.003 & 0.018 & 0.018 & 0.957 & 3.132 & 0.019 & 0.020 & 0.961 \\
              & 1.0 & 0.275 & 0.020 & 0.020 & 0.948 & 2.508 & 0.037 & 0.039 & 0.958 & 2.543 & 0.043 & 0.045 & 0.959 \\
marginal-OR0 & 0.1 & 0.205 & 0.001 & 0.001 & 0.905 & 6.316 & 0.003 & 0.003 & 0.656 & 31.134 & 0.003 & 0.003 & 0.644 \\
              & 0.5 & 0.643 & 0.007 & 0.007 & 0.909 & 12.197 & 0.019 & 0.020 & 0.583 & 24.559 & 0.020 & 0.021 & 0.562 \\ 
              & 1.0 & 1.033 & 0.020 & 0.021 & 0.913 & 14.628 & 0.038 & 0.039 & 0.588 & 20.799 & 0.042 & 0.043 & 0.577 \\
frailty-OR1  & 0.1 & 0.171 & 0.000 & 0.000 & 0.911 & 1.249 & 0.002 & 0.002 & 0.926 & 4.939 & 0.002 & 0.002 & 0.938 \\
              & 0.5 & 1.165 & 0.006 & 0.007 & 0.864 & 4.067 & 0.018 & 0.019 & 0.924 & 4.240 & 0.019 & 0.020 & 0.949 \\
              & 1.0 & 2.284 & 0.017 & 0.019 & 0.827 & 6.478 & 0.039 & 0.042 & 0.919 & 3.009 & 0.043 & 0.046 & 0.967 \\
frailty-OR0  & 0.1 & 0.317 & 0.001 & 0.001 & 0.865 & 2.626 & 0.003 & 0.003 & 0.908 & 14.273 & 0.003 & 0.003 & 0.890 \\
              & 0.5 & 1.589 & 0.007 & 0.007 & 0.797 & 1.085 & 0.018 & 0.019 & 0.949 & 6.568 & 0.019 & 0.020 & 0.925 \\
              & 1.0 & 3.070 & 0.020 & 0.021 & 0.736 & 2.188 & 0.040 & 0.041 & 0.953 & 4.183 & 0.044 & 0.045 & 0.936 \\
KM           & 0.1 & 0.101 & 0.001 & 0.001 & 0.920 & 0.761 & 0.003 & 0.003 & 0.940 & 2.035 & 0.004 & 0.003 & 0.938 \\
              & 0.5 & 0.432 & 0.006 & 0.007 & 0.922 & 1.205 & 0.024 & 0.024 & 0.944 & 3.480 & 0.025 & 0.025 & 0.946 \\
              & 1.0 & 0.694 & 0.018 & 0.019 & 0.922 & 1.303 & 0.050 & 0.051 & 0.941 & 3.215 & 0.054 & 0.054 & 0.941 \\
\bottomrule
\end{tabular}
}
\end{table}

\begin{table}[ht!]
\centering
\caption{Simulation results for individual-level RMST \(\mu_I^{(a)}(t)\) and causal effects \(\Delta_I^{\text{RMST}}(t)\) on different scales under Scenario 3, estimated under 13 methods at time points \( t = \{0.1, 0.5, 1\} \). The number of clusters is \( M = 50 \), with censoring rate of approximately \(25\%\). Reported metrics include PBias (percentage bias), MCSD (Monte Carlo standard deviation), AESE (average estimated standard error from the jackknife procedure), and CP (empirical coverage probability of the \(95\%\) confidence interval).}
\label{M50_C25_ICS_E1_C1_rmsti}
\resizebox{\textwidth}{!}{
\begin{tabular}{@{}l r rrrr rrrr rrrr@{}}
\toprule
Method & \(t\) & \multicolumn{4}{c}{$\mu_I^{(1)}(t)$} & \multicolumn{4}{c}{$\mu_I^{(0)}(t)$} & \multicolumn{4}{c}{$\Delta_C^{\mathrm{RMST}}(t)$} \\
\cmidrule(lr){3-6} \cmidrule(lr){7-10} \cmidrule(lr){11-14}
 & & PBias & MCSD & AESE & CP  
   & PBias & MCSD & AESE & CP  
   & PBias & MCSD & AESE & CP \\
\midrule
\multicolumn{14}{@{}c}{\textbf{\textit{Doubly robust estimator (Marginal Cox)}}}\\
marginal-o1c1  & 0.1 & 0.058 & 0.001 & 0.001 & 0.920 & 0.490 & 0.003 & 0.003 & 0.953 & 1.790 & 0.003 & 0.003 & 0.949 \\
               & 0.5 & 0.239 & 0.007 & 0.007 & 0.926 & 0.681 & 0.018 & 0.018 & 0.956 & 1.315 & 0.019 & 0.020 & 0.953 \\
               & 1.0 & 0.421 & 0.018 & 0.019 & 0.927 & 0.741 & 0.034 & 0.035 & 0.956 & 1.301 & 0.039 & 0.041 & 0.950 \\
marginal-o1c0  & 0.1 & 0.060 & 0.001 & 0.001 & 0.920 & 0.484 & 0.003 & 0.003 & 0.955 & 1.777 & 0.003 & 0.003 & 0.949 \\
               & 0.5 & 0.245 & 0.007 & 0.007 & 0.922 & 0.659 & 0.017 & 0.018 & 0.957 & 1.303 & 0.019 & 0.020 & 0.956 \\
               & 1.0 & 0.428 & 0.018 & 0.019 & 0.924 & 0.708 & 0.033 & 0.035 & 0.957 & 1.289 & 0.038 & 0.040 & 0.952 \\
marginal-o0c1  & 0.1 & 0.044 & 0.001 & 0.001 & 0.926 & 0.305 & 0.003 & 0.003 & 0.954 & 1.147 & 0.003 & 0.003 & 0.953 \\
               & 0.5 & 0.205 & 0.007 & 0.007 & 0.934 & 0.624 & 0.019 & 0.020 & 0.958 & 1.175 & 0.019 & 0.021 & 0.962 \\
               & 1.0 & 0.363 & 0.019 & 0.020 & 0.941 & 0.842 & 0.036 & 0.038 & 0.957 & 1.276 & 0.039 & 0.042 & 0.960 \\
marginal-o0c0  & 0.1 & 0.031 & 0.001 & 0.001 & 0.922 & 0.152 & 0.003 & 0.003 & 0.953 & 0.609 & 0.003 & 0.003 & 0.953 \\
               & 0.5 & 0.151 & 0.007 & 0.007 & 0.926 & 0.181 & 0.019 & 0.020 & 0.958 & 0.539 & 0.019 & 0.021 & 0.954 \\
               & 1.0 & 0.289 & 0.019 & 0.020 & 0.929 & 0.217 & 0.037 & 0.039 & 0.960 & 0.673 & 0.039 & 0.042 & 0.952 \\
\midrule
\multicolumn{14}{@{}c}{\textbf{\textit{Doubly robust estimator (Frailty Cox)}}}\\
frailty-o1c1  & 0.1 & 0.060 & 0.000 & 0.000 & 0.927 & 0.482 & 0.003 & 0.003 & 0.960 & 1.772 & 0.003 & 0.003 & 0.959 \\
               & 0.5 & 0.285 & 0.006 & 0.007 & 0.943 & 0.670 & 0.017 & 0.018 & 0.960 & 1.403 & 0.018 & 0.020 & 0.966 \\
               & 1.0 & 0.489 & 0.017 & 0.018 & 0.942 & 0.735 & 0.033 & 0.035 & 0.962 & 1.416 & 0.038 & 0.041 & 0.957 \\
frailty-o1c0  & 0.1 & 0.068 & 0.000 & 0.000 & 0.928 & 0.483 & 0.003 & 0.003 & 0.959 & 1.811 & 0.003 & 0.003 & 0.957 \\
               & 0.5 & 0.303 & 0.006 & 0.007 & 0.944 & 0.693 & 0.017 & 0.018 & 0.961 & 1.468 & 0.019 & 0.020 & 0.962 \\
               & 1.0 & 0.513 & 0.017 & 0.018 & 0.944 & 0.766 & 0.033 & 0.035 & 0.963 & 1.481 & 0.038 & 0.041 & 0.957 \\
frailty-o0c1  & 0.1 & 0.041 & 0.000 & 0.001 & 0.937 & 0.103 & 0.003 & 0.003 & 0.952 & 0.496 & 0.003 & 0.003 & 0.950 \\
               & 0.5 & 0.225 & 0.006 & 0.007 & 0.944 & 0.012 & 0.020 & 0.021 & 0.954 & 0.503 & 0.020 & 0.022 & 0.960 \\
               & 1.0 & 0.405 & 0.018 & 0.019 & 0.949 & 0.078 & 0.039 & 0.043 & 0.966 & 0.652 & 0.041 & 0.046 & 0.968 \\
frailty-o0c0  & 0.1 & 0.032 & 0.000 & 0.001 & 0.936 & 0.104 & 0.003 & 0.003 & 0.955 & 0.462 & 0.003 & 0.003 & 0.955 \\
               & 0.5 & 0.212 & 0.006 & 0.007 & 0.942 & 0.012 & 0.020 & 0.021 & 0.956 & 0.474 & 0.020 & 0.022 & 0.965 \\
               & 1.0 & 0.390 & 0.018 & 0.020 & 0.944 & 0.062 & 0.040 & 0.043 & 0.966 & 0.639 & 0.041 & 0.046 & 0.968 \\
\midrule
\multicolumn{14}{@{}c}{\textbf{\textit{Outcome regression \& KM}}}\\
marginal-OR1  & 0.1 & 0.020 & 0.001 & 0.001 & 0.924 & 0.905 & 0.003 & 0.003 & 0.949 & 2.945 & 0.003 & 0.003 & 0.947 \\
               & 0.5 & 0.004 & 0.007 & 0.007 & 0.945 & 1.145 & 0.017 & 0.018 & 0.959 & 1.332 & 0.019 & 0.020 & 0.953 \\
               & 1.0 & 0.014 & 0.019 & 0.019 & 0.943 & 1.111 & 0.033 & 0.035 & 0.961 & 0.866 & 0.039 & 0.040 & 0.959 \\
marginal-OR0  & 0.1 & 0.021 & 0.000 & 0.001 & 0.937 & 3.026 & 0.003 & 0.003 & 0.907 & 9.653 & 0.003 & 0.003 & 0.901 \\
               & 0.5 & 0.105 & 0.007 & 0.007 & 0.956 & 4.482 & 0.019 & 0.019 & 0.907 & 5.018 & 0.020 & 0.020 & 0.908 \\
               & 1.0 & 0.294 & 0.019 & 0.020 & 0.960 & 3.593 & 0.035 & 0.036 & 0.928 & 2.206 & 0.039 & 0.040 & 0.939 \\
frailty-OR1   & 0.1 & 0.184 & 0.000 & 0.000 & 0.902 & 1.689 & 0.003 & 0.003 & 0.925 & 4.572 & 0.003 & 0.003 & 0.931 \\
               & 0.5 & 1.163 & 0.006 & 0.006 & 0.851 & 5.578 & 0.018 & 0.019 & 0.916 & 4.004 & 0.019 & 0.020 & 0.948 \\
               & 1.0 & 2.233 & 0.016 & 0.018 & 0.802 & 8.937 & 0.036 & 0.038 & 0.914 & 2.845 & 0.040 & 0.043 & 0.967 \\
frailty-OR0   & 0.1 & 0.075 & 0.001 & 0.001 & 0.931 & 1.224 & 0.003 & 0.003 & 0.933 & 3.555 & 0.003 & 0.003 & 0.940 \\
               & 0.5 & 0.667 & 0.007 & 0.007 & 0.922 & 9.205 & 0.018 & 0.018 & 0.790 & 9.325 & 0.019 & 0.019 & 0.852 \\
               & 1.0 & 1.481 & 0.019 & 0.020 & 0.899 & 17.914 & 0.037 & 0.038 & 0.630 & 10.969 & 0.041 & 0.043 & 0.795 \\
KM             & 0.1 & 0.101 & 0.000 & 0.000 & 0.912 & 0.511 & 0.003 & 0.003 & 0.932 & 2.035 & 0.004 & 0.003 & 0.938 \\
               & 0.5 & 0.393 & 0.006 & 0.007 & 0.921 & 0.790 & 0.022 & 0.022 & 0.942 & 1.776 & 0.023 & 0.023 & 0.936 \\
               & 1.0 & 0.633 & 0.018 & 0.018 & 0.925 & 0.901 & 0.043 & 0.043 & 0.941 & 1.794 & 0.047 & 0.047 & 0.930 \\
\bottomrule
\end{tabular}
}
\end{table}

%\import{./resultsRev/}{M50_C75_ICS_E1_C1.tex}
\begin{table}[ht!]
\centering
\caption{Simulation results for cluster-level survival probabilities \(S_C^{(a)}(t)\) and causal effects \(\Delta_C^{\text{SPCE}}(t)\) on different scales under Scenario 3, estimated under 13 methods at time points \( t = \{0.1, 0.5, 1\} \). The number of clusters is \( M = 50 \), with censoring rate of approximately \(75\%\). Reported metrics include PBias (percentage bias), MCSD (Monte Carlo standard deviation), AESE (average estimated standard error from the jackknife procedure), and CP (empirical coverage probability of the \(95\%\) confidence interval).}
\label{M50_C75_ICS_E1_C1_Sc}
\resizebox{\textwidth}{!}{
\begin{tabular}{@{}l r 
  rrrr   rrrr   rrrr@{}}
\toprule
Method & \(t\) 
  & \multicolumn{4}{c}{$S_C^{(1)}(t)$} 
  & \multicolumn{4}{c}{$S_C^{(0)}(t)$} 
  & \multicolumn{4}{c}{$\Delta_C^{\mathrm{SPCE}}(t)$} \\
\cmidrule(lr){3-6} \cmidrule(lr){7-10} \cmidrule(lr){11-14}
 & & PBias & MCSD & AESE & CP 
     & PBias & MCSD & AESE & CP 
     & PBias & MCSD & AESE & CP \\
\midrule
\multicolumn{14}{@{}c}{\textbf{\textit{Doubly robust estimator (Marginal Cox)}}}\\
marginal-o1c1 & 0.1 & 2.394 & 0.064 & 0.032 & 0.951 
                      & 1.440 & 0.061 & 0.050 & 0.966 
                      & 3.530 & 0.074 & 0.060 & 0.953 \\
              & 0.5 & 4.317 & 0.087 & 0.061 & 0.938 
                      & 2.794 & 0.067 & 0.064 & 0.972 
                      & 4.455 & 0.103 & 0.094 & 0.964 \\
              & 1.0 & 5.322 & 0.088 & 0.076 & 0.943 
                      & 1.873 & 0.063 & 0.059 & 0.961 
                      & 6.539 & 0.109 & 0.103 & 0.958 \\
marginal-o1c0 & 0.1 & 11.853 & 0.255 & 0.174 & 0.973 
                      & 12.070 & 0.226 & 0.196 & 0.973 
                      & 16.322 & 0.260 & 0.257 & 0.969 \\
              & 0.5 & 18.106 & 0.310 & 0.301 & 0.960 
                      & 7.187 & 0.244 & 0.262 & 0.964 
                      & 0.610 & 0.336 & 0.414 & 0.914 \\
              & 1.0 & 19.468 & 0.335 & 0.370 & 0.929 
                      & 5.602 & 0.252 & 0.264 & 0.959 
                      & 1.726 & 0.365 & 0.482 & 0.880 \\
marginal-o0c1 & 0.1 & 2.095 & 0.074 & 0.035 & 0.970 
                      & 0.418 & 0.074 & 0.062 & 0.944 
                      & 7.413 & 0.084 & 0.072 & 0.935 \\
              & 0.5 & 3.478 & 0.091 & 0.059 & 0.968 
                      & 0.899 & 0.091 & 0.083 & 0.957 
                      & 4.856 & 0.118 & 0.108 & 0.963 \\
              & 1.0 & 5.136 & 0.101 & 0.079 & 0.960 
                      & 0.798 & 0.074 & 0.072 & 0.948 
                      & 7.597 & 0.120 & 0.116 & 0.960 \\
marginal-o0c0 & 0.1 & 11.552 & 0.264 & 0.173 & 0.966 
                      & 10.102 & 0.180 & 0.164 & 0.833 
                      & 37.740 & 0.178 & 0.177 & 0.806 \\
              & 0.5 & 20.018 & 0.318 & 0.306 & 0.953 
                      & 40.194 & 0.262 & 0.263 & 0.881 
                      & 41.184 & 0.278 & 0.317 & 0.812 \\
              & 1.0 & 24.561 & 0.333 & 0.380 & 0.937 
                      & 66.151 & 0.284 & 0.279 & 0.885 
                      & 39.474 & 0.317 & 0.376 & 0.768 \\
\midrule
\multicolumn{14}{@{}c}{\textbf{\textit{Doubly robust estimator (Frailty Cox)}}}\\
frailty-o1c1 & 0.1 & 0.353 & 0.017 & 0.017 & 0.962 
                      & 0.792 & 0.038 & 0.039 & 0.958 
                      & 0.883 & 0.041 & 0.043 & 0.971 \\
              & 0.5 & 0.031 & 0.033 & 0.036 & 0.974 
                      & 0.505 & 0.045 & 0.048 & 0.961 
                      & 0.462 & 0.055 & 0.062 & 0.970 \\
              & 1.0 & 0.430 & 0.041 & 0.047 & 0.970 
                      & 0.373 & 0.043 & 0.045 & 0.960 
                      & 0.555 & 0.063 & 0.068 & 0.964 \\
frailty-o1c0 & 0.1 & 0.178 & 0.014 & 0.015 & 0.953 
                      & 0.858 & 0.035 & 0.037 & 0.963 
                      & 1.733 & 0.038 & 0.040 & 0.962 \\
              & 0.5 & 0.090 & 0.034 & 0.037 & 0.963 
                      & 0.559 & 0.047 & 0.049 & 0.966 
                      & 0.784 & 0.058 & 0.063 & 0.967 \\
              & 1.0 & 0.433 & 0.048 & 0.052 & 0.966 
                      & 0.947 & 0.045 & 0.049 & 0.966 
                      & 0.093 & 0.066 & 0.074 & 0.971 \\
frailty-o0c1 & 0.1 & 0.569 & 0.018 & 0.015 & 0.871 
                      & 2.777 & 0.043 & 0.045 & 0.896 
                      & 5.642 & 0.046 & 0.048 & 0.933 \\
              & 0.5 & 2.395 & 0.033 & 0.035 & 0.867 
                      & 6.700 & 0.055 & 0.058 & 0.918 
                      & 2.091 & 0.064 & 0.069 & 0.967 \\
              & 1.0 & 3.362 & 0.042 & 0.048 & 0.911 
                      & 11.334 & 0.055 & 0.058 & 0.918 
                      & 1.948 & 0.071 & 0.077 & 0.968 \\
frailty-o0c0 & 0.1 & 0.994 & 0.013 & 0.013 & 0.807 
                      & 6.574 & 0.037 & 0.041 & 0.792 
                      & 14.700 & 0.039 & 0.043 & 0.866 \\
              & 0.5 & 2.394 & 0.033 & 0.035 & 0.868 
                      & 15.890 & 0.056 & 0.061 & 0.819 
                      & 11.655 & 0.065 & 0.070 & 0.916 \\
              & 1.0 & 2.548 & 0.055 & 0.052 & 0.907 
                      & 24.226 & 0.060 & 0.064 & 0.821 
                      & 11.926 & 0.079 & 0.084 & 0.936 \\
\midrule
\multicolumn{14}{@{}c}{\textbf{\textit{Outcome regression \& KM}}}\\
marginal-OR1  & 0.1 & 0.365 & 0.015 & 0.016 & 0.957 
                      & 1.549 & 0.034 & 0.037 & 0.959 
                      & 2.967 & 0.037 & 0.040 & 0.964 \\
              & 0.5 & 1.057 & 0.034 & 0.039 & 0.958 
                      & 2.619 & 0.043 & 0.046 & 0.957 
                      & 0.570 & 0.055 & 0.061 & 0.963 \\
              & 1.0 & 1.658 & 0.045 & 0.052 & 0.947 
                      & 2.948 & 0.041 & 0.044 & 0.949 
                      & 0.788 & 0.061 & 0.069 & 0.966 \\
marginal-OR0  & 0.1 & 1.446 & 0.011 & 0.012 & 0.700 
                      & 0.122 & 0.044 & 0.047 & 0.938 
                      & 5.168 & 0.045 & 0.048 & 0.949 \\
              & 0.5 & 3.106 & 0.033 & 0.036 & 0.848 
                      & 4.769 & 0.052 & 0.056 & 0.931 
                      & 11.313 & 0.061 & 0.066 & 0.892 \\
              & 1.0 & 2.830 & 0.049 & 0.054 & 0.924 
                      & 5.972 & 0.047 & 0.050 & 0.919 
                      & 8.759 & 0.066 & 0.073 & 0.920 \\
frailty-OR1   & 0.1 & 0.138 & 0.012 & 0.014 & 0.959 
                      & 0.070 & 0.033 & 0.035 & 0.965 
                      & 0.724 & 0.036 & 0.038 & 0.969 \\
              & 0.5 & 0.427 & 0.028 & 0.032 & 0.960 
                      & 2.913 & 0.044 & 0.046 & 0.958 
                      & 2.164 & 0.051 & 0.057 & 0.968 \\
              & 1.0 & 1.195 & 0.037 & 0.042 & 0.959 
                      & 6.067 & 0.043 & 0.045 & 0.961 
                      & 2.086 & 0.056 & 0.063 & 0.966 \\
frailty-OR0   & 0.1 & 1.670 & 0.010 & 0.010 & 0.603 
                      & 2.900 & 0.036 & 0.039 & 0.910 
                      & 1.788 & 0.037 & 0.040 & 0.958 \\
              & 0.5 & 5.091 & 0.027 & 0.030 & 0.671 
                      & 9.840 & 0.048 & 0.051 & 0.892 
                      & 0.142 & 0.053 & 0.058 & 0.971 \\
              & 1.0 & 7.251 & 0.039 & 0.043 & 0.764 
                      & 20.173 & 0.050 & 0.053 & 0.829 
                      & 1.455 & 0.060 & 0.068 & 0.975 \\
KM            & 0.1 & 1.861 & 0.010 & 0.010 & 0.540 
                      & 16.400 & 0.034 & 0.035 & 0.164 
                      & 39.031 & 0.036 & 0.037 & 0.282 \\
              & 0.5 & 5.450 & 0.028 & 0.029 & 0.602 
                      & 38.806 & 0.058 & 0.060 & 0.257 
                      & 29.309 & 0.065 & 0.067 & 0.559 \\
              & 1.0 & 7.242 & 0.041 & 0.043 & 0.713 
                      & 55.015 & 0.065 & 0.067 & 0.328 
                      & 24.942 & 0.077 & 0.080 & 0.729 \\
\bottomrule
\end{tabular}
}
\end{table}

\begin{table}[ht!]
\centering
\caption{Simulation results for cluster-level survival probabilities \(S_I^{(a)}(t)\) and causal effects \(\Delta_I^{\text{SPCE}}(t)\) on different scales under Scenario 3, estimated under 13 methods at time points \( t = \{0.1, 0.5, 1\} \). The number of clusters is \( M = 50 \), with censoring rate of approximately \(75\%\). Reported metrics include PBias (percentage bias), MCSD (Monte Carlo standard deviation), AESE (average estimated standard error from the jackknife procedure), and CP (empirical coverage probability of the \(95\%\) confidence interval).}
\label{M50_C75_ICS_E1_C1_Si}
\resizebox{\textwidth}{!}{
\begin{tabular}{@{}l r 
  rrrr   rrrr   rrrr@{}}
\toprule
Method & \(t\) 
  & \multicolumn{4}{c}{$S_I^{(1)}(t)$} 
  & \multicolumn{4}{c}{$S_I^{(0)}(t)$} 
  & \multicolumn{4}{c}{$\Delta_I^{\mathrm{SPCE}}(t)$} \\
\cmidrule(lr){3-6} \cmidrule(lr){7-10} \cmidrule(lr){11-14}
 & & PBias & MCSD & AESE & CP 
     & PBias & MCSD & AESE & CP 
     & PBias & MCSD & AESE & CP \\
\midrule
\multicolumn{14}{@{}c}{\textbf{\textit{Doubly robust estimator (Marginal Cox)}}}\\
marginal-o1c1 & 0.1 & 3.564 & 0.073 & 0.038 & 0.937 
                      & 1.765 & 0.070 & 0.059 & 0.962 
                      & 5.693 & 0.086 & 0.073 & 0.959 \\
              & 0.5 & 6.712 & 0.094 & 0.069 & 0.916 
                      & 3.802 & 0.067 & 0.064 & 0.970 
                      & 7.514 & 0.107 & 0.101 & 0.962 \\
              & 1.0 & 8.321 & 0.095 & 0.084 & 0.903 
                      & 2.779 & 0.059 & 0.050 & 0.967 
                      & 9.418 & 0.111 & 0.105 & 0.950 \\
marginal-o1c0 & 0.1 & 12.863 & 0.257 & 0.179 & 0.969 
                      & 13.218 & 0.207 & 0.186 & 0.979 
                      & 5.135 & 0.250 & 0.256 & 0.972 \\
              & 0.5 & 19.598 & 0.293 & 0.282 & 0.964 
                      & 7.409 & 0.204 & 0.204 & 0.964 
                      & 10.629 & 0.313 & 0.381 & 0.936 \\
              & 1.0 & 21.444 & 0.308 & 0.331 & 0.943 
                      & 6.966 & 0.210 & 0.191 & 0.960 
                      & 13.755 & 0.330 & 0.429 & 0.922 \\
marginal-o0c1 & 0.1 & 3.088 & 0.085 & 0.041 & 0.971 
                      & 1.855 & 0.090 & 0.074 & 0.907 
                      & 10.641 & 0.106 & 0.089 & 0.901 \\
              & 0.5 & 5.598 & 0.098 & 0.068 & 0.960 
                      & 2.468 & 0.091 & 0.085 & 0.956 
                      & 9.069 & 0.132 & 0.121 & 0.936 \\
              & 1.0 & 7.999 & 0.102 & 0.086 & 0.935 
                      & 5.676 & 0.066 & 0.067 & 0.966 
                      & 11.370 & 0.127 & 0.122 & 0.932 \\
marginal-o0c0 & 0.1 & 11.723 & 0.267 & 0.179 & 0.967 
                      & 14.645 & 0.183 & 0.173 & 0.792 
                      & 36.903 & 0.189 & 0.200 & 0.827 \\
              & 0.5 & 18.160 & 0.298 & 0.279 & 0.963 
                      & 51.925 & 0.245 & 0.244 & 0.876 
                      & 38.187 & 0.275 & 0.326 & 0.873 \\
              & 1.0 & 22.601 & 0.306 & 0.338 & 0.955 
                      & 86.482 & 0.257 & 0.233 & 0.893 
                      & 38.004 & 0.300 & 0.372 & 0.866 \\
\midrule
\multicolumn{14}{@{}c}{\textbf{\textit{Doubly robust estimator (Frailty Cox)}}}\\
frailty-o1c1 & 0.1 & 0.605 & 0.021 & 0.019 & 0.963 
                      & 0.679 & 0.043 & 0.044 & 0.962 
                      & 0.458 & 0.047 & 0.049 & 0.970 \\
              & 0.5 & 0.602 & 0.037 & 0.042 & 0.964 
                      & 0.182 & 0.043 & 0.045 & 0.957 
                      & 1.065 & 0.056 & 0.064 & 0.970 \\
              & 1.0 & 0.276 & 0.046 & 0.054 & 0.971 
                      & 1.553 & 0.035 & 0.037 & 0.956 
                      & 0.859 & 0.060 & 0.068 & 0.966 \\
frailty-o1c0 & 0.1 & 0.358 & 0.014 & 0.017 & 0.954 
                      & 0.946 & 0.039 & 0.040 & 0.958 
                      & 0.709 & 0.042 & 0.044 & 0.967 \\
              & 0.5 & 0.417 & 0.034 & 0.040 & 0.961 
                      & 0.108 & 0.042 & 0.043 & 0.952 
                      & 0.599 & 0.054 & 0.060 & 0.963 \\
              & 1.0 & 0.181 & 0.044 & 0.052 & 0.969 
                      & 1.961 & 0.035 & 0.037 & 0.958 
                      & 0.924 & 0.057 & 0.066 & 0.969 \\
frailty-o0c1 & 0.1 & 0.691 & 0.020 & 0.015 & 0.847 
                      & 5.950 & 0.052 & 0.053 & 0.861 
                      & 8.833 & 0.055 & 0.056 & 0.902 \\
              & 0.5 & 2.622 & 0.037 & 0.038 & 0.864 
                      & 15.728 & 0.056 & 0.060 & 0.887 
                      & 5.099 & 0.067 & 0.072 & 0.957 \\
              & 1.0 & 3.433 & 0.047 & 0.053 & 0.910 
                      & 27.335 & 0.051 & 0.055 & 0.870 
                      & 4.782 & 0.070 & 0.078 & 0.959 \\
frailty-o0c0 & 0.1 & 1.269 & 0.011 & 0.012 & 0.746 
                      & 10.729 & 0.041 & 0.045 & 0.683 
                      & 15.879 & 0.042 & 0.046 & 0.780 \\
              & 0.5 & 3.123 & 0.031 & 0.034 & 0.835 
                      & 27.408 & 0.049 & 0.055 & 0.705 
                      & 11.220 & 0.058 & 0.064 & 0.885 \\
              & 1.0 & 3.614 & 0.047 & 0.051 & 0.870 
                      & 45.334 & 0.047 & 0.053 & 0.676 
                      & 10.857 & 0.066 & 0.073 & 0.911 \\
\midrule
\multicolumn{14}{@{}c}{\textbf{\textit{Outcome regression \& KM}}}\\
marginal-OR1  & 0.1 & 0.658 & 0.018 & 0.021 & 0.950 
                      & 1.727 & 0.039 & 0.042 & 0.953 
                      & 1.281 & 0.043 & 0.047 & 0.960 \\
              & 0.5 & 1.802 & 0.044 & 0.050 & 0.952 
                      & 2.287 & 0.040 & 0.044 & 0.946 
                      & 1.515 & 0.059 & 0.067 & 0.967 \\
              & 1.0 & 2.655 & 0.057 & 0.065 & 0.953 
                      & 1.961 & 0.032 & 0.036 & 0.951 
                      & 2.895 & 0.066 & 0.076 & 0.956 \\
marginal-OR0  & 0.1 & 1.049 & 0.012 & 0.013 & 0.805 
                      & 8.755 & 0.050 & 0.053 & 0.798 
                      & 12.926 & 0.050 & 0.054 & 0.859 \\
              & 0.5 & 1.470 & 0.040 & 0.044 & 0.913 
                      & 15.628 & 0.055 & 0.059 & 0.914 
                      & 6.892 & 0.066 & 0.074 & 0.962 \\
              & 1.0 & 0.032 & 0.059 & 0.066 & 0.951 
                      & 24.010 & 0.047 & 0.051 & 0.924 
                      & 8.371 & 0.074 & 0.084 & 0.961 \\
frailty-OR1   & 0.1 & 0.291 & 0.014 & 0.016 & 0.952 
                      & 0.086 & 0.037 & 0.039 & 0.960 
                      & 0.974 & 0.040 & 0.043 & 0.968 \\
              & 0.5 & 0.005 & 0.033 & 0.038 & 0.951 
                      & 4.621 & 0.041 & 0.043 & 0.961 
                      & 2.722 & 0.052 & 0.058 & 0.967 \\
              & 1.0 & 0.564 & 0.044 & 0.050 & 0.952 
                      & 9.841 & 0.035 & 0.037 & 0.964 
                      & 2.654 & 0.056 & 0.063 & 0.961 \\
frailty-OR0   & 0.1 & 1.357 & 0.010 & 0.011 & 0.716 
                      & 11.869 & 0.040 & 0.043 & 0.616 
                      & 17.706 & 0.040 & 0.044 & 0.737 \\
              & 0.5 & 3.755 & 0.030 & 0.033 & 0.816 
                      & 35.411 & 0.049 & 0.053 & 0.492 
                      & 14.941 & 0.055 & 0.063 & 0.825 \\
              & 1.0 & 4.813 & 0.044 & 0.050 & 0.878 
                      & 62.895 & 0.050 & 0.054 & 0.417 
                      & 15.333 & 0.064 & 0.074 & 0.863 \\
KM            & 0.1 & 2.281 & 0.008 & 0.008 & 0.320 
                      & 24.301 & 0.041 & 0.042 & 0.138 
                      & 37.651 & 0.042 & 0.043 & 0.235 \\
              & 0.5 & 6.676 & 0.025 & 0.025 & 0.407 
                      & 58.917 & 0.059 & 0.063 & 0.174 
                      & 24.177 & 0.064 & 0.068 & 0.555 \\
              & 1.0 & 8.786 & 0.038 & 0.039 & 0.579 
                      & 86.730 & 0.059 & 0.062 & 0.218 
                      & 18.250 & 0.071 & 0.074 & 0.751 \\
\bottomrule
\end{tabular}
}
\end{table}

\begin{table}[ht!]
\centering
\caption{Simulation results for cluster-level RMST \(\mu_C^{(a)}(t)\) and causal effects \(\Delta_C^{\text{RMST}}(t)\) on different scales under Scenario 3, estimated under 13 methods at time points \( t = \{0.1, 0.5, 1\} \). The number of clusters is \( M = 50 \), with censoring rate of approximately \(75\%\). Reported metrics include PBias (percentage bias), MCSD (Monte Carlo standard deviation), AESE (average estimated standard error from the Jackknife procedure), and CP (empirical coverage probability of the \(95\%\) confidence interval).}
\label{M50_C75_ICS_E1_C1_rmstc}
\resizebox{\textwidth}{!}{
\begin{tabular}{@{}l r 
  rrrr   rrrr   rrrr@{}}
\toprule
Method & \(t\) 
  & \multicolumn{4}{c}{$\mu_C^{(1)}(t)$} 
  & \multicolumn{4}{c}{$\mu_C^{(0)}(t)$} 
  & \multicolumn{4}{c}{$\Delta_C^{\mathrm{RMST}}(t)$} \\
\cmidrule(lr){3-6} \cmidrule(lr){7-10} \cmidrule(lr){11-14}
 & & PBias & MCSD & AESE & CP 
     & PBias & MCSD & AESE & CP 
     & PBias & MCSD & AESE & CP \\
\midrule
\multicolumn{14}{@{}c}{\textbf{\textit{Doubly robust estimator (Marginal Cox)}}}\\
marginal-o1c1 & 0.1 & 1.566 & 0.004 & 0.002 & 0.956 
                      & 1.173 & 0.004 & 0.004 & 0.971 
                      & 1.777 & 0.005 & 0.004 & 0.962 \\
              & 0.5 & 3.071 & 0.030 & 0.020 & 0.936 
                      & 2.003 & 0.025 & 0.026 & 0.970 
                      & 3.718 & 0.033 & 0.033 & 0.958 \\
              & 1.0 & 3.983 & 0.069 & 0.051 & 0.934 
                      & 2.259 & 0.050 & 0.053 & 0.969 
                      & 4.841 & 0.075 & 0.076 & 0.960 \\
marginal-o1c0 & 0.1 & 8.685 & 0.020 & 0.012 & 0.976 
                      & 9.836 & 0.019 & 0.014 & 0.975 
                      & 28.261 & 0.021 & 0.017 & 0.972 \\
              & 0.5 & 14.333 & 0.123 & 0.101 & 0.968 
                      & 10.175 & 0.097 & 0.100 & 0.972 
                      & 7.010 & 0.122 & 0.139 & 0.949 \\
              & 1.0 & 16.573 & 0.259 & 0.249 & 0.963 
                      & 6.740 & 0.196 & 0.210 & 0.973 
                      & 1.460 & 0.259 & 0.322 & 0.938 \\
marginal-o0c1 & 0.1 & 1.480 & 0.005 & 0.002 & 0.971 
                      & 0.270 & 0.006 & 0.005 & 0.923 
                      & 7.545 & 0.006 & 0.005 & 0.919 \\
              & 0.5 & 2.472 & 0.035 & 0.021 & 0.972 
                      & 0.142 & 0.035 & 0.033 & 0.954 
                      & 5.369 & 0.042 & 0.040 & 0.948 \\
              & 1.0 & 3.407 & 0.078 & 0.053 & 0.968 
                      & 0.183 & 0.070 & 0.069 & 0.958 
                      & 5.983 & 0.092 & 0.090 & 0.955 \\
marginal-o0c0 & 0.1 & 8.298 & 0.021 & 0.012 & 0.968 
                      & 4.019 & 0.015 & 0.012 & 0.782 
                      & 28.982 & 0.015 & 0.012 & 0.763 \\
              & 0.5 & 14.644 & 0.126 & 0.104 & 0.970 
                      & 19.598 & 0.083 & 0.087 & 0.863 
                      & 40.679 & 0.082 & 0.097 & 0.811 \\
              & 1.0 & 18.376 & 0.268 & 0.258 & 0.954 
                      & 32.171 & 0.194 & 0.201 & 0.874 
                      & 40.827 & 0.200 & 0.239 & 0.807 \\
\midrule
\multicolumn{14}{@{}c}{\textbf{\textit{Doubly robust estimator (Frailty Cox)}}}\\
frailty-o1c1 & 0.1 & 0.275 & 0.001 & 0.001 & 0.966 
                      & 0.650 & 0.003 & 0.003 & 0.960 
                      & 1.587 & 0.003 & 0.003 & 0.971 \\
              & 0.5 & 0.218 & 0.011 & 0.012 & 0.976 
                      & 0.788 & 0.019 & 0.020 & 0.959 
                      & 0.857 & 0.022 & 0.024 & 0.967 \\
              & 1.0 & 0.022 & 0.028 & 0.032 & 0.975 
                      & 0.550 & 0.039 & 0.042 & 0.959 
                      & 0.670 & 0.048 & 0.054 & 0.968 \\
frailty-o1c0 & 0.1 & 0.108 & 0.001 & 0.001 & 0.965 
                      & 0.652 & 0.003 & 0.003 & 0.964 
                      & 2.476 & 0.003 & 0.003 & 0.968 \\
              & 0.5 & 0.056 & 0.010 & 0.011 & 0.967 
                      & 0.754 & 0.019 & 0.020 & 0.967 
                      & 1.248 & 0.021 & 0.023 & 0.963 \\
              & 1.0 & 0.120 & 0.028 & 0.032 & 0.968 
                      & 0.418 & 0.041 & 0.043 & 0.969 
                      & 0.806 & 0.050 & 0.055 & 0.966 \\
frailty-o0c1 & 0.1 & 0.211 & 0.001 & 0.001 & 0.870 
                      & 1.931 & 0.003 & 0.003 & 0.881 
                      & 7.883 & 0.003 & 0.003 & 0.907 \\
              & 0.5 & 1.314 & 0.011 & 0.011 & 0.868 
                      & 3.876 & 0.022 & 0.024 & 0.915 
                      & 3.443 & 0.024 & 0.026 & 0.958 \\
              & 1.0 & 2.056 & 0.029 & 0.031 & 0.888 
                      & 5.735 & 0.048 & 0.052 & 0.914 
                      & 2.567 & 0.056 & 0.061 & 0.964 \\
frailty-o0c0 & 0.1 & 0.538 & 0.001 & 0.001 & 0.818 
                      & 4.444 & 0.003 & 0.003 & 0.728 
                      & 17.976 & 0.003 & 0.003 & 0.795 \\
              & 0.5 & 1.546 & 0.010 & 0.010 & 0.838 
                      & 9.243 & 0.021 & 0.024 & 0.791 
                      & 12.777 & 0.023 & 0.026 & 0.895 \\
              & 1.0 & 1.999 & 0.029 & 0.031 & 0.882 
                      & 13.318 & 0.048 & 0.054 & 0.800 
                      & 12.235 & 0.055 & 0.061 & 0.918 \\
\midrule
\multicolumn{14}{@{}c}{\textbf{\textit{Outcome regression \& KM}}}\\
marginal-OR1  & 0.1 & 0.191 & 0.001 & 0.001 & 0.961 
                      & 1.007 & 0.003 & 0.003 & 0.959 
                      & 3.677 & 0.003 & 0.003 & 0.963 \\
              & 0.5 & 0.623 & 0.011 & 0.013 & 0.959 
                      & 1.873 & 0.018 & 0.020 & 0.958 
                      & 1.706 & 0.021 & 0.024 & 0.961 \\
              & 1.0 & 0.987 & 0.030 & 0.035 & 0.957 
                      & 2.250 & 0.038 & 0.042 & 0.955 
                      & 0.608 & 0.049 & 0.055 & 0.961 \\
marginal-OR0  & 0.1 & 0.800 & 0.001 & 0.001 & 0.662 
                      & 1.096 & 0.003 & 0.003 & 0.911 
                      & 0.604 & 0.003 & 0.003 & 0.929 \\
              & 0.5 & 2.164 & 0.010 & 0.010 & 0.777 
                      & 1.698 & 0.023 & 0.025 & 0.948 
                      & 9.357 & 0.025 & 0.027 & 0.929 \\
              & 1.0 & 2.560 & 0.030 & 0.033 & 0.877 
                      & 3.162 & 0.047 & 0.051 & 0.937 
                      & 9.783 & 0.054 & 0.060 & 0.912 \\
frailty-OR1   & 0.1 & 0.106 & 0.001 & 0.001 & 0.963 
                      & 0.159 & 0.002 & 0.003 & 0.963 
                      & 0.148 & 0.003 & 0.003 & 0.969 \\
              & 0.5 & 0.062 & 0.009 & 0.010 & 0.959 
                      & 0.942 & 0.018 & 0.020 & 0.961 
                      & 1.577 & 0.020 & 0.022 & 0.971 \\
              & 1.0 & 0.386 & 0.025 & 0.028 & 0.963 
                      & 2.231 & 0.040 & 0.042 & 0.967 
                      & 1.943 & 0.046 & 0.051 & 0.970 \\
frailty-OR0   & 0.1 & 0.886 & 0.001 & 0.001 & 0.602 
                      & 2.162 & 0.003 & 0.003 & 0.877 
                      & 5.162 & 0.003 & 0.003 & 0.927 \\
              & 0.5 & 2.999 & 0.008 & 0.009 & 0.621 
                      & 4.775 & 0.019 & 0.021 & 0.910 
                      & 0.310 & 0.020 & 0.023 & 0.971 \\
              & 1.0 & 4.486 & 0.024 & 0.027 & 0.688 
                      & 8.476 & 0.043 & 0.046 & 0.884 
                      & 0.551 & 0.047 & 0.053 & 0.971 \\
KM            & 0.1 & 0.975 & 0.001 & 0.001 & 0.545 
                      & 10.032 & 0.002 & 0.002 & 0.148 
                      & 41.981 & 0.002 & 0.002 & 0.235 \\
              & 0.5 & 3.287 & 0.008 & 0.008 & 0.539 
                      & 23.108 & 0.022 & 0.022 & 0.198 
                      & 33.629 & 0.023 & 0.024 & 0.404 \\
              & 1.0 & 4.763 & 0.025 & 0.026 & 0.626 
                      & 32.066 & 0.052 & 0.054 & 0.245 
                      & 29.700 & 0.058 & 0.060 & 0.543 \\
\bottomrule
\end{tabular}
}
\end{table}

\begin{table}[ht!]
\centering
\caption{Simulation results for individual-level RMST \(\mu_I^{(a)}(t)\) and causal effects \(\Delta_I^{\text{RMST}}(t)\) on different scales under Scenario 3, estimated under 13 methods at time points \( t = \{0.1, 0.5, 1\} \). The number of clusters is \( M = 50 \), with censoring rate of approximately \(75\%\). Reported metrics include PBias (percentage bias), MCSD (Monte Carlo standard deviation), AESE (average estimated standard error from the jackknife procedure), and CP (empirical coverage probability of the \(95\%\) confidence interval).}
\label{M50_C75_ICS_E1_C1_rmsti}
\resizebox{\textwidth}{!}{
\begin{tabular}{@{}l r 
  rrrr   rrrr   rrrr@{}}
\toprule
Method & \(t\) 
  & \multicolumn{4}{c}{$\mu_I^{(1)}(t)$} 
  & \multicolumn{4}{c}{$\mu_I^{(0)}(t)$} 
  & \multicolumn{4}{c}{$\Delta_I^{\mathrm{RMST}}(t)$} \\
\cmidrule(lr){3-6} \cmidrule(lr){7-10} \cmidrule(lr){11-14}
 & & PBias & MCSD & AESE & CP 
     & PBias & MCSD & AESE & CP 
     & PBias & MCSD & AESE & CP \\
\midrule
\multicolumn{14}{@{}c}{\textbf{\textit{Doubly robust estimator (Marginal Cox)}}}\\
marginal-o1c1 & 0.1 & 2.350 & 0.005 & 0.002 & 0.939 
                      & 1.490 & 0.005 & 0.004 & 0.965 
                      & 3.738 & 0.006 & 0.005 & 0.954 \\
              & 0.5 & 4.699 & 0.034 & 0.023 & 0.906 
                      & 2.568 & 0.027 & 0.028 & 0.968 
                      & 6.198 & 0.038 & 0.038 & 0.956 \\
              & 1.0 & 6.126 & 0.076 & 0.059 & 0.900 
                      & 2.980 & 0.049 & 0.053 & 0.969 
                      & 7.569 & 0.081 & 0.082 & 0.963 \\
marginal-o1c0 & 0.1 & 9.674 & 0.021 & 0.013 & 0.980 
                      & 10.984 & 0.019 & 0.014 & 0.978 
                      & 15.961 & 0.021 & 0.018 & 0.978 \\
              & 0.5 & 15.451 & 0.122 & 0.101 & 0.968 
                      & 11.052 & 0.086 & 0.087 & 0.972 
                      & 3.773 & 0.118 & 0.135 & 0.957 \\
              & 1.0 & 18.119 & 0.250 & 0.238 & 0.959 
                      & 7.688 & 0.168 & 0.170 & 0.972 
                      & 9.436 & 0.248 & 0.303 & 0.948 \\
marginal-o0c1 & 0.1 & 2.174 & 0.006 & 0.003 & 0.977 
                      & 1.212 & 0.007 & 0.006 & 0.880 
                      & 10.999 & 0.008 & 0.007 & 0.860 \\
              & 0.5 & 3.901 & 0.040 & 0.024 & 0.963 
                      & 1.806 & 0.038 & 0.037 & 0.942 
                      & 9.288 & 0.050 & 0.047 & 0.931 \\
              & 1.0 & 5.321 & 0.085 & 0.061 & 0.947 
                      & 2.414 & 0.071 & 0.071 & 0.952 
                      & 9.851 & 0.105 & 0.101 & 0.935 \\
marginal-o0c0 & 0.1 & 8.820 & 0.021 & 0.013 & 0.962 
                      & 6.625 & 0.016 & 0.013 & 0.727 
                      & 31.353 & 0.016 & 0.015 & 0.746 \\
              & 0.5 & 13.960 & 0.125 & 0.101 & 0.967 
                      & 24.856 & 0.084 & 0.088 & 0.858 
                      & 37.680 & 0.089 & 0.107 & 0.862 \\
              & 1.0 & 17.081 & 0.258 & 0.241 & 0.960 
                      & 39.357 & 0.187 & 0.188 & 0.884 
                      & 37.991 & 0.206 & 0.251 & 0.857 \\
\midrule
\multicolumn{14}{@{}c}{\textbf{\textit{Doubly robust estimator (Frailty Cox)}}}\\
frailty-o1c1 & 0.1 & 0.443 & 0.002 & 0.001 & 0.963 
                      & 0.624 & 0.003 & 0.003 & 0.958 
                      & 0.213 & 0.003 & 0.004 & 0.964 \\
              & 0.5 & 0.598 & 0.013 & 0.014 & 0.965 
                      & 0.574 & 0.019 & 0.021 & 0.960 
                      & 0.611 & 0.023 & 0.026 & 0.972 \\
              & 1.0 & 0.530 & 0.033 & 0.037 & 0.963 
                      & 0.130 & 0.037 & 0.040 & 0.958 
                      & 0.813 & 0.049 & 0.056 & 0.960 \\
frailty-o1c0 & 0.1 & 0.214 & 0.001 & 0.001 & 0.956 
                      & 0.715 & 0.003 & 0.003 & 0.955 
                      & 1.375 & 0.003 & 0.003 & 0.961 \\
              & 0.5 & 0.372 & 0.011 & 0.013 & 0.962 
                      & 0.705 & 0.019 & 0.020 & 0.958 
                      & 0.019 & 0.022 & 0.024 & 0.967 \\
              & 1.0 & 0.337 & 0.030 & 0.035 & 0.965 
                      & 0.189 & 0.037 & 0.039 & 0.956 
                      & 0.447 & 0.048 & 0.054 & 0.963 \\
frailty-o0c1 & 0.1 & 0.248 & 0.002 & 0.001 & 0.859 
                      & 3.898 & 0.004 & 0.004 & 0.816 
                      & 11.212 & 0.004 & 0.004 & 0.858 \\
              & 0.5 & 1.485 & 0.013 & 0.012 & 0.848 
                      & 8.378 & 0.024 & 0.026 & 0.879 
                      & 6.556 & 0.027 & 0.029 & 0.939 \\
              & 1.0 & 2.222 & 0.032 & 0.034 & 0.874 
                      & 12.595 & 0.049 & 0.053 & 0.874 
                      & 5.595 & 0.058 & 0.064 & 0.950 \\
frailty-o0c0 & 0.1 & 0.691 & 0.001 & 0.001 & 0.743 
                      & 7.219 & 0.003 & 0.003 & 0.606 
                      & 19.932 & 0.003 & 0.003 & 0.688 \\
              & 0.5 & 2.007 & 0.009 & 0.010 & 0.798 
                      & 14.917 & 0.021 & 0.024 & 0.682 
                      & 13.100 & 0.023 & 0.025 & 0.837 \\
              & 1.0 & 2.668 & 0.028 & 0.030 & 0.848 
                      & 21.925 & 0.044 & 0.049 & 0.672 
                      & 11.919 & 0.052 & 0.057 & 0.865 \\
\midrule
\multicolumn{14}{@{}c}{\textbf{\textit{Outcome regression \& KM}}}\\
marginal-OR1  & 0.1 & 0.356 & 0.001 & 0.001 & 0.958 
                      & 1.141 & 0.003 & 0.003 & 0.954 
                      & 2.124 & 0.003 & 0.003 & 0.960 \\
              & 0.5 & 1.094 & 0.014 & 0.016 & 0.953 
                      & 1.855 & 0.019 & 0.021 & 0.953 
                      & 0.204 & 0.024 & 0.027 & 0.964 \\
              & 1.0 & 1.651 & 0.039 & 0.045 & 0.947 
                      & 1.989 & 0.036 & 0.040 & 0.949 
                      & 1.395 & 0.054 & 0.061 & 0.967 \\
marginal-OR0  & 0.1 & 0.597 & 0.001 & 0.001 & 0.763 
                      & 6.597 & 0.004 & 0.004 & 0.699 
                      & 18.369 & 0.004 & 0.004 & 0.748 \\
              & 0.5 & 1.313 & 0.011 & 0.013 & 0.882 
                      & 10.171 & 0.025 & 0.027 & 0.856 
                      & 9.053 & 0.027 & 0.030 & 0.926 \\
              & 1.0 & 1.054 & 0.036 & 0.040 & 0.927 
                      & 13.268 & 0.050 & 0.054 & 0.892 
                      & 8.200 & 0.060 & 0.067 & 0.957 \\
frailty-OR1   & 0.1 & 0.192 & 0.001 & 0.001 & 0.962 
                      & 0.233 & 0.003 & 0.003 & 0.956 
                      & 0.065 & 0.003 & 0.003 & 0.967 \\
              & 0.5 & 0.194 & 0.011 & 0.012 & 0.958 
                      & 1.347 & 0.019 & 0.020 & 0.964 
                      & 1.999 & 0.022 & 0.024 & 0.968 \\
              & 1.0 & 0.006 & 0.030 & 0.034 & 0.956 
                      & 3.235 & 0.037 & 0.039 & 0.962 
                      & 2.440 & 0.047 & 0.053 & 0.962 \\
frailty-OR0   & 0.1 & 0.727 & 0.001 & 0.001 & 0.698 
                      & 7.624 & 0.003 & 0.003 & 0.576 
                      & 21.074 & 0.003 & 0.003 & 0.653 \\
              & 0.5 & 2.317 & 0.009 & 0.010 & 0.770 
                      & 17.926 & 0.021 & 0.022 & 0.540 
                      & 15.952 & 0.022 & 0.024 & 0.774 \\
              & 1.0 & 3.250 & 0.027 & 0.030 & 0.827 
                      & 28.000 & 0.044 & 0.048 & 0.472 
                      & 15.501 & 0.050 & 0.057 & 0.801 \\
KM            & 0.1 & 1.205 & 0.000 & 0.000 & 0.322 
                      & 14.829 & 0.003 & 0.003 & 0.113 
                      & 41.852 & 0.003 & 0.003 & 0.178 \\
              & 0.5 & 4.029 & 0.007 & 0.007 & 0.340 
                      & 33.374 & 0.024 & 0.025 & 0.140 
                      & 30.317 & 0.025 & 0.026 & 0.347 \\
              & 1.0 & 5.827 & 0.023 & 0.023 & 0.432 
                      & 46.403 & 0.053 & 0.056 & 0.149 
                      & 24.914 & 0.058 & 0.061 & 0.512 \\
\bottomrule
\end{tabular}
}
\end{table}

\clearpage
\section{Example Code for the R Package}

We introduce the \texttt{DRsurvCRT} package at \url{https://github.com/fancy575/DRsurvCRT}, which provides the implementation of our proposed method. The main function, \texttt{DRsurvfit()}, implements two estimands: the survival probability causal estimand (SPCE), and the restricted mean survival time (RMST). To quantify uncertainty, the package offers leave-one-cluster-out jackknife variance estimation. This enables construction of confidence intervals based on the \(t\)-distribution with degrees of freedom \(\mathrm{df} = M - 1\).

The function has the following declaration:
\begin{verbatim}
DRsurvfit <- function(data,
                      formula,
                      cens_formula = NULL,
                      intv,
                      method   = c("marginal", "frailty"),
                      estimand = c("SPCE", "RMST"),
                      tau = NULL,
                      trt_prob = NULL,
                      variance = c("none","jackknife"),
                      fit_controls = NULL,
                      strata = NULL)
\end{verbatim}
with arguments:
\begin{itemize}
  \item \texttt{data}: A \texttt{data.frame} of individual-level observations.
  \item \texttt{formula}: Outcome model of the form \texttt{Surv(time, event) \textasciitilde\ covariates + cluster(clusterID)}.
  \item \texttt{cens\_formula}: Optional censoring model. If \texttt{NULL}, the RHS of \texttt{formula} is reused.
        \texttt{Surv(time, event == 0)}.
  \item \texttt{intv}: Name of the cluster-level treatment column (0/1), constant within cluster.
  \item \texttt{method}: \texttt{``marginal''} (Cox PH via \texttt{survival::coxph}) or \texttt{``frailty''} (gamma frailty via \texttt{frailtyEM::emfrail}).
  \item \texttt{estimand}: \texttt{``SPCE''} (survival probability causal effects) or \texttt{``RMST''} (restricted mean survival time at target times \(\tau\)).
  \item \texttt{tau}: Numeric vector of RMST target times; required if \texttt{estimand = ``RMST''}.
 computed from the cluster-level treatment. If NULL, the \(25\%\), \(50\%\), and \(75\%\) quantiles of times will be reported.
  \item \texttt{variance}: \texttt{``none''} or \texttt{``jackknife''} for leave-one-cluster-out (LOCO) standard errors.
  \item \texttt{fit\_controls}: Optional \texttt{frailtyEM::emfrail\_control()} list (used only for \texttt{method = ``frailty''}).
\end{itemize}

\subsection*{Example: SPCE with CIs from jackknife variance}

We demonstrate \texttt{DRsurvfit()} using the example dataset \texttt{dat}. We fit marginal Cox models for both outcome and censoring, compute LOCO jackknife variances, and summarize SPCE at \(t \in \{0.5, 1.0, 1.5, 2.0\}\) with two-sided $t$-intervals (\(\alpha=0.05\), \(\mathrm{df}=M-1\)).

\begin{verbatim}
library(DRsurvCRT)
data(dat)

fit_spce <- DRsurvfit(
  data     = dat,
  formula  = Surv(time, event) ~ W1 + W2 + Z1 + Z2 + cluster(M),
  intv     = "A",
  method   = "marginal",
  estimand = "SPCE",
  variance = "jackknife"
)

summary(fit_spce, digits = 3, times = c(0.5, 1, 1.5, 2))
\end{verbatim}

Example output:
\begin{verbatim}
DRsurvfit: method = marginal, estimand = SPCE
Treatment probs (p0, p1): 0.5, 0.5
Outcome model:   Surv(time, event) ~ W1 + W2 + Z1 + Z2 + cluster(M)
Censoring model: Surv(time, event == 0) ~ W1 + W2 + Z1 + Z2 + cluster(M)
Cluster id col:  M
Clusters (M):    50
Obs (N):         2495

Cluster-level SPCE:
        S1 (LCL, UCL)        S0 (LCL, UCL)        S1-S0 (LCL, UCL)
t=0.499 0.659 (0.569, 0.748) 0.793 (0.718, 0.868) -0.134 (-0.228, -0.0404)
t=   1  0.544 (0.448, 0.64)  0.711 (0.612, 0.809) -0.167 (-0.272, -0.0612)
t=1.49  0.461 (0.362, 0.56)  0.651 (0.552, 0.751) -0.19 (-0.296, -0.0842)
t=   2  0.416 (0.313, 0.518) 0.618 (0.513, 0.723) -0.202 (-0.31, -0.0945)
  t-intervals with df = 49, alpha = 0.050

Individual-level SPCE:
        S1 (LCL, UCL)        S0 (LCL, UCL)        S1-S0 (LCL, UCL)
t=0.499 0.645 (0.54, 0.749)  0.762 (0.68, 0.845)  -0.117 (-0.225, -0.00995)
t=   1  0.526 (0.417, 0.634) 0.661 (0.554, 0.768) -0.135 (-0.25, -0.0195)
t=1.49  0.444 (0.337, 0.551) 0.604 (0.489, 0.719) -0.16 (-0.276, -0.0437)
t=   2   0.4 (0.293, 0.507)  0.566 (0.447, 0.685) -0.166 (-0.278, -0.0531)
  t-intervals with df = 49, alpha = 0.050
\end{verbatim}

\end{document}